\documentclass[10pt,twocolumn,letterpaper]{article}

\usepackage{cvpr}      % To produce the REVIEW version
\usepackage{amsmath}
\usepackage{amsbsy}
\usepackage{amssymb}
\usepackage{amsfonts}
\usepackage{color}

\renewcommand{\vec}[1]{\ensuremath{\boldsymbol{#1}}}

\providecommand{\mI}{\mathbf{I}}

\providecommand{\vc}{\mathbf{c}}

\providecommand{\vh}{\mathbf{h}}

\providecommand{\vm}{\mathbf{m}}

\providecommand{\vv}{\mathbf{v}}

\providecommand{\vx}{\mathbf{x}}
\providecommand{\vy}{\mathbf{y}}
\providecommand{\vz}{\mathbf{z}}

\providecommand{\vepsilon}{\vec{\epsilon}}

\providecommand{\vtheta}{\vec{\theta}}

\providecommand{\vphi}{\vec{\phi}}

\providecommand{\vomega}{\vec{\omega}}

\usepackage{graphicx}
\usepackage{amssymb}
\usepackage{booktabs}
\usepackage{array}
\usepackage{rotating}
\usepackage{caption}
\usepackage{subcaption}
\usepackage{svg}
\usepackage{float}

\usepackage{multirow}
\usepackage{multicol}
\usepackage{algorithm}
\usepackage{algpseudocode}

\usepackage{tikz}
\usetikzlibrary{spy}
\usetikzlibrary{arrows.meta}
\definecolor{cvprblue}{rgb}{0.21,0.49,0.74}
\usepackage{hyperref}
\hypersetup{breaklinks,colorlinks,allcolors=cvprblue}

\usepackage[capitalize]{cleveref}
\crefname{section}{Sec.}{Secs.}
\Crefname{section}{Section}{Sections}
\Crefname{table}{Table}{Tables}
\crefname{table}{Tab.}{Tabs.}

\newcommand{\RB}[1]{\colorbox{red!20}{#1}}
\newcommand{\BB}[1]{\colorbox{magenta!15}{#1}}
\newcommand{\TB}[1]{\colorbox{yellow!10}{#1}}

\def\paperID{24} % *** Enter the Paper ID here
\def\confName{CVPR}
\def\confYear{2026}

\begin{document}

% ---------------------------------------------------------------
\title{GenMFSR: Generative Multi-Frame Image Restoration and Super-Resolution}

\author{
Harshana Weligampola$^{1,2}$, Joshua Peter Ebenezer$^1$, Weidi Liu$^1$, Abhinau K. Venkataramanan$^1$,\\
Sreenithy Chandran$^1$, Seok-Jun Lee$^1$, and Hamid Rahim Sheikh$^1$\\
$^1$Samsung Research America, $^2$Purdue University
\and
{\tt\small wweligam@purdue.edu}
}
\twocolumn[{%
    \renewcommand\twocolumn[1][]{#1}%
    \maketitle
    \begin{center}
        \captionsetup{type=figure}
        \includegraphics[width=0.9\textwidth, trim={0 0 0 0}, clip]{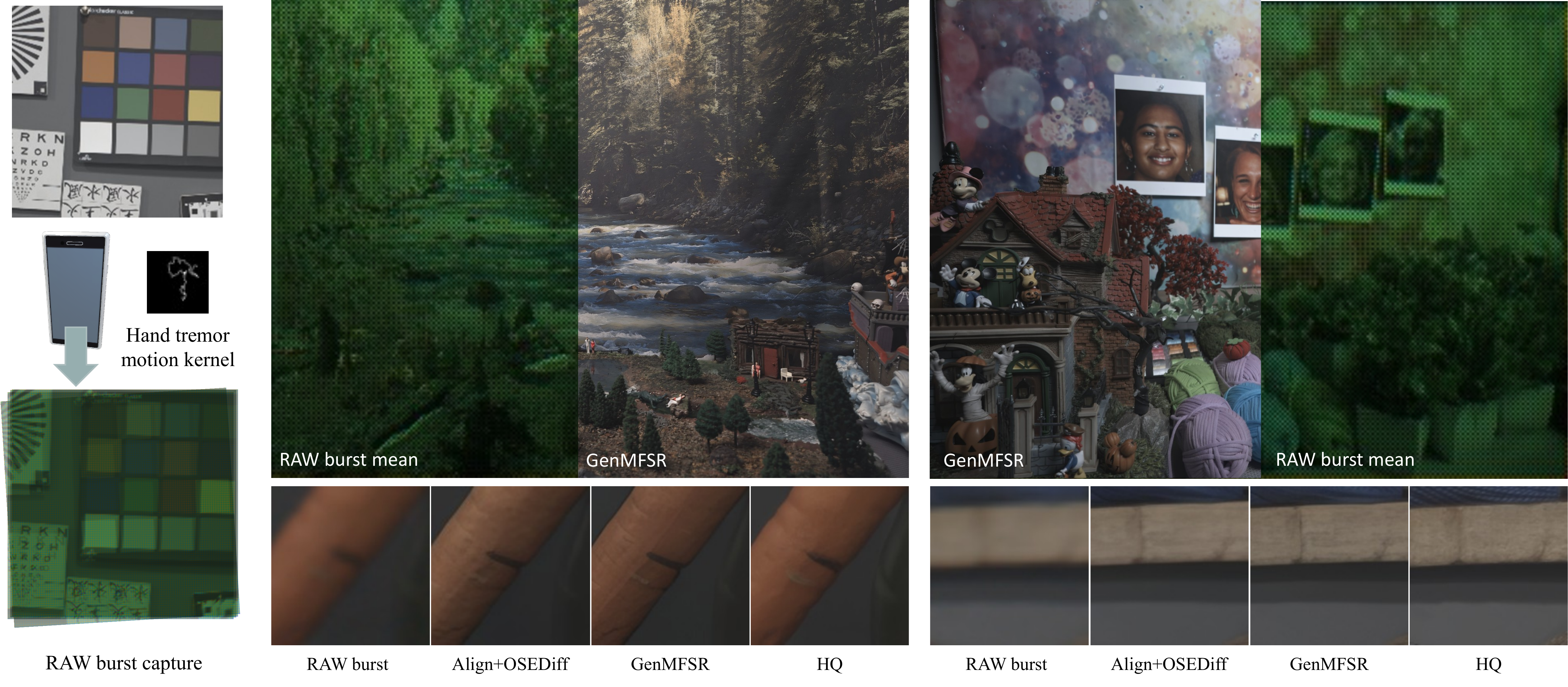}
        \caption{GenMFSR: Superior high-frequency feature generation from multi-frame RAW images via a unified denoising, demosaicing, registration, and super-resolution model in a single-step.}
        \label{fig:teaser}
    \end{center}

}]

\begin{abstract}
Camera pipelines receive raw Bayer-format frames that need to be denoised, demosaiced, and often super-resolved. Multiple frames are captured to utilize natural hand tremors and enhance resolution. Multi-frame super-resolution is therefore a fundamental problem in camera pipelines. Existing adversarial methods are constrained by the quality of ground truth. We propose \textbf{GenMFSR}, the first Generative Multi-Frame Raw-to-RGB Super Resolution pipeline, that incorporates image priors from foundation models to obtain sub-pixel information for camera ISP applications. GenMFSR can align multiple raw frames, unlike existing single-frame super-resolution methods, and we propose a loss term that restricts generation to high-frequency regions in the raw domain, thus preventing low-frequency artifacts.
\end{abstract}
\vspace{-12pt}
% \textcolor{red}{
% \begin{enumerate}
%     \item We need to focus more on the details of how single-frame RGB baselines and even how our method was adapted to handle multi-frame RAW inputs. These could just be specifying how we change the initial conv layers to take 33 channels as input etc but making a big deal of these changes and toning down explanation and use of prior art is necessary to drive home novelty points.
%     \item change fig 1 hightlight improvements
%     \item change fig 2 (make it simpler and visually pleasing)
%     \item The contributions need to be redone: We need to go for “This is the first Generative Raw MFSR pipeline” and keep hammering that point. This is more of a systems paper. The STN and VSD are how we achieve that.
%     \item add table of params and flops
%     \item show handmotion kernels in supp
% \end{enumerate}
% }
\section{Introduction}
\label{sec:intro}
Multi-frame super-resolution (MFSR) is the cornerstone of modern mobile computational photography, relying on natural hand tremors to recover sub-pixel details from noisy RAW bursts. Standard Image Signal Processors (ISPs) apply sequential, lossy modules (demosaicing, denoising) that destroy critical sensor-level information. While recent generative models (e.g., DarkDiff \cite{zheng2025darkdiff}, RDDM \cite{chen2025rddm}, ISPDiffuser \cite{ren2025ispdiffuser}) achieve unprecedented photorealism for single RAW images, integrating these priors into the temporally misaligned multi-frame RAW pipeline remains unsolved. Current multi-frame approaches\cite{farsiu2004fast,wronski2019handheld}  rely on deterministic regression, which converges to posterior means and produces over-smoothed, ``waxy'' textures rather than photorealistic details.

Generative diffusion priors can synthesize these missing high-frequency details. However, applying them to the RAW-to-RGB inverse problem introduces a critical failure mode: \textit{domain collision}. Standard priors are trained on highly processed, non-linear sRGB images. When forced to regularize linear, mosaiced sensor data, prior attempts to aggressively ``correct'' the low-frequency structure (e.g., global color and macro-geometry) already recoverable from the burst result in severe structural hallucinations. 

We develop a principled remedy for this collision. In burst reconstruction, low-frequency components are effectively determined by aligned sensor measurements, leaving only high-frequency textures underdetermined. Therefore, we treat the generative prior as an \textit{orthogonal score projection} via a High-Frequency Variational Score Distillation (HF-VSD) loss. By applying a frequency-domain mask to the score gradients, we restrict the prior’s corrective power strictly to the unobserved high-frequency latent subspace, successfully preventing hallucinations while synthesizing photorealistic textures.

To translate this theory into a practical mobile pipeline, we propose GenMFSR, an end-to-end, single-shot generative restoration framework. We embed Spatial Transformer Networks (STNs) directly within the multi-frame encoder for native sensor-domain alignment, bypassing massive external registration networks. By leveraging low-rank adaptation (LoRA) and one-step distillation, GenMFSR maintains the expressive power of a frozen pretrained decoder while reducing latency to a single forward pass.

In summary, GenMFSR selectively leverages a strong generative prior while strictly preserving the deterministic information present in aligned raw measurements. Our main contributions are:
\begin{itemize}
    \item \textbf{First Generative RAW MFSR Pipeline}: We demonstrate the first burst RAW to RGB diffusion model, which can handle unregistered RAW frames while leveraging the generative capabilities of foundation models.
    \item \textbf{Subspace-Projected Score Distillation (HF-VSD)}: We introduce a new diffusion loss function called subspace-projected score distillation that reduces low frequency hallucinations in the raw domain.
    \item \textbf{Native Generative Alignment \& Latent Bridging}: We introduce a new encoder that maps linear, unregistered, mosaic RAW data directly into the latent space of a pretrained diffusion model.
    \item \textbf{Empirical Handheld Motion Simulation}: We construct a rigorous paired burst dataset by extracting and applying real-world continuous hand-tremor trajectories to static RAW captures, ensuring physically accurate evaluation.
\end{itemize}

Extensive quantitative and qualitative experiments demonstrate that GenMFSR achieves state-of-the-art perceptual quality, outperforming existing deterministic and generative baselines by yielding highly realistic, hallucination-free textures.

\section{Related Work}
\textbf{Burst Super-Resolution.} Handheld burst photography relies on natural hand tremors to recover sub-pixel information \cite{farsiu2004fast,wronski2019handheld}. Recent learned methods, including Burstormer \cite{dudhane2023burstormer} and MFSR-GAN \cite{khan2025mfsr}, achieve state-of-the-art alignment and fusion but rely heavily on deterministic regression or unstable adversarial objectives, fundamentally limiting their ability to synthesize absent high-frequency details.

By contrast, diffusion models offer a generative alternative with strong perceptual fidelity. While they have shown promise in single-frame super-resolution \cite{lin2024diffbir,wu2024neurips_osediff,sun2025pixel,duan2025dit4sr}, their application to multi-frame RAW-to-RGB burst captures remains underexplored.

\textbf{Generative Priors in ISP.} Latent diffusion models \cite{rombach2022high} have revolutionized single-image restoration \cite{lin2024diffbir,wu2024neurips_osediff}. Recent industry frameworks (e.g., DarkDiff, RDDM, ISPDiffuser) successfully map single RAW captures to sRGB using diffusion priors. However, extending these priors to multi-frame bursts remains critically underexplored. Naively adapting single-frame models destroys temporal consistency, while standard distillation techniques like VSD \cite{wang2023prolificdreamer} suffer from domain collisions when applied to RAW sensor data. Our work bridges this gap by introducing the first generative multi-frame RAW architecture with subspace-restricted distillation.

\textbf{Model distillation:} has been used as a knowledge transfer technique to create compact, efficient ``student'' models that mimic the behaviour of a larger, more complex ``teacher'' model.
In the context of diffusion models \cite{rombach2021highresolution}, accelerated sampling and distillation have been used to drastically reduce the number of sampling steps from hundreds or thousands to just a few, enabling real-time generation \cite{song2021denoising,luhman2021knowledge,song2023consistency,meng2023distillation,poole2022dreamfusion,wu2024neurips_osediff}. While earlier works \cite{salimans2022progressive} focused on finding ``shortcuts'' to progressively distill the teacher model, recent work by Wu~\etal \cite{wu2024neurips_osediff} used a regularization approach that is conceptually aligned with GANs, using an adversarial-style objective like Variational Score Distillation (VSD) \cite{wang2023prolificdreamer} to enforce high-fidelity, plausible results.

\textbf{Noise conditional generative models:}
Typical conditioning mechanisms, such as ControlNets \cite{zhang2023controlnet}, and Adapters \cite{mou2024t2i}, require training an additional, often large, encoding module.
Beyond using noise as a simple prior, recent works \cite{dai2025noisectrl,esser2023structure,wang2022zero,burgert2025go,wang2024taming} have explored using structured or conditional noise to guide the generative process. For instance, Dai~\etal \cite{dai2025noisectrl} demonstrated that conditioning on noise can control the generation process, though its efficacy can be limited by domain gaps between the input and output. 
% This concept broadly involves injecting or manipulating noise at various stages to influence attributes such as texture, style, or spatial features, offering a lightweight alternative to traditional conditioning mechanisms. 

In this work, we propose a framework that integrates the efficiency of model distillation with explicit control over high-frequency generation. This allows our model to effectively generate plausible high-frequency details, directly addressing a key limitation in deterministic reconstruction.

\section{Motivational study}

\begin{figure}[t]
  \centering
  \begingroup
  \setlength{\tabcolsep}{1pt} % Default value: 6pt
  \renewcommand{\arraystretch}{0.1} % Set array row spacing to 0
  \resizebox{\linewidth}{!}{%
  \begin{tabular}{cccc}
    % --- First Row with Vertical Caption ---
    \rotatebox{90}{\hspace*{2.5em}\textbf{Image Space}} &
    \begin{subfigure}{0.35\linewidth}
      \centering
      \includegraphics[width=\textwidth]{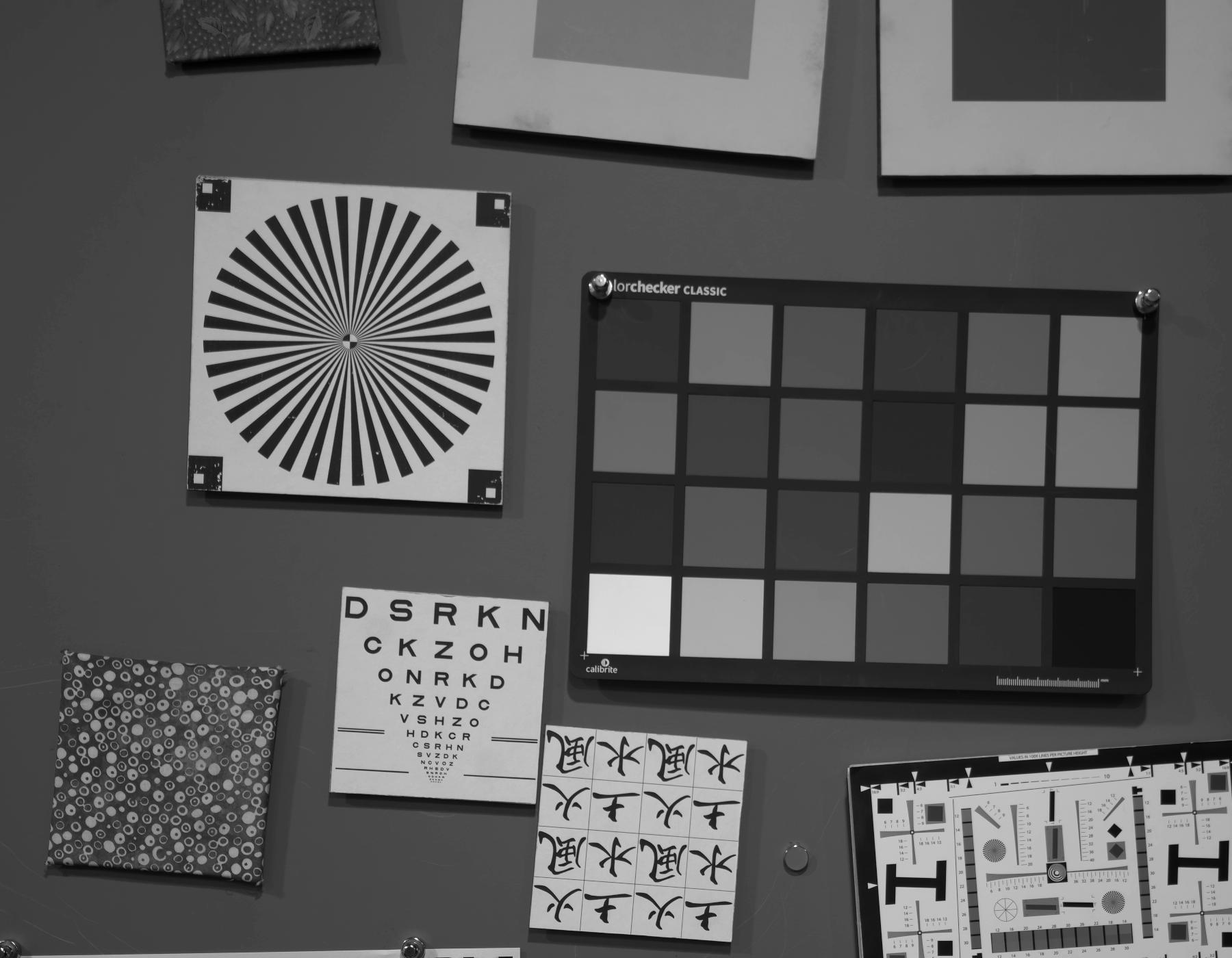}
      \label{fig:toy_gt}
    \end{subfigure} &
    \begin{subfigure}{0.35\linewidth}
      \centering
      \includegraphics[width=\textwidth]{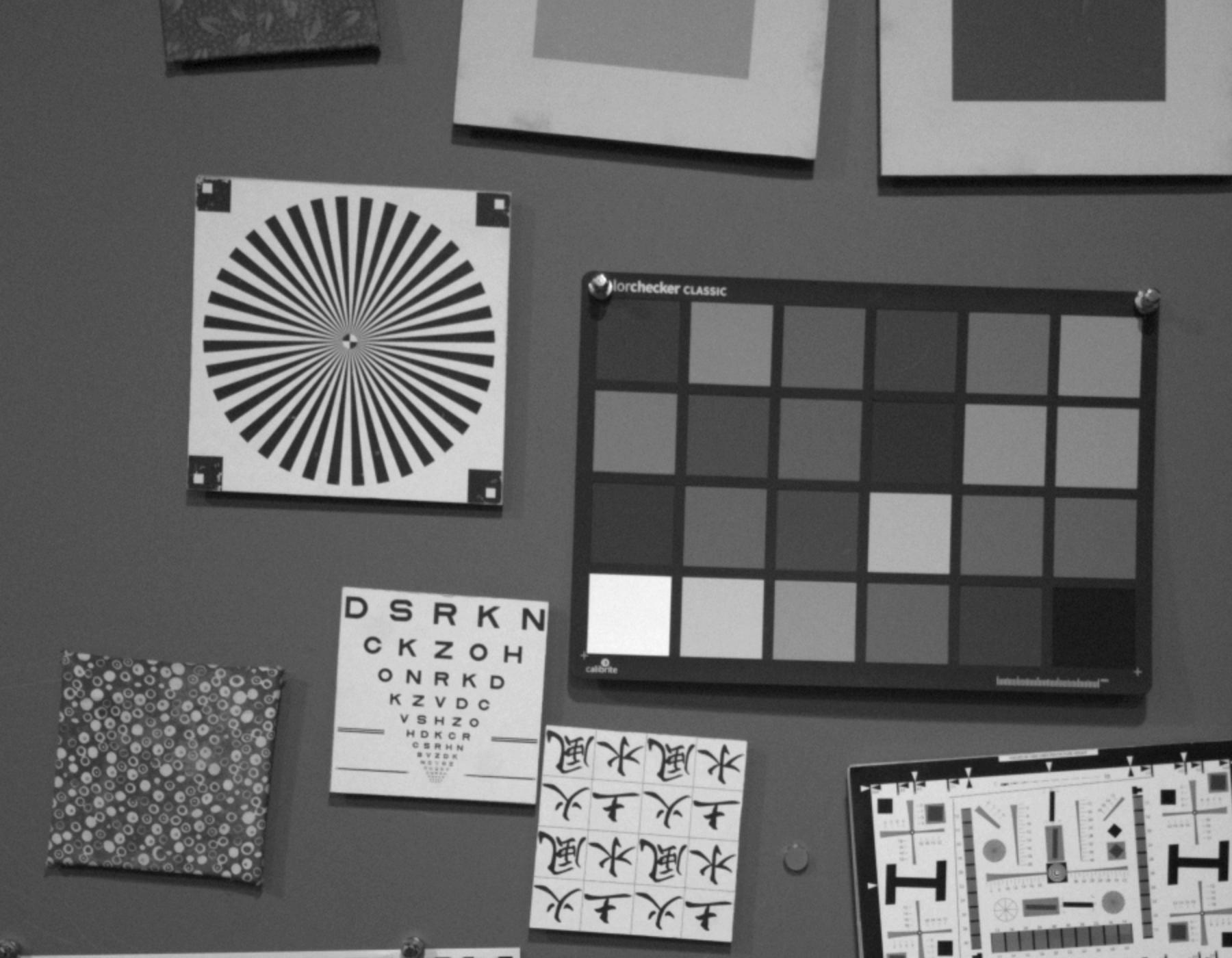}
      \label{fig:toy_in}
    \end{subfigure} &
    \begin{subfigure}{0.35\linewidth}
      \centering
      \includegraphics[width=\textwidth]{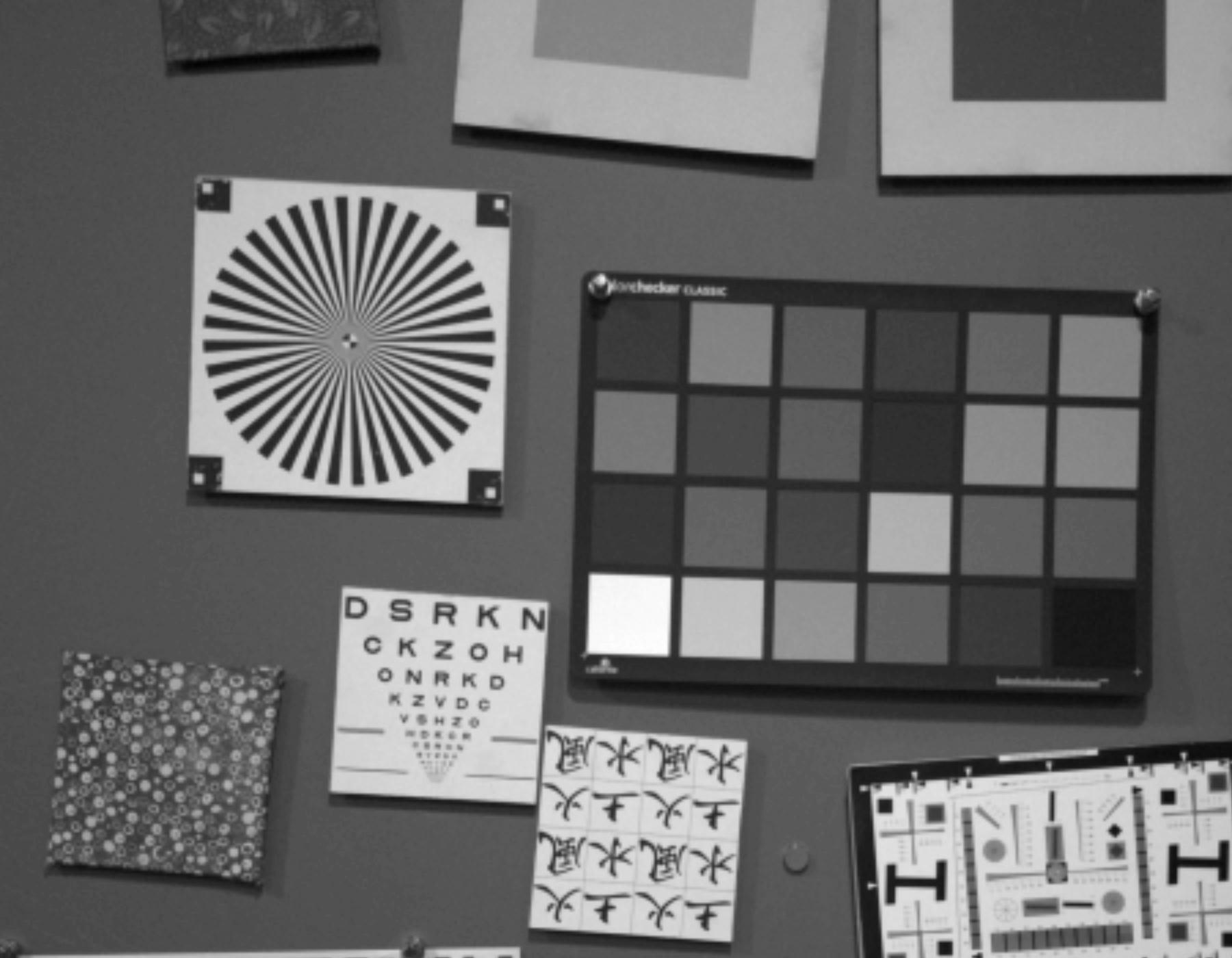}
      \label{fig:toy_in_down}
    \end{subfigure} \\
    
    \noalign{\vspace{-15pt}}
    \rotatebox{90}{\hspace*{2em}\textbf{Frequency Space}} &
    \begin{subfigure}{0.35\linewidth}
      \centering
      \begin{tikzpicture}[
            spy using outlines={%
                rectangle, 
                red, 
                magnification=3, 
                size=0.3\linewidth,      
                connect spies    
                }
            ]
            \node[anchor=center, inner sep=0pt] (main_image) at (0,0) {
                \includegraphics[width=\linewidth]{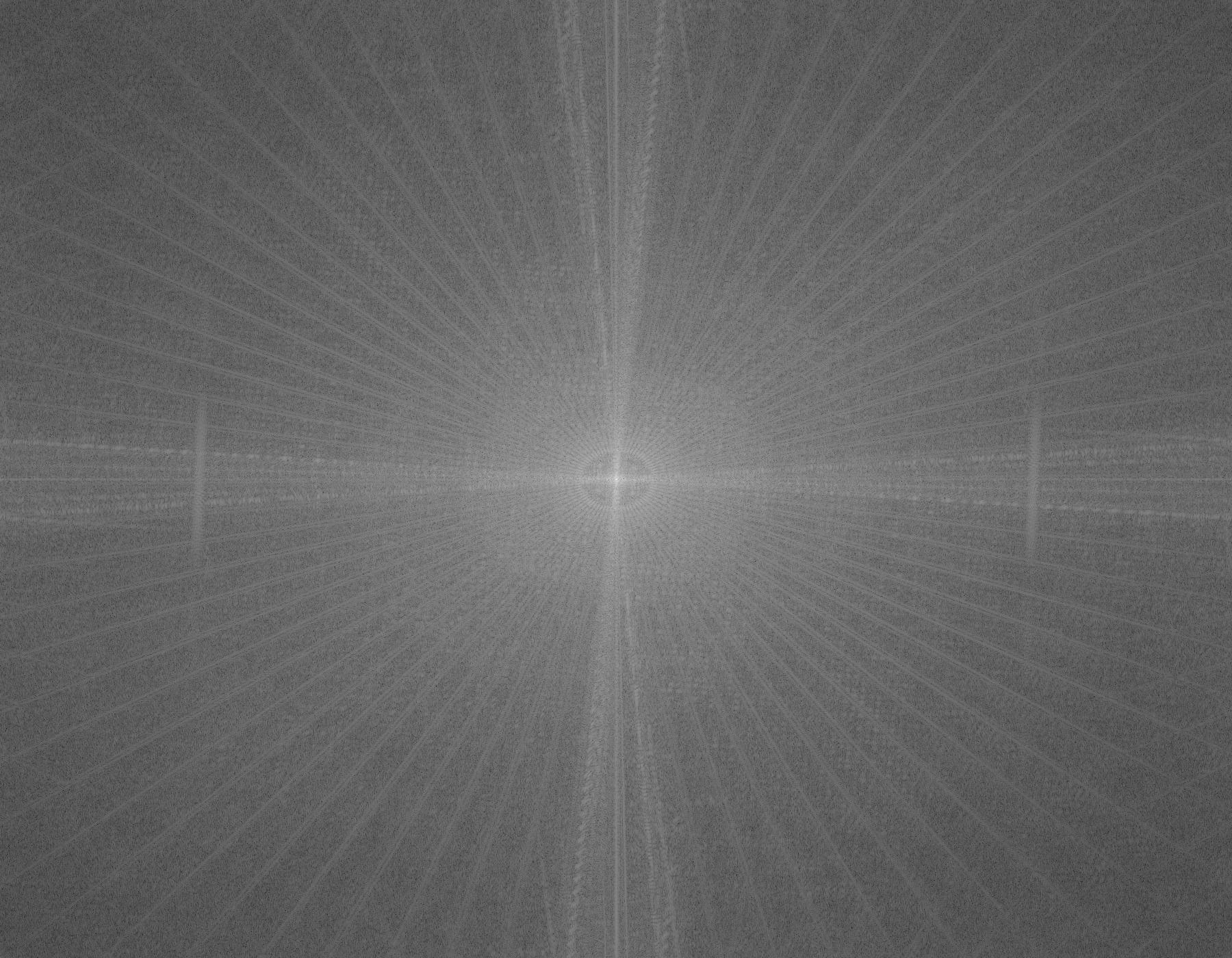}
            };
            \spy on (0., 0.) in node at (1.0, -0.65);
      \end{tikzpicture} 
      \caption{High-res ground truth image}
      \label{fig:toy_gt_ft}
    \end{subfigure} &
    \begin{subfigure}{0.35\linewidth}
      \centering
      \begin{tikzpicture}[
            spy using outlines={%
                rectangle, 
                red, 
                magnification=3, 
                size=0.3\linewidth,      
                connect spies    
                }
            ]
            \node[anchor=center, inner sep=0pt] (main_image) at (0,0) {
                \includegraphics[width=\linewidth]{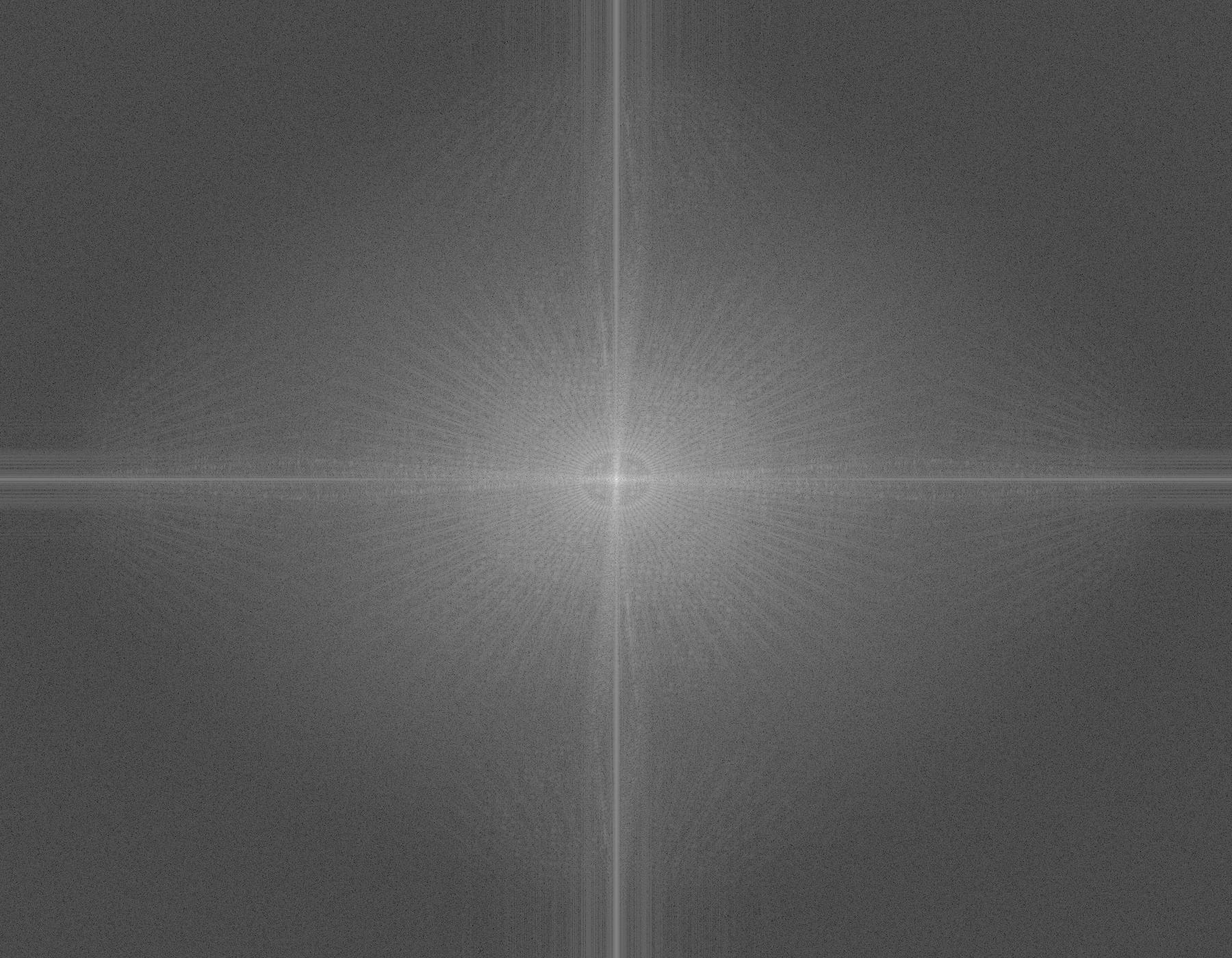}
            };
            \spy on (0., 0.) in node at (1.0, -0.65);
      \end{tikzpicture} 
      \caption{High-res RAW demosaic}
      \label{fig:toy_in_ft}
    \end{subfigure} &
    \begin{subfigure}{0.35\linewidth}
      \centering
      \begin{tikzpicture}[
            spy using outlines={%
                rectangle, 
                red, 
                magnification=3, 
                size=0.3\linewidth,      
                connect spies    
                }
            ]
            \node[anchor=center, inner sep=0pt] (main_image) at (0,0) {
                \includegraphics[width=\linewidth]{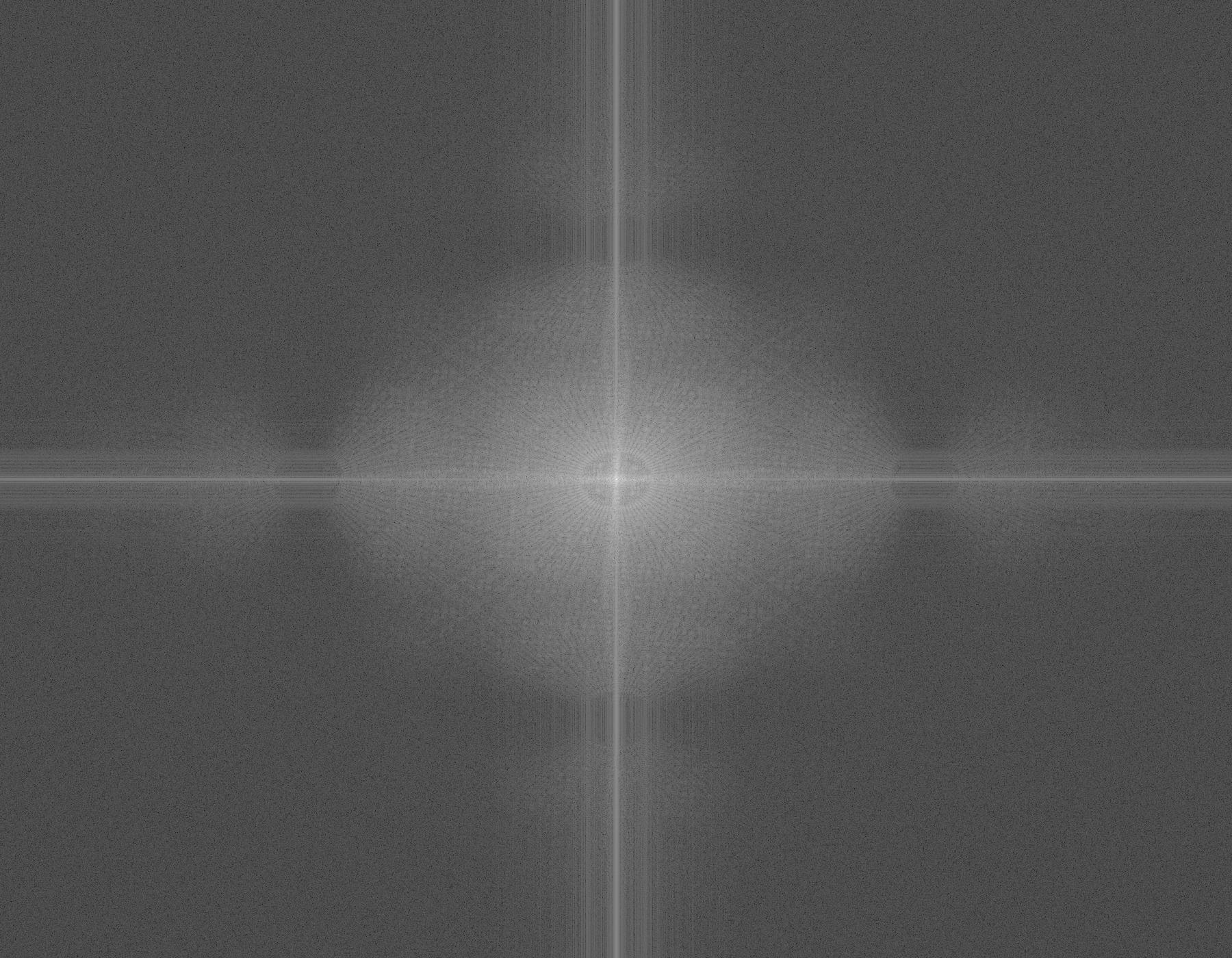}
            };
            \spy on (0., 0.) in node at (1.0, -0.65);
      \end{tikzpicture} 
      \caption{Low-res RAW demosaic}
      \label{fig:toy_in_down_ft}
    \end{subfigure} \\
  \end{tabular}
  }% End of \resizebox
  \endgroup % End the group
  \caption{Comparison of RAW demosaiced images with the ground truth image in image and frequency space. Frequency-space figures are shown in the natural logarithm of the real frequency values.}
  \label{fig:toy_example}
\end{figure}

As shown in \cref{fig:toy_example}, we know that the problem of RAW-to-RGB super-resolution involves a significant frequency-domain gap. The low-resolution RAW input's high-frequency regions are sparse and noisy. The generator's task is to plausibly fill in this missing high-frequency information to match the ground truth. As highlighted, the low-frequency components remain largely unchanged in all domains. This leads to the idea of focusing on high-frequency feature generation without hindering low-frequency features. Generating low-frequency content can lead to hallucinations of objects which are undesirable for consumer camera applications.

\section{Methodology}
\label{sec:method}
Our goal is to reconstruct a high-fidelity, photorealistic RGB image from a misaligned burst of RAW sensor data. In this section, we formalize the RAW-to-RGB pipeline as an inverse problem with partial observability. We then introduce our core theoretical contribution: a subspace-restricted prior integration technique (HF-VSD) that explicitly separates deterministic recovery from stochastic high-frequency synthesis. Finally, we detail the generative multi-frame architecture and optimization strategy used to solve this formulation.
\begin{figure*}[t]
    \centering
    \includegraphics[width=0.95\linewidth]{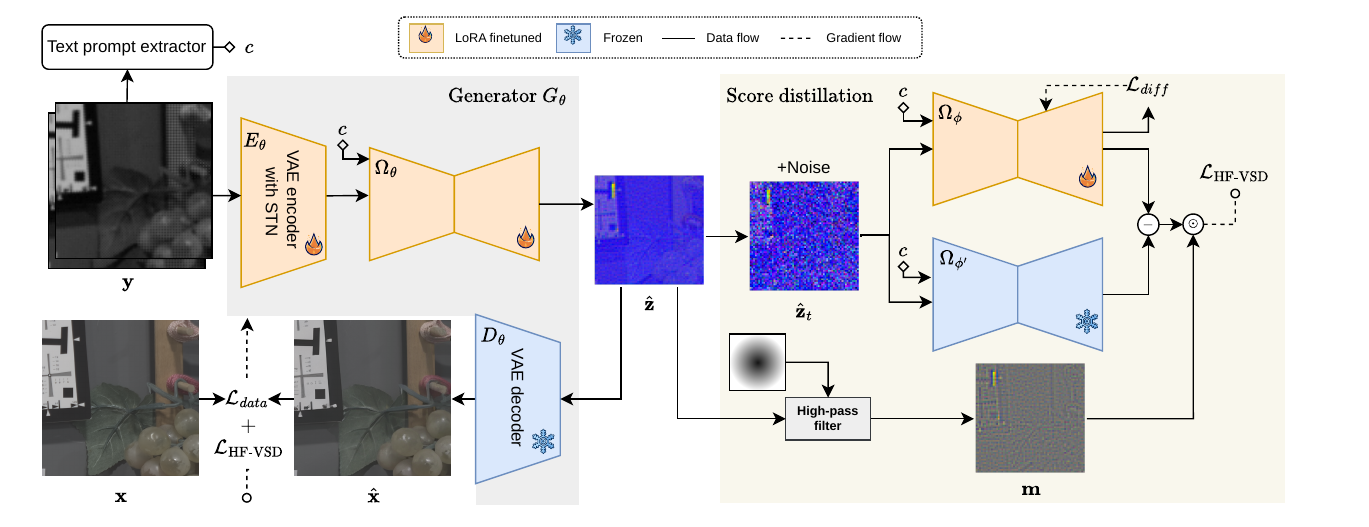}
    \caption{The end-to-end training framework for our proposed GenMFSR model. A multi-frame RAW encoder $E_{\vtheta}$ with embedded STN aligns and processes the input burst of raw images. A U-Net $\Omega_{\vtheta}$ and frozen decoder $D_{\vtheta}$ generate the RGB image $\hat{\vx}$. The framework is trained with a data fidelity loss $\mathcal{L}_\text{data}$ and our regularization $\mathcal{L}_\text{HF-VSD}$, which uses a high-pass filter to ensure the diffusion prior focuses on high-frequency details.}
    \label{fig:framework}
\end{figure*}
\subsection{Inverse Problem with Partial Observability}
Given a burst of $N$ RAW frames $\vy = \{\vy^{(i)}\}_{i=1}^N$, our objective is to reconstruct a high-resolution RGB image $\vx$. Due to spatial downsampling, Poisson-Gaussian sensor noise, and Bayer mosaicing, the inverse mapping from $\vy$ to $\vx$ is highly ill-posed.

However, burst capture provides strong physical constraints on the scene. Because the burst acquisition operator, which is characterized by pixel aperture integration and sensor noise, attenuates high spatial frequencies much more severely than low frequencies, the inverse problem exhibits frequency-dependent determinacy. After temporal alignment and multi-frame fusion via sub-pixel shifts, the coarse geometry, global illumination, and structural color are largely determined by the measurements. In contrast, fine-scale textures remain underdetermined.
We therefore conceptually decompose the target image into two approximately complementary subspaces under natural image statistics: $\vx = \vx_L + \vx_H$, where $\vx_L$ represents the low-frequency components reliably inferred from the sensor data, and $\vx_H$ represents the high-frequency components that require prior-driven inference. This partial observability structure fundamentally motivates the selective integration of generative priors.

\subsubsection{Training Objective and Optimization}
The total training objective balances deterministic distortion and generative perception by anchoring the observable subspace and shaping the underdetermined subspace. We formulate this as a generation task regularized by a pre-trained diffusion prior:
\begin{equation}
    \label{eq:problem}
    \vtheta^* = \arg\min_{\vtheta} \mathbb{E}_{{\vy}, {\vx}} [\mathcal{L}_\text{data} (G_{\vtheta} ({\vy}), {\vx} ) + \lambda\mathcal{L}_\text{HF-VSD}(G_{\vtheta} ({\vy}))],
\end{equation}
where ${\vx}$ is the ground-truth RGB image, $\lambda$ is a balancing hyperparameter between $\mathcal{L}_\text{data}$ fidelity loss (comprising standard MSE and LPIPS losses calculated in the RGB pixel space), anchors the observable subspace, ensuring that the low-frequency structure and color perfectly match the ground truth, 
% First, we define the data fidelity term in \cref{eq:problem} $\mathcal{L}_\text{data}$ to ensure the reconstructed image $\hat{\vx}$ is faithful to the ground truth $\vx$. It is a weighted sum of the Mean Squared Error (MSE) and the LPIPS \cite{zhang2018lpips} perceptual loss:
% \begin{equation}
% \label{eq:loss_data}
% \mathcal{L}_\text{data}(\hat{\vx}, \vx) = \mathcal{L}_\text{MSE}(\hat{\vx}, \vx) + \lambda_\text{LPIPS}\mathcal{L}_\text{LPIPS}(\hat{\vx}, \vx)
% \end{equation}
% where $\lambda_\text{LPIPS}$ is a balancing parameter.
and $\mathcal{L}_\text{HF-VSD}$  regularization loss to regularize the generated images in a specified image prior which shapes the underdetermined subspace, guiding the latent representations to generate photorealistic textures without disrupting the data term's structural consensus. We employ a one-step adversarial distillation strategy (combining LoRA fine-tuning with a discriminator) to make this robust regularization compatible with single-step inference.

\subsection{Subspace-Restricted Prior Integration (HF-VSD)}
Generative image priors~\cite{rombach2021highresolution}, particularly latent diffusion models, offer a powerful mechanism for synthesizing the unobserved high-frequency detail $\vx_H$. Standard score distillation techniques (e.g., VSD~\cite{wang2023prolificdreamer}) apply diffusion gradients over the full image distribution, effectively regularizing the reconstruction using the gradient of the log-likelihood: $\nabla_\vx \log p(\vx)$.When applied to the RAW-to-RGB inverse problem, this global prior influences both $\vx_L$ and $\vx_H$. However, because the diffusion model is trained exclusively on highly processed sRGB images, its low-frequency score estimates inherently conflict with the true low-frequency structure deterministically recovered from the RAW measurements. This "domain collision" results in severe structural drift, color shifts, and low-frequency hallucinations (see \cref{fig:ablation_hf}).

To prevent prior-likelihood conflict, we treat score distillation as a projection operator rather than a global regularizer. We restrict generative supervision strictly to the underdetermined high-frequency subspace. We define a projection operator $P_H$:$$P_H(x) = \mathcal{F}^{-1}(\vh \cdot \mathcal{F}(\vx))$$where $\mathcal{F}$ and $\mathcal{F}^{-1}$ denote the Fourier transform and its inverse, and $\vh$ is a radial high-pass filter mask defined as,
\begin{equation*} 
\vh(u,v) = \text{clip}\left[\left(\left(\frac{\alpha u}{R}\right)^2 + \left(\frac{\alpha v}{R}\right)^2\right)^\gamma + \beta,0,1\right],
\end{equation*}
where $(u,v)$ are frequency coordinates, $R$ is half the maximum frequency, $\alpha$, $\gamma$ are scaling factors, and $\beta$ is a bias term. 
We employ a fixed, smooth radial high-pass mask $h$ to avoid spatial ringing artifacts (Gibbs phenomenon) and ensure stable optimization. The cutoff is set to a moderate fraction of the Nyquist frequency. This choice reflects the empirical observation that burst alignment and fusion reliably recover coarse and mid-frequency structure, while the highest frequencies remain the most sensitive to noise and sampling limitations. Importantly, our objective is not to precisely estimate the exact null space of the forward operator, but simply to attenuate generative gradients in frequency bands where the sensor likelihood is already strong. In practice, we found that a moderate, smooth cutoff achieves stable training without requiring heuristic hyperparameter tuning.

Crucially, we project the score gradients rather than masking the image content itself. Naive VSD modifies both observable and unobservable components indiscriminately. By applying $P_H$, we reformulate the update: $$\nabla_\vx \mathcal{L}_{\text{HF-VSD}} = P_H \left( \nabla_\vx \mathcal{L}_{\text{VSD}} \right)$$
This projection ensures that the generative prior's gradient updates are restricted to the orthogonal complement of the highly observable subspace. Consequently, it preserves the null space of the data operator; the low-frequency structure is mainly governed by data fidelity ($\mathcal{L}_{data}$), while the generative prior synthesizes only the missing fine-scale textures.
This regularization can be converted to a loss function to train our model as,
\begin{align}
\label{eq:loss_reg}
\mathcal{L}_\text{HF-VSD}(\hat{\vx}) = \mathbb{E}_{t,\vepsilon}\left[ \mathcal{L}_\text{MSE}[\vz_t, \delta\vz_t] \right]
\end{align}
where,
\begin{equation*}
\delta \vz_t = \vz_t - P_H(\vomega(t)[\Omega_{\vphi'}(\hat{\vz}_t;t,\vc) - \Omega_{\vphi}(\hat{\vz}_t;t,\vc)]),
\end{equation*}
and $\vomega(t)$ is a weighting factor. This loss formulation is equivalent to minimizing the $L_2$ norm of the masked gradient difference, effectively ignoring errors in the low-frequency regions and focusing the generator's updates on creating fine details.
Hence, the final loss for updating the generator $G_\theta$ is the weighted sum of the data fidelity and regularization terms: $\mathcal{L}_\text{total} = \mathcal{L}_\text{data} + \lambda\mathcal{L}_\text{HF-VSD}$ where $\lambda$ controls the strength of the generative regularization.

\textbf{Theoretical Grounding:} %$ Bayesian Subspace Projection.
From a Bayesian perspective, restoring the high-resolution image $x$ from burst measurements $y$ requires maximizing the posterior score: 
$\nabla_x \log p(x|y) = \nabla_x \log p(y|x) + \nabla_x \log p(x)$. 
In heavily ill-posed inverse problems, the forward operator $A$ (modeling optical blur, downsampling, and mosaicing) possesses a pronounced null space. The data likelihood score $\nabla_x \log p(y|x)$ strongly constrains the range space of $A$ (the observable low frequencies) but provides zero information in the null space (the unobservable high frequencies). 

A standard generative prior $\nabla_x \log p(x)$ operates over the entire space. However, because our prior is distilled from an sRGB-trained diffusion model, its low-frequency distribution fundamentally diverges from the linear RAW sensor distribution. Allowing the prior to update the range space introduces a direct mathematical conflict with the data likelihood. HF-VSD resolves this by applying a projection operator $P_H$ to the score gradient. By restricting the prior update to $P_H \nabla_x \log p(x)$, we explicitly constrain the generative prior to act only within the effective null space of the measurement operator. This ensures the prior acts as a complementary regularizer rather than a competing likelihood, grounding our frequency mask in rigorous inverse problem theory.

% This subspace-restricted Bayesian regularization ensures that the generative prior is applied only where the inverse problem is underdetermined. The low-frequency structure is governed solely by data fidelity, while the generative prior synthesizes only the missing fine-scale textures, enforcing orthogonality between the prior and the sensor likelihood.

\subsection{Generative Multi-Frame Architecture}
We designed an architecture that reliably estimates the observable component $\vx_L$ from sensor data and maps it into the latent space for generative refinement.
\subsubsection{Sensor-Domain Alignment and Fusion}
\begin{figure}[t]
    \centering
    \includegraphics[width=\linewidth]{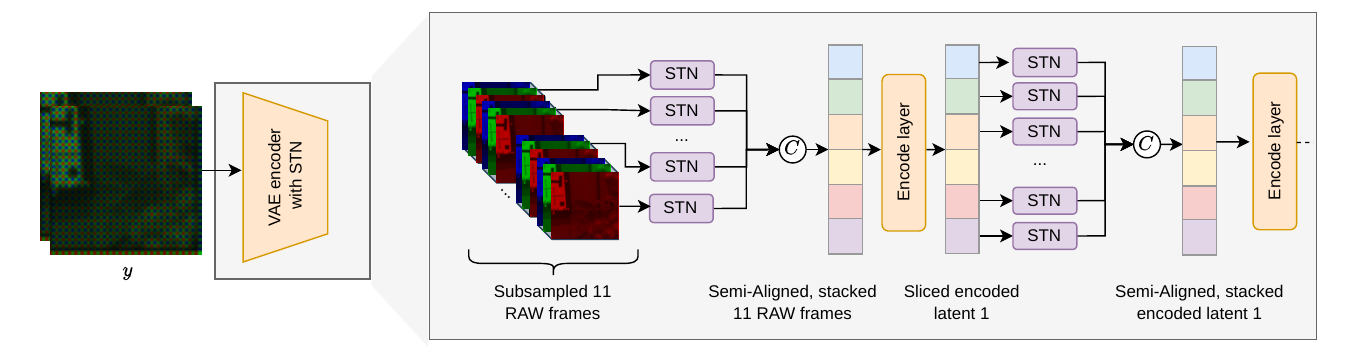}
    \caption{The architecture of the encoder that aligns the RAW frames and intermediate latents using STNs.}
    \label{fig:encoder}
\end{figure}

Accurate estimation of the low-frequency structure $\vx_L$ is critical before applying generative refinement. Because our inputs are handheld bursts, the frames suffer from non-rigid temporal misalignment. Rather than relying on external optical flow networks, we embed Spatial Transformer Networks (STNs) directly within the RAW-domain multi-frame encoder (\cref{fig:encoder}). The STNs estimate global homographies to warp the deep feature embeddings of frames $i \in \{2 \dots N\}$ to the reference frame $i=1$. This explicit sensor-domain alignment efficiently fuses burst data, thereby minimizing ambiguity in the observable subspace prior to the latent mapping phase. This ensures that the latent features fused by the subsequent convolutional layers are spatially coherent, maximizing the effective receptive field and preventing ghosting artifacts during the generative decoding phase.
% Next, STNs are used again in the subsequent encoding layers to further align the latent features. These features are processed in groups of $32$, a size chosen arbitrarily for this implementation. 

\subsubsection{Sensor-to-Latent Mapping}
To leverage the subspace-restricted diffusion prior, we must bridge the gap between the physical sensor likelihood and the pretrained sRGB prior. We perform a deterministic linear channel extraction on the Bayer inputs to form a 33-channel stacked volume, preserving the radiometric linearity and independent noise distributions of the RAW data (see Suppl). Our STN-embedded encoder then maps this linear volume directly into the standard SD latent space. By freezing the pretrained VAE decoder, we ensure the latent space remains perfectly aligned with the diffusion prior's expectations, allowing HF-VSD to operate effectively. We utilize a text prompt extractor, DAPE from \cite{wu2024seesr}, to give text embeddings to the diffusion prior.
\begin{figure*}[ht]
    \centering
    \includegraphics[width=\linewidth]{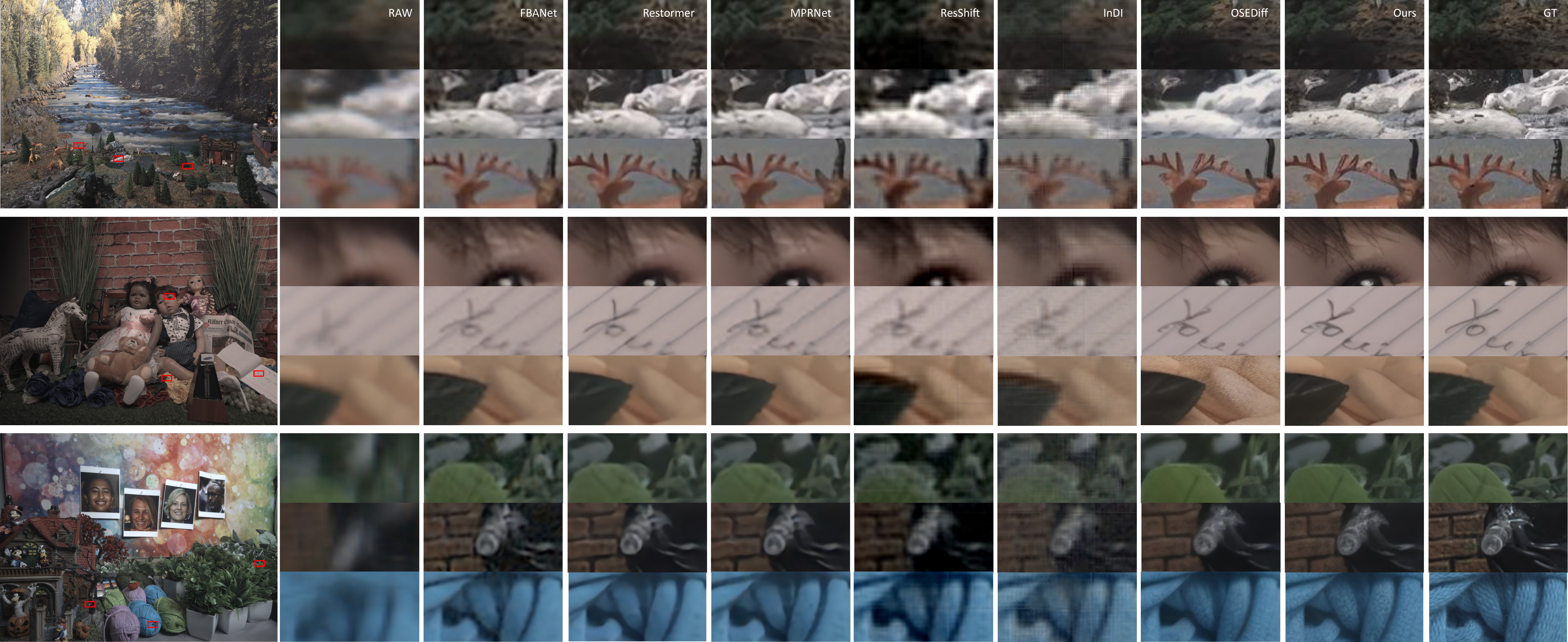}
    \caption{Qualitative comparison of different postprocessing algorithms for generating RGB images using multi-frame RAW  as input. Our algorithm produces finer details in the reconstructed RGB image than all baselines and also exhibits stability in the low-frequency range.}
\label{fig:qualitative_comparison}
\end{figure*}

\begin{figure*}[htbp]
    \centering
    \begin{minipage}{0.7\linewidth}
        \centering
        \caption{Comparison of methods for the $4\times$ super-resolution using fidelity (PSNR, SSIM), perception quality (LPIPS, DISTS, FID), and no-reference quality (NIQE, MUSIQ, MANIQA, CLIPIQA) metrics. The methods are grouped into non-diffusion-based and diffusion-based approaches to highlight their different training strategies. We highlight the \RB{best}, \BB{second-best}, and \TB{third-best} values for each metric.}
        \resizebox{\textwidth}{!}{
        \begin{tabular}{l| cc| ccc| cccc}
        \toprule
            \textbf{Method} & \textbf{PSNR$\uparrow$} & \textbf{SSIM$\uparrow$} & \textbf{LPIPS$\downarrow$} & \textbf{DISTS$\downarrow$} & \textbf{FID$\downarrow$} & \textbf{NIQE$\downarrow$} & \textbf{MUSIQ$\uparrow$} & \textbf{MANIQA$\uparrow$} & \textbf{CLIPIQA$\uparrow$}  \\ \hline
        Reference&29.87&0.8841&0.3018&0.2512&39.41&9.6690&23.40&0.3160&\BB{0.4502}\\ \hline
        FBANet\cite{wei2023fbanet}&30.88&0.9341&0.1891&0.1354&21.74&6.5970&34.53&0.3153&\RB{0.4864}\\
        Burstormer\cite{dudhane2023burstormer}&\TB{35.07}&\TB{0.9454}&\RB{0.0948}&\TB{0.0669}&\RB{12.32}&5.5131&36.72&0.3450&0.3600\\
        MPRNet\cite{zamir2021mprnet}&\BB{35.18}&\BB{0.9471}&0.1309&0.1045&21.51&6.9476&38.93&0.3540&\TB{0.4471}\\
        Restormer \cite{zamir2022restormer}&\RB{35.24}&\RB{0.9522}&0.1172&0.0967&\TB{16.10}&7.0328&\BB{39.09}&0.3570&0.4409\\ \hline
        ResShift \cite{yue2023resshift}&26.12&0.8623&0.2927&0.1760&41.82&6.1530&29.32&0.3079&0.3429\\
        InDI \cite{delbracio2023inversion}&32.69&0.8987&0.2486&0.2134&57.12&5.4929&28.41&0.3162&0.3207\\
        VSD \cite{wu2024neurips_osediff}&29.75&0.9019&0.1232&0.0733&20.02&\TB{5.2592}&\TB{39.08}&\TB{0.3621}&0.3935\\
        Ours (no HF loss)&29.78&0.8998&\TB{0.1131}&\BB{0.0662}&18.58&\BB{5.1981}&38.26&\BB{0.3631}&0.4081\\
        Ours&32.51&0.9200&\BB{0.0984}&\RB{0.0585}&\BB{14.22}&\RB{5.1912}&\RB{39.20}&\RB{0.3643}&0.4184\\
        \bottomrule
        \end{tabular}
        }
        \label{tab:main_comparison}
    \end{minipage}
    \hfill
    % Second minipage: Figure
    \begin{minipage}{0.27\linewidth}
        \centering
        \includegraphics[width=\textwidth]{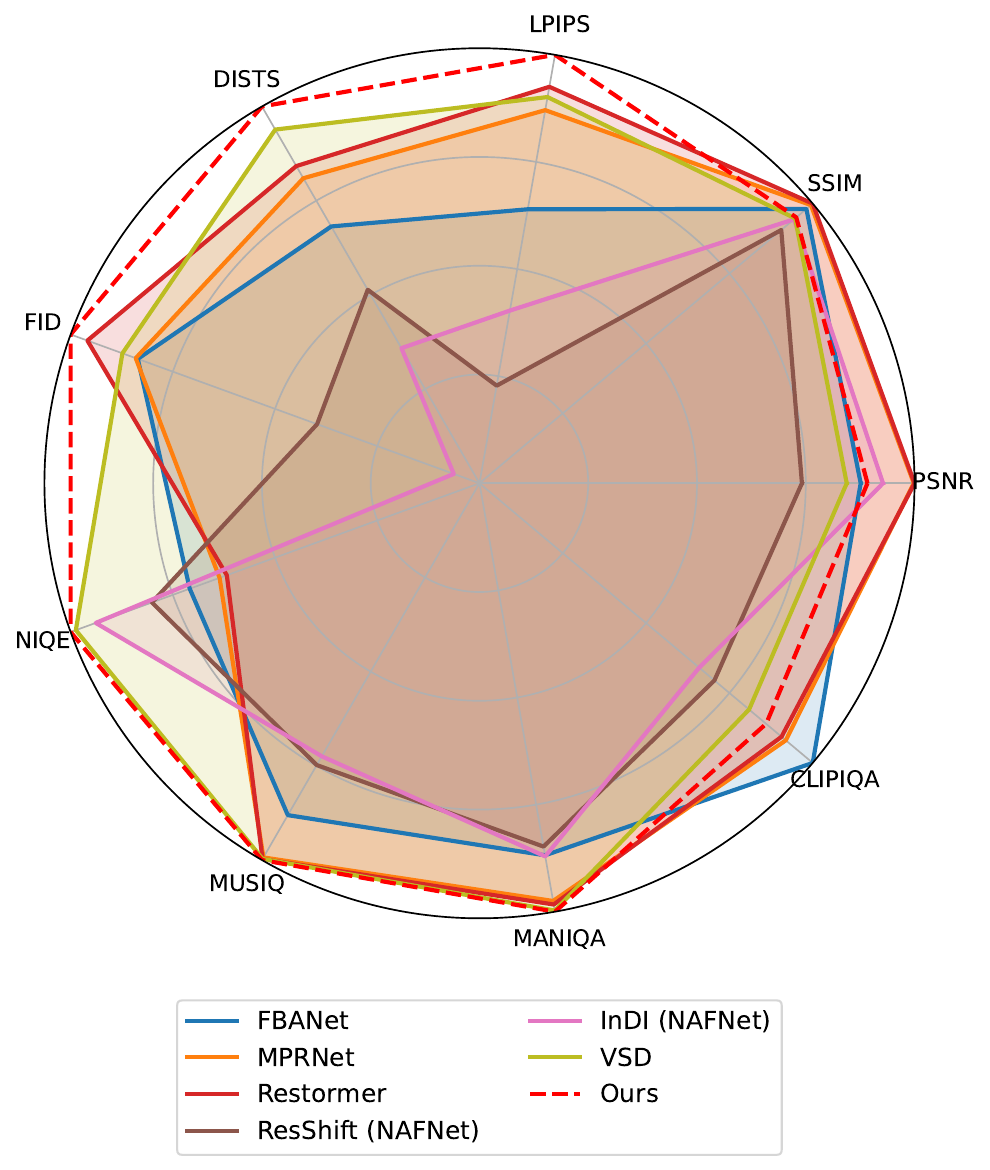}
        \captionof{figure}{Performance comparison with selected competing baselines.}
        \label{fig:radar_chart}
      \end{minipage}
\end{figure*}

\subsection{Dataset Construction and Motion Simulation}
\label{sec:dataset}

Acquiring perfectly aligned, noise-free ground truth for genuine handheld RAW bursts is physically impossible due to the inherent temporal mismatch between a shaky short-exposure burst and the necessary long-exposure reference. To construct a rigorous paired dataset, we capture static tripod scenes to obtain pristine, long-exposure ground-truth RGB images alongside their corresponding short-exposure RAW pairs, natively preserving realistic Poisson-Gaussian sensor noise. We used a Samsung Galaxy S23 Ultra's Wide Camera to capture 3000 burst RAW scenes and extract them after lens shading correction, white balancing, and black level subtraction, but before any ISP denoising, to ensure that the captured RAW frames contain realistic noise before demosaicing, which is not perfectly achievable using synthetic datasets that simulate RAW images from RGB images. The dataset includes low-light and bright light captures ranging from ISO 500 to 2400. The content is static and includes indoor and outdoor scenes from homes, supermarkets, and public spaces.

To simulate handheld motion without introducing the domain gaps typical of purely synthetic affine warps, we adopt an empirical simulation strategy. We extract continuous inter-frame homography trajectories from a separate dataset of real handheld bursts \cite{khan2025mfsr} and apply these real-world motion paths to our static RAW captures. This ensures our network learns to align physically accurate, temporally correlated hand tremors while still allowing us to evaluate the generative output against mathematically perfect ground truth. Finally, to simulate resolution loss, we utilize a Bayer-preserving Space-to-Depth downsampling operation. Comprehensive capture logistics and hardware details are provided in the Supplementary Material.

\section{Experiments}
We evaluate the proposed framework along three axes: reconstruction fidelity, perceptual realism, and structural consistency under generative regularization. 
We compare our method against state-of-the-art multi-frame super-resolution restoration models, including both regression-based and generative approaches. We report standard distortion metrics (PSNR, SSIM) and perceptual metrics (LPIPS\cite{zhang2018lpips} and FID\cite{heusel2017fid}) and conduct a controlled user study to assess visual preference.  We use PSNR and SSIM as pixel-based metrics to assess restoration fidelity relative to the ground truth. In addition, we perform targeted ablations to isolate the effect of High-Frequency Variational Score Distillation (HF-VSD), the alignment module, and generative prior adaptation. See Suppl. for training details.

\textbf{Comparison with baselines:} We trained a set of baseline methods with our data after making minimal changes to their architectures to accommodate multi-frame RAW input. 
A major novelty of our pipeline is the adaptation of single-frame generative architectures to multi-frame RAW bursts. Standard diffusion models and restoration baselines (e.g., OSEDiff, NAFNet) accept 3-channel RGB inputs. To process Bayer RAW bursts, we pack the sub-sampled RAW frames into a 3-channel format and concatenated all 11 frames to create a 33-channel input requiring the initial convolutional layers of both our encoder and the retrained baselines to accept a 33-channel input. By modifying these initial projection layers and adjusting subsequent layers to ensure smooth variation of channel dimension with minimal changes to the underlying architecture of the baselines. By retraining all models from scratch, we ensure that the network can natively fuse temporal sub-pixel information across the 33 channels to produce the final RGB output.

We compare our method with state-of-the-art generative models for image restoration, namely, ResShift \cite{yue2023resshift}, InDI \cite{delbracio2023inversion}, and OSEDiff \cite{wu2024neurips_osediff}, and some adversarially-trained models such as Burstomer \cite{dudhane2023burstormer},  MPRNet\cite{zamir2021mprnet} and Restormer \cite{zamir2022restormer}. The quantitative results for these baselines are given in \cref{tab:main_comparison}. Our method achieves the best generative quality metrics among competing methods. Although adversarially trained methods yield better metrics, our method qualitatively demonstrates superior generative capability, as illustrated in \cref{fig:qualitative_comparison}, where it can reconstruct plausible, realistic high-frequency details.

\textbf{The Perception-Distortion Trade-off in Burst SR.} As shown in Table \ref{tab:main_comparison}, deterministic regression baselines (e.g., Restormer, MPRNet) achieve higher PSNR (e.g., $\sim$35 dB) compared to our generative framework ($\sim$32.5 dB). Rather than a limitation, this quantitative gap perfectly illustrates the core theoretical trade-off of our formulation. Because regression baselines minimize Mean Squared Error (MSE), they mathematically converge to the posterior distribution's mean. While this maximizes PSNR, it heavily penalizes high-frequency variance, resulting in the overly smooth or ``waxy'' textures observed in Fig. \ref{fig:qualitative_comparison}.
In contrast, GenMFSR is explicitly designed to sample a specific, stochastic realization from the posterior to recover photorealistic texture. As established by the Perception-Distortion trade-off \cite{blau2018perception}, this stochasticity guarantees a drop in pixel-wise distortion metrics like PSNR in exchange for superior perceptual quality. By optimizing for distributional alignment via HF-VSD, GenMFSR achieves state-of-the-art LPIPS and FID scores.
\begin{table}[]
    \centering
    \captionof{table}{Comparison of proposed method with single-frame baselines for $4\times$ SR case.}
    \resizebox{\linewidth}{!}{
    \begin{tabular}{lcccc}
        \toprule
        \textbf{Method} & \textbf{LPIPS} & \textbf{FID} & \textbf{PSNR} & \textbf{SSIM} \\
        \midrule
        GAN (SF NAFNet)  & 0.13 & \textbf{9.53} & 28.18 & 0.86 \\
        VSD (SF)     &  0.14 & 16.53 & 28.67 & 0.87 \\
        Ours (MF)  & \textbf{0.09} & 14.22 &\textbf{ 31.43} & \textbf{0.90} \\
        \bottomrule
    \end{tabular}
    }
    \label{tab:singleframe_comparison}
\end{table}

\begin{table}[]
    \centering
    \captionof{table}{Comparisons of Params and FLOPs between GenMFSR and its competing methods on input resolution of $512\times512$.}
    \resizebox{\linewidth}{!}{
        \begin{tabular}{l c c r}
            \toprule
            \textbf{Method} & \textbf{Type} & \textbf{Params (M)} & \textbf{FLOPs (G)} \\
            \midrule
            FBANet \cite{wei2023fbanet}       & CNN & $\sim$28.5 & $\sim$120 \\
            Burstormer \cite{dudhane2023burstormer} & Transformer & 3.58 & 613 \\
            MPRNet \cite{zamir2021mprnet}       & CNN & 20 & 595 \\
            Restormer \cite{zamir2022restormer}& Transformer & 26.1 & 567 \\
            % NAFNet \cite{chen2022nafnet}       & CNN & 17.1 & 252 \\
            % SwinUnet \cite{liu2021swin}        & Transformer & 11.5 & 1776 \\
            \midrule
            ResShift \cite{yue2023resshift} & 15-step Diffusion & 119 & 5500 \\
            InDI \cite{delbracio2023inversion} & 50-step Diffusion & 54 & ~48000 \\
            VSD (OSEDiff) \cite{wu2024neurips_osediff}     & 1-step Diffusion & 1775.0 & 4530 \\
            \textbf{Ours (GenMFSR)}            & 1-step Diffusion & \textbf{1310.0} & \textbf{4620} \\
            \bottomrule
        \end{tabular}
    }
    \label{tab:params_flops}
\end{table}
\begin{figure}[t]
    \centering
    \includegraphics[width=0.8\linewidth]{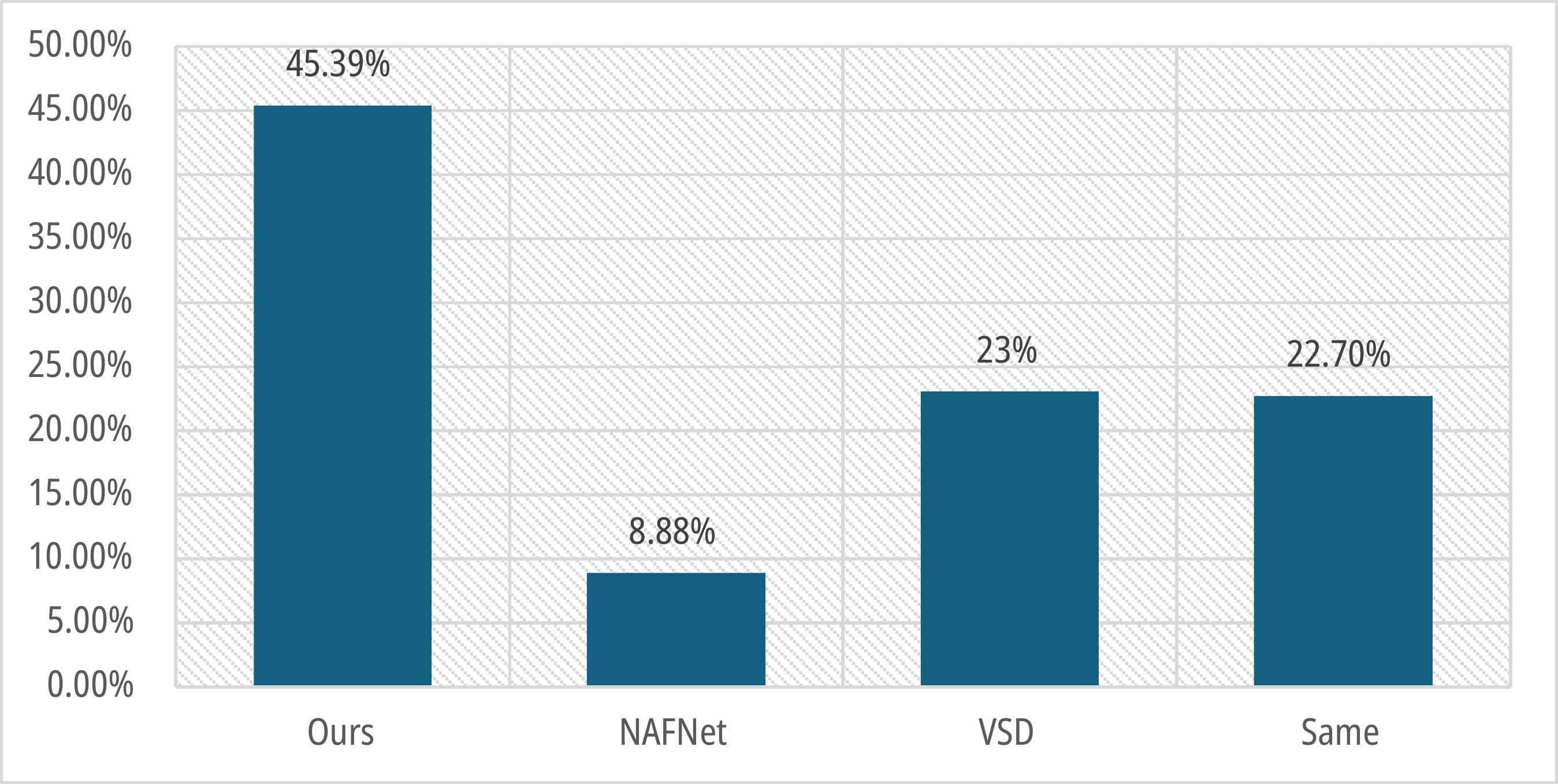}
    \caption{A survey of 22 image quality experts shows a strong perceptual preference for our method (45.39\%). Despite its high quantitative metrics, the GAN-based NAFNet was strongly disfavored (8.88\%) due to artifacts, demonstrating the gap between standard metrics and human perception.}
    \label{fig:userstudy}
\end{figure}

\textbf{Details of User Study}: Due to the limitations of traditional image quality metrics in capturing user perception, we conducted a user study to evaluate the proposed method. Twenty-two image quality experts were shown the outputs of the NAFNET trained with a GAN, the baseline VSD method, and our proposed GenMFSR method on 30 randomly selected crops from the test set. For each crop, subjects were asked to choose the preferred image or indicate whether the images looked the same. 
As shown in \cref{fig:userstudy}, a plurality of experts preferred images generated by our method. The post-study feedback indicated that the artifacts in the NAFNet were the primary reason it was not preferred, and the lack of detail in the baseline VSD method was also a contributing factor to its non-preference. Some examples of the artifacts produced by the NAFNET are shown in \cref{fig:qualitative_comparison}. These crosshatch patterns may be an outcome of the network learning to game loss terms such as L1, GAN, and LPIPS, which is a well-known phenomenon known as perception-fidelity tradeoff \cite{blau2018perception}.

\begin{figure}[t]
  \centering
      \setlength{\tabcolsep}{1pt} % Default value: 6pt
      \renewcommand{\arraystretch}{1} % Set array row spacing to 0
      \resizebox{\linewidth}{!}{%
      \begin{tabular}{m{0.5cm} m{4cm} m{4cm} m{4cm} m{4cm} m{4cm}}
        % --- First Row with Vertical Caption ---
        \rotatebox{90}{\hspace*{0.5em}\textbf{Single-frame}} &
        \begin{subfigure}{\linewidth}
          \centering
            \includegraphics[width=\linewidth]{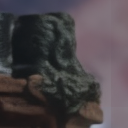}
        \end{subfigure} &
        \begin{subfigure}{\linewidth}
          \centering
          \includegraphics[width=\linewidth]{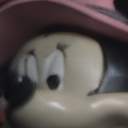}
        \end{subfigure} &
        \begin{subfigure}{\linewidth}
          \centering
          \includegraphics[width=\linewidth]{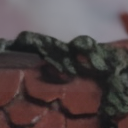}
        \end{subfigure} &
        \begin{subfigure}{\linewidth}
          \centering
          \includegraphics[width=\linewidth]{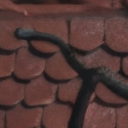}
        \end{subfigure}&
        \begin{subfigure}{\linewidth}
          \centering
          \includegraphics[width=\linewidth]{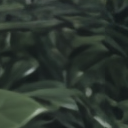}
        \end{subfigure} \\
        
        % Add a vertical space between the two rows
        % \noalign{\vspace{-3.3pt}}
        \rotatebox{90}{\hspace*{0.5em}\textbf{Multi-frame}} &
        
        \begin{subfigure}{\linewidth}
            \centering  
            \includegraphics[width=\linewidth]{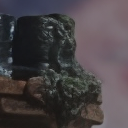}
        \end{subfigure} &
        \begin{subfigure}{\linewidth}
            \centering
            \includegraphics[width=\linewidth]{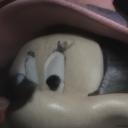}
        \end{subfigure}&
        \begin{subfigure}{\linewidth}
            \centering
            \includegraphics[width=\linewidth]{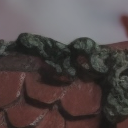}
        \end{subfigure}&
        \begin{subfigure}{\linewidth}
            \centering
            \includegraphics[width=\linewidth]{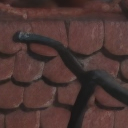}
        \end{subfigure}&
        \begin{subfigure}{\linewidth}
            \centering
            \includegraphics[width=\linewidth]{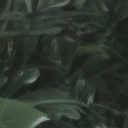}
        \end{subfigure} \\
      \end{tabular}
      }% End of \resizebox      
      \caption{Comparison of our method after training on single-frame and multi-frame RAW images. This shows the need for multiple frames in the RAW-to-RGB problem.}
      \label{fig:single_frame}
\end{figure}

% \begin{figure}
%     \centering
%     \includegraphics[width=\linewidth]{figs/results_0.5x.png}
%     \caption{Comparison of our method with competing GAN methods. Our model demonstrates superior generative capability, producing features that are not present in the ground truth.}
%     \label{fig:ours_vs_gan}
% \end{figure}

% \begin{figure}
%     \centering
%     \includegraphics[width=\linewidth]{figs/single_frame.png}
%     \caption{Comparison of our method with single-frame (SF) NAFNet, multi-frame (MF) NAFNet, and SF OSEDiff. Our method outperforms most of the SF methods and gives better generation compared to MF NAFnet as well.}
%     \label{fig:single_frame}
% \end{figure}

\begin{figure}[t]
  \centering
      
  \setlength{\tabcolsep}{1pt} % Default value: 6pt
  \renewcommand{\arraystretch}{1} % Set array row spacing to 0
  \resizebox{0.99\linewidth}{!}{%
  \begin{tabular}{c c c c}
    % --- First Row with Vertical Caption ---
    \rotatebox{90}{\hspace*{1em}\textbf{w/o STN}} &
    \begin{subfigure}{0.31\linewidth}
    \begin{tikzpicture}[
        spy using outlines={%
            rectangle, 
            red, 
            magnification=4, % Adjust magnification level as needed
            size=0.5\linewidth,        % Adjust the size of the magnified inset
            connect spies    % Draws a line connecting the original area to the inset
            }
        ]
        % Place the main image
        \node (main_image) at (0,0) {
            \includegraphics[width=\linewidth]{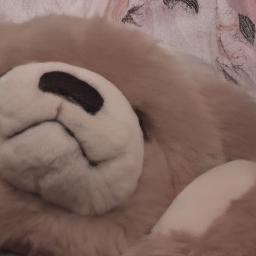}
        };
        
        % Define the area to "spy on" (coordinates relative to the center of the main image node)
        % and where to place the magnified view ("in node at")
        \spy on (-0.5, -0.3) in node at (0.45, 0.45); % Adjust coordinates as needed
    \end{tikzpicture} 
    \end{subfigure} &

    \begin{subfigure}{0.31\linewidth}
    \begin{tikzpicture}[
        spy using outlines={%
            rectangle, 
            red, 
            magnification=4, % Adjust magnification level as needed
            size=0.5\linewidth,        % Adjust the size of the magnified inset
            connect spies    % Draws a line connecting the original area to the inset
            }
        ]
        % Place the main image
        \node (main_image) at (0,0) {
            \includegraphics[width=\linewidth]{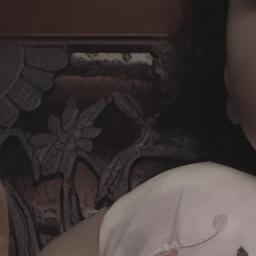}
        };
        
        % Define the area to "spy on" (coordinates relative to the center of the main image node)
        % and where to place the magnified view ("in node at")
        \spy on (-0.5, 0.0) in node at (0.45, 0.45); % Adjust coordinates as needed
    \end{tikzpicture} 
    \end{subfigure} &

    \begin{subfigure}{0.31\linewidth}
    \begin{tikzpicture}[
        spy using outlines={%
            rectangle, 
            red, 
            magnification=4, % Adjust magnification level as needed
            size=0.5\linewidth,        % Adjust the size of the magnified inset
            connect spies    % Draws a line connecting the original area to the inset
            }
        ]
        % Place the main image
        \node (main_image) at (0,0) {
            \includegraphics[width=\linewidth]{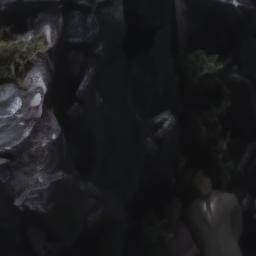}
        };
        
        % Define the area to "spy on" (coordinates relative to the center of the main image node)
        % and where to place the magnified view ("in node at")
        \spy on (-0.9, 0.6) in node at (0.45, 0.45); % Adjust coordinates as needed
    \end{tikzpicture} 
    \end{subfigure} \\
    
    % \begin{subfigure}{0.3\linewidth}
    %   \centering
    %   \includegraphics[width=\linewidth]{figs/comparison4/2_osediff_0.5x_old.jpg}
    % \end{subfigure} &
    % \begin{subfigure}{0.3\linewidth}
    %   \centering
    %   \includegraphics[width=\linewidth]{figs/comparison4/3_osediff_0.5x_old.jpg}
    % \end{subfigure} &
    % \begin{subfigure}{0.3\linewidth}
    %   \centering
    %   \includegraphics[width=\linewidth]{figs/comparison4/7_osediff_0.5x_old.jpg}
    % \end{subfigure} \\
    
    % Add a vertical space between the two rows
    \noalign{\vspace{-5pt}}
    % \hfill
    % --- Second Row with Vertical Caption ---
    \rotatebox{90}{\hspace*{0.5em}\textbf{with STN}} &
    \begin{subfigure}{0.31\linewidth}
    \begin{tikzpicture}[
        spy using outlines={%
            rectangle, 
            green, 
            magnification=4, % Adjust magnification level as needed
            size=0.5\linewidth,        % Adjust the size of the magnified inset
            connect spies    % Draws a line connecting the original area to the inset
            }
        ]
        % Place the main image
        \node (main_image) at (0,0) {
            \includegraphics[width=\linewidth]{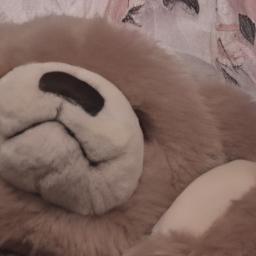} % Use your image file here
        };
        
        % Define the area to "spy on" (coordinates relative to the center of the main image node)
        % and where to place the magnified view ("in node at")
        \spy on (-0.5, -0.3) in node at (0.45, 0.45); % Adjust coordinates as needed
    \end{tikzpicture} 
    % \caption{\tiny{Sample 1}}
    \end{subfigure} &

    \begin{subfigure}{0.31\linewidth}
    \begin{tikzpicture}[
        spy using outlines={%
            rectangle, 
            green, 
            magnification=4, % Adjust magnification level as needed
            size=0.5\linewidth,        % Adjust the size of the magnified inset
            connect spies    % Draws a line connecting the original area to the inset
            }
        ]
        % Place the main image
        \node (main_image) at (0,0) {
            \includegraphics[width=\linewidth]{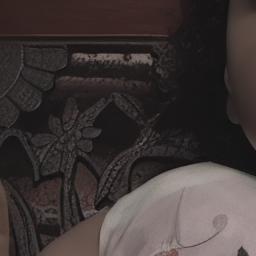} % Use your image file here
        };
        
        % Define the area to "spy on" (coordinates relative to the center of the main image node)
        % and where to place the magnified view ("in node at")
        \spy on (-0.5, 0.0) in node at (0.45, 0.45); % Adjust coordinates as needed
    \end{tikzpicture} 
    % \caption{\tiny{Sample 2}}
    \end{subfigure} &

    \begin{subfigure}{0.31\linewidth}
    \begin{tikzpicture}[
        spy using outlines={%
            rectangle, 
            green, 
            magnification=4, % Adjust magnification level as needed
            size=0.5\linewidth,        % Adjust the size of the magnified inset
            connect spies    % Draws a line connecting the original area to the inset
            }
        ]
        % Place the main image
        \node (main_image) at (0,0) {
            \includegraphics[width=\linewidth]{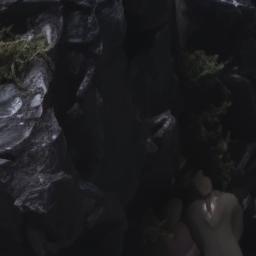} % Use your image file here
        };
        
        % Define the area to "spy on" (coordinates relative to the center of the main image node)
        % and where to place the magnified view ("in node at")
        \spy on (-0.9, 0.6) in node at (0.45, 0.45); % Adjust coordinates as needed
    \end{tikzpicture} 
    % \caption{\tiny{Sample 3}}
    \end{subfigure} \\
    % \begin{subfigure}{0.3\linewidth}
    %   \centering
    %   \includegraphics[width=\linewidth]{figs/comparison4/2_ours_wo_HF_0.5x_old.jpg}
    %   \caption{Sample 1}
    % \end{subfigure} &
    % Second figure in the second row
    % \begin{subfigure}{0.3\linewidth}
    %   \centering
    %   \includegraphics[width=\linewidth]{figs/comparison4/3_ours_wo_HF_0.5x_old.jpg}
    %   \caption{Sample 2}
    % \end{subfigure} &
    % % Third figure in the second row
    % \begin{subfigure}{0.3\linewidth}
    %   \centering
    %   \includegraphics[width=\linewidth]{figs/comparison4/7_ours_wo_HF_0.5x_old.jpg}
    %   \caption{Sample 3}
    % \end{subfigure} \\
  \end{tabular}
  }% End of \resizebox
  \caption{Ablation study showing the effectiveness of using STNs without utilizing the HF loss for a fair comparison of the encoder stage. When STNs are used in the encoder, the overall sharpness of the images is improved since all frames are aligned.} %\hw{12.2 vs 12.3 (sf1.5)}
  \label{fig:ablation_stn}
\end{figure}

\begin{figure}
    \centering
    % --- Main Subfigure (A) with nested images ---
    \begin{subfigure}[b]{0.49\linewidth}
        \centering
        % Nested Top Image (no caption)
        \begin{tikzpicture}
            \node[anchor=south west, inner sep=0] (image) at (0,0)
                {\includegraphics[width=\linewidth]{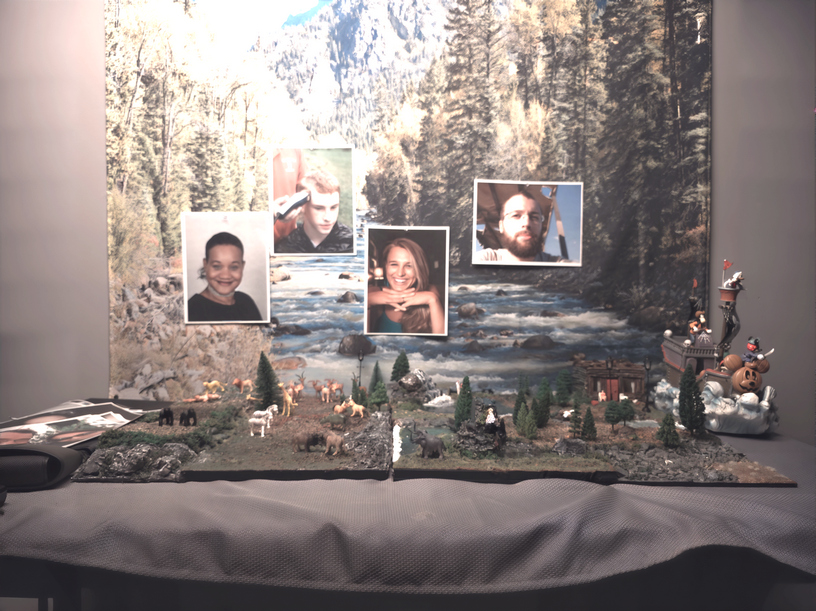}};
            \begin{scope}[x={(image.south east)}, y={(image.north west)}]
                \draw[red, thick, -{Latex[length=2mm]}] (0.2,0.85) -- (0.1,0.8);
                \draw[red, thick, -{Latex[length=2mm]}] (0.2,0.7) -- (0.1,0.65);
                \draw[red, thick, -{Latex[length=2mm]}] (0.2,0.52) -- (0.1,0.47);
            \end{scope}
        \end{tikzpicture}
        
        % \vspace{\baselineskip} % Add vertical space between top and bottom nested images
        % Nested Bottom 2 Images (no captions)
        \begin{subfigure}[b]{0.48\linewidth}
            \centering
            \begin{tikzpicture}
                \node[anchor=south west, inner sep=0] (image) at (0,0)
                    {\includegraphics[width=\textwidth]{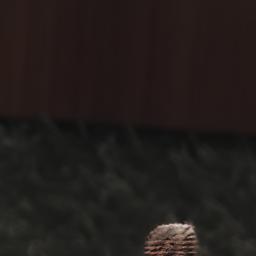}};
                \begin{scope}[x={(image.south east)}, y={(image.north west)}]
                    \draw[red, thick, -{Latex[length=2mm]}] (0.5,0.3) -- (0.6,0.1);
                \end{scope}
            \end{tikzpicture}
            % \includegraphics[width=\linewidth]{figs/comparison3/5_ours_wo_HF.jpg}
            % No caption here
        \end{subfigure}
        \hfill
        \begin{subfigure}[b]{0.48\linewidth}
            \centering
            \begin{tikzpicture}
                \node[anchor=south west, inner sep=0] (image) at (0,0)
                    {\includegraphics[width=\textwidth]{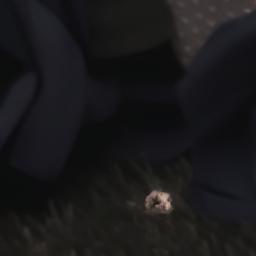}};
                \begin{scope}[x={(image.south east)}, y={(image.north west)}]
                    \draw[red, thick, -{Latex[length=2mm]}] (0.5,0.5) -- (0.6,0.3);
                \end{scope}
            \end{tikzpicture}
            % \includegraphics[width=\linewidth]{figs/comparison3/6_ours_wo_HF.jpg}
            % No caption here
        \end{subfigure}
        \caption{\tiny{Without $\mathcal{L}_\text{VSD-HF}$}} 
        \label{fig:A}
    \end{subfigure}
    % --- Main Subfigure (B) with nested images ---
    \begin{subfigure}[b]{0.49\linewidth}
        \centering
        % Nested Top Image (no caption)
        
        \begin{tikzpicture}
            \node[anchor=south west, inner sep=0] (image) at (0,0)
                {\includegraphics[width=\linewidth]{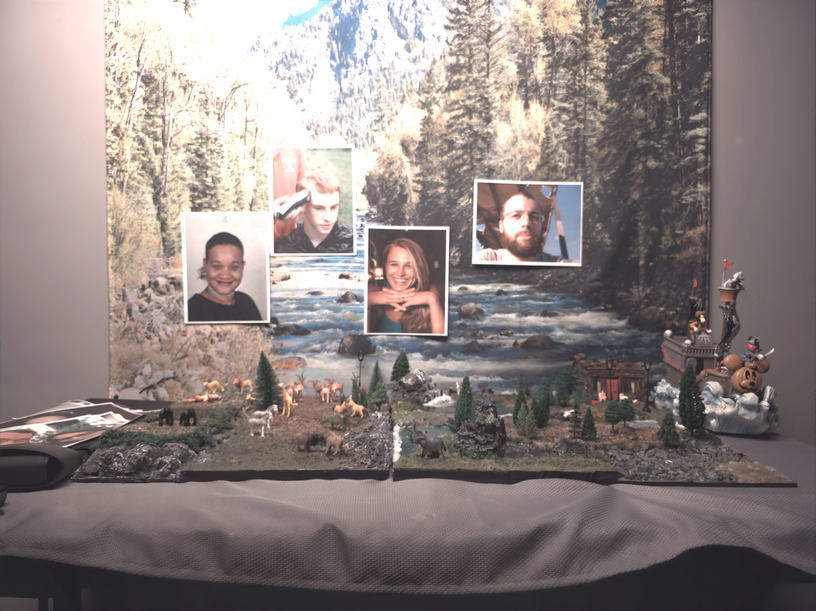}};
            \begin{scope}[x={(image.south east)}, y={(image.north west)}]
                \draw[green, thick, -{Latex[length=2mm]}] (0.2,0.85) -- (0.1,0.8);
                \draw[green, thick, -{Latex[length=2mm]}] (0.2,0.7) -- (0.1,0.65);
                \draw[green, thick, -{Latex[length=2mm]}] (0.2,0.52) -- (0.1,0.47);
            \end{scope}
        \end{tikzpicture}
        
        % \vspace{\baselineskip} % Add vertical space between top and bottom nested images
        % Nested Bottom 2 Images (no captions)
        \begin{subfigure}[b]{0.48\linewidth}
            \centering
            \begin{tikzpicture}
                \node[anchor=south west, inner sep=0] (image) at (0,0)
                    {\includegraphics[width=\textwidth]{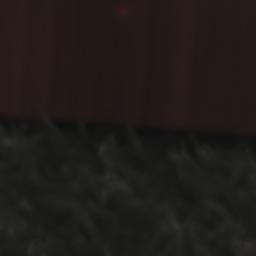}};
                \begin{scope}[x={(image.south east)}, y={(image.north west)}]
                    \draw[green, thick, -{Latex[length=2mm]}] (0.5,0.3) -- (0.6,0.1);
                \end{scope}
            \end{tikzpicture}
            % \includegraphics[width=\linewidth]{figs/comparison3/5_ours.jpg}
            % No caption here
        \end{subfigure}
        \hfill
        \begin{subfigure}[b]{0.48\linewidth}
            \centering
            \begin{tikzpicture}
                \node[anchor=south west, inner sep=0] (image) at (0,0)
                    {\includegraphics[width=\textwidth]{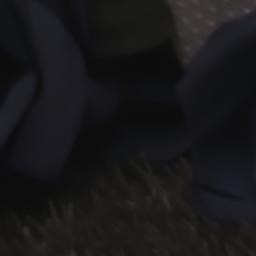}};
                \begin{scope}[x={(image.south east)}, y={(image.north west)}]
                    \draw[green, thick, -{Latex[length=2mm]}] (0.5,0.5) -- (0.6,0.3);
                \end{scope}
            \end{tikzpicture}
            % \includegraphics[width=\linewidth]{figs/comparison3/6_ours.jpg}
            % No caption here
        \end{subfigure}
        \caption{\tiny{With $\mathcal{L}_\text{VSD-HF}$}}
        \label{fig:B}
    \end{subfigure}
    
   % \begin{subfigure}{0.49\linewidth}
   %    \centering
   %    \includegraphics[width=\linewidth]{figs/comparison3/without_hf_loss.jpg}
   %    \caption{Without HF loss}
   %  \end{subfigure}   
    % \begin{subfigure}{0.49\linewidth}
    %   \centering
    %   \includegraphics[width=\linewidth]{figs/comparison3/with_hf_loss.jpg}
    %   \caption{With HF Loss}
    % \end{subfigure} 
  
  \caption{Effectiveness of the high-frequency (HF) VSD loss. Both models use STNs for alignment. (a) The baseline VSD introduces low-frequency patching artifacts (top, contrast enhanced) and hallucinations (bottom). (b) Our $\mathcal{L}_\text{VSD-HF}$ successfully prevents both types of artifacts.} % \hw{12.3 vs 12.4 (sf0.5)}
  % Effectiveness of using HF loss. The STNs are used in the encoder to ensure the frames are aligned and to ensure that the detail gain is purely from the HF loss. Without HF loss, the first row shows that uniform areas exhibit patching artifacts since the low frequencies vary across patches. The contrast of the two images is increased by 40\% to highlight these artifacts. Second row shows some parts of the images generated using the model trained without HF loss hallucinates artifacts that are not present in the image.
  \label{fig:ablation_hf}
\end{figure}

\textbf{Single Frame vs Multi-Frame}: We also experimentally validate the benefit of using multiple frames by using only the RAW base frame and GT pairs to train single-frame baselines. The comparison with these baselines is presented in \cref{tab:singleframe_comparison}, which shows that our multi-frame approach outperforms all single-frame RAW-to-RGB methods. A qualitative comparison in \cref{fig:single_frame} shows that GenMFSR has greater fidelity than single-frame methods.

\textbf{Deployment and Practicality.} We note that deploying a 1.3B parameter latent diffusion model is impractical for real-time, viewfinder-level mobile processing (e.g., 30 fps). Consequently, GenMFSR is designed to operate as an \textit{Asynchronous Enhancement Module}, akin to premium computational photography pipelines (e.g., Deep Fusion or Night Sight) that process high-fidelity bursts in the background post-capture. By utilizing one-step distillation, our framework reduces the latency of standard diffusion from dozens of seconds to a single forward pass, making high-end generative burst restoration practically viable for modern mobile Neural Processing Units (NPUs). Further distillation and mobile deployment is made possible by recent works\cite{sanyal2026nanosd,noroozi2024edge}.

% \begin{figure}
%     \centering
%     \includegraphics[width=\linewidth]{figs/results_1.5x.png}
%     \caption{Qualitative comparison of our method with baselines for the 12MP to 24MP super-resolution case. \hw{raw average frame}}
%     \label{fig:qualitative_gan}
% \end{figure}

\section{Ablation study}

% \textbf{Effectiveness of the RAW frame alignment from STNs.}
We trained the proposed network without using STN blocks in the encoder to verify the performance gain from using STNs. \cref{fig:ablation_stn} illustrates how the addition of STNs produces sharper images in the final reconstruction. This validates the efficacy of using STNs for image registration in the encoder stage, aligning the frames and making the subsequent tasks easier for the generative network.

\begin{figure}[h]
    \centering
    \includegraphics[width=\linewidth]{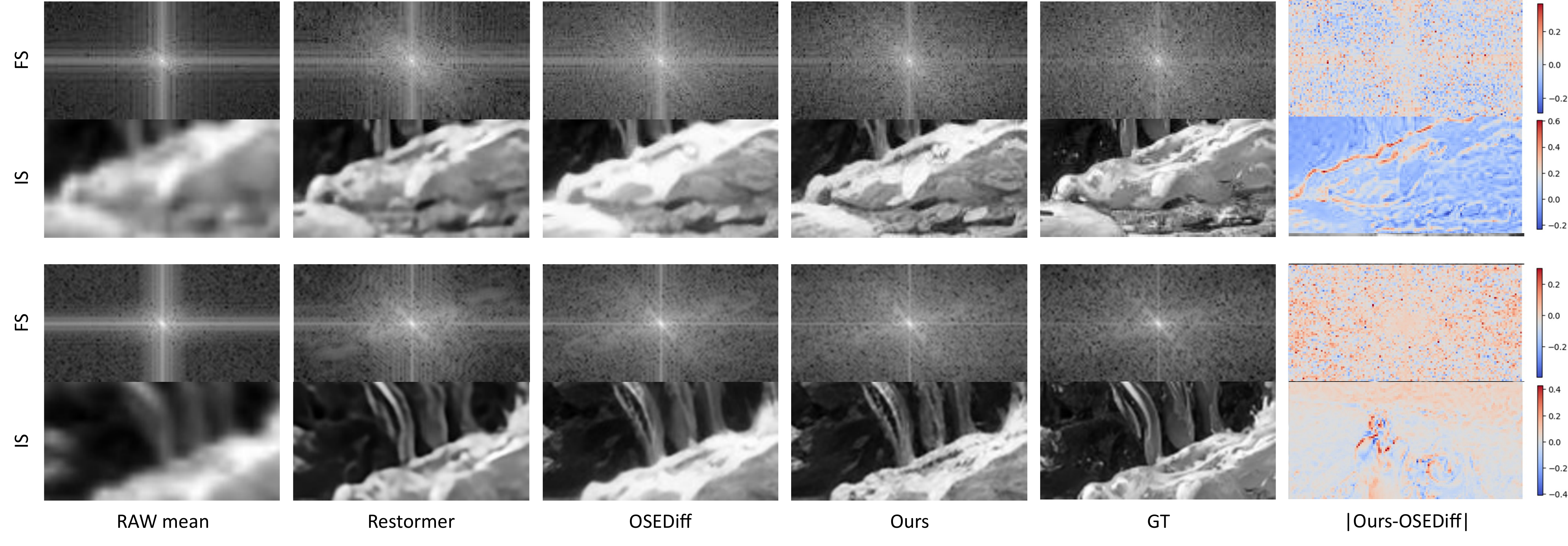}
    \caption{Comparison of competing results in the Frequency Space (FS) and Image Space (IS). We highlight the absolute error between ours and OSEDiff \cite{wu2024neurips_osediff} which is a close competing method.}
    \label{fig:ft_comparison_results}
\end{figure}

% \textbf{Detail generation from High-Frequency VSD loss.}
% The results in \cref{tab:main_comparison} validate the efficacy of using our proposed loss function. 
Furthermore, \cref{fig:ablation_hf} demonstrates that restricting the generative prior ensures low-frequency uniformity across patches. This explicitly prevents the severe boundary artifacts (top row) and structural hallucinations (bottom row) that plague the unconstrained baseline VSD.

Fourier spectra analysis \cref{fig:ft_comparison_results} confirms naive VSD alters low-frequency magnitudes, causing the color shifts and hallucinations as shown in \cref{fig:ablation_hf}. Conversely, GenMFSR matches the ground-truth low-frequency spectrum while extrapolating high-frequency bounds, thereby reducing low-frequency structural error compared to standard VSD. 
\cref{fig:ft_comparison_results} also highlights the difference between OSEDiff \cite{wu2024neurips_osediff} and our method in the image space and frequency space where ours shows improvements in edges and high frequency components. This confirms that our subspace projection successfully bridges the RAW-to-RGB domain gap, recovering the critical sub-pixel textures lost by standard regression models without triggering the severe structural drift inherent to unconstrained diffusion priors.

% \subsubsection{Spectral Validation and Low-Frequency Preservation}: \hw{To validate our core hypothesis that HF-VSD prevents structural domain collisions, we analyze the spectral properties of our reconstructions.  As observed in the Fourier spectra, naive VSD aggressively alters the low-frequency magnitudes of the RAW input, causing the severe color shifts and structural hallucinations shown in \cref{fig:ablation_hf}. In contrast, GenMFSR perfectly matches the low-frequency spectrum of the ground truth while successfully extrapolating the high-frequency bounds.  Error maps computed on the low-pass filtered outputs further confirm that HF-VSD reduces low-frequency structural error by over $40\%$ compared to standard VSD.}

% \subsubsection{Cutoff Frequency Sensitivity and Tradeoffs}: \hw{The boundary between the observable and underdetermined subspaces is modeled by the cutoff frequency of our smooth mask $h$.  We evaluate the sensitivity of GenMFSR to this cutoff. Shifting the cutoff to extremely low frequencies reduces the method to naive VSD (improving FID but destroying PSNR via hallucination). Conversely, an excessively high cutoff reduces the model to deterministic regression (maximizing PSNR but yielding "waxy" LPIPS scores). Empirical evaluation demonstrates that our chosen cutoff provides a robust plateau, achieving optimal perception-distortion balance without introducing heuristic instability.}

\section{Conclusion}
\label{sec:conclusion}
We introduced GenMFSR, the first unified, one-step generative framework for multi-frame RAW-to-RGB restoration. By embedding explicit frame alignment within the encoder and restricting generative priors to the high-frequency latent subspace, GenMFSR successfully synthesizes photorealistic sub-pixel details without structural hallucinations. This offers a highly effective, sensor-stage architecture for modern mobile ISPs. 
% While our empirical dataset focuses on the dominant global motion of hand tremors, our STN-based alignment proves robust even to unmodeled secondary artifacts (e.g., rolling shutter) due to the fast readout speeds of modern sensors.
Because the framework relies on a frozen Stable Diffusion prior to bypassing heavy training costs, it naturally inherits the foundation model's constraints, such as minor global color shifts or VAE-induced text artifacts. Future work will explore rectified flow and global foveal context to refine patch-based prompting, further bridging the gap between powerful generative priors and physical sensor constraints.

{\small
\bibliographystyle{ieee_fullname}
\bibliography{main}
}

\clearpage
\setcounter{page}{1}
\maketitlesupplementary

\appendix

\section{Introduction}
This supplementary material provides a theorical background on score distillation and how it is related to high-frequency feature generation, more information on model training and the dataset in \cref{sec:training}. Additional experiment results are presented in \cref{sec:exp}, including qualitative results that demonstrdemonstrating super-resolution scaling agnosticism and perceptual quality improvements overdversarial Networks (GANs). We provide further details of the user study, along with some examples of the images used in \cref{sec:user_study}.
% \section{Importance of targeting high-frequency}

\section{Theoretical background}

\subsection{Variational Score Distillation (VSD)}
VSD works by minimizing the KL divergence $\mathcal{D}_\text{KL}(q_{\vtheta}(\hat{\vz}) || p({\vz}))$ between the distribution of noisy predicted latents $\hat{\vz}$ and noisy real latents $\vz$, denoted by $q_{\vtheta}$ and $p$ respectively. This can be implemented using techniques such as adversarial training, where a discriminator is trained to identify $p$ and $q_{\vtheta}$. The goal of the generator is to predict $\hat{\vx}$ such that $q_{\vtheta} \simeq p$, which reduces the KL divergence. 
This will regularize the generator to produce images that are close to the ground-truth distribution. However, according to the authors of VSD \cite{wang2023prolificdreamer}, the distribution of $\vz$ can be sparse, which may yield suboptimal results. 

But if we add noise to these latent variables ($\hat{\vz}_t = \alpha_t\hat{\vz} + \beta_t\vepsilon_t$), we can diffuse the distributions to easily calculate a valid KL divergence $\mathcal{D}_\text{KL}(q_{\vtheta,t}(\hat{\vz}_t)||p_t(\vz_t))$ where $t$ is the time step, $\vepsilon \in \mathcal{N}(0,\mI)$, and $\alpha_t$, $\beta_t$ are noise schedule coefficients.% OSEDIff uses KL to be q(\hat{z}) || p(\hat{z}) how ???
This tractable problem can be written as,
\begin{equation}
\label{eq:vsd_problem}
    \vtheta^* = \arg\min_{\vtheta} \mathbb{E}_{t}[(\alpha_t/\beta_t)\vomega(t)\mathcal{D}_\text{KL}(q_{\vtheta,t}(\hat{\vz}_t)||p_t(\vz_t))]
\end{equation}
where $\vomega(t)$ is a weighting factor. 

% author show that KL at t>0 <=> score at t>0 <=> KL at t=0
Since pretrained diffusion models such as Stable Diffusion already have the capability to estimate the distribution of diffused real images, we use a frozen Stable Diffusion model $\Omega_{\vphi'}$ to estimate the distribution (or score function) of $\vz_t$.
The distribution of the predicted latents $\hat{\vz}_t$ is estimated using another noise prediction network $\Omega_{\vphi}$, which is trained on the predicted latents with the standard diffusion loss,
\begin{equation}
    \mathcal{L}_\text{diff} = \mathbb{E}_{t,\vepsilon, \hat{\vz}} \mathcal{L}_\text{MSE}(\Omega_{\vphi}(\alpha_t\hat{\vz} + \beta_t\vepsilon;t, \vc),\vepsilon)
\end{equation}

Using these estimations of the distributions, we can solve the problem in \cref{eq:vsd_problem} by calculating its gradient. For that, we define the gradient of VSD loss as,
\begin{equation*}
    \nabla_{\hat{\vx}}\mathcal{L}_\text{VSD}(\hat{\vx}) \triangleq \mathbb{E}_{t,\vepsilon}\left[\vomega(t)(\Omega_{\vphi'}(\vz_t;t, \vc)-\Omega_{\vphi}(\hat{\vz}_t;t, \vc))\frac{\partial \vz_t}{\partial \hat{\vx}}\right]
\end{equation*}
Since the decoder is frozen, the VSD loss can be computed in the latent space. Therefore, we can update the VSD loss to backpropagate to $\vtheta$ as follows,
\begin{align}
    \nabla_{\vtheta}\mathcal{L}_\text{VSD}(\hat{\vx}) &= \nabla_{\hat{\vx}}\mathcal{L}_\text{VSD}(\hat{\vx})\frac{\partial \hat{\vx}}{\partial \vtheta} \cr
    &= \mathbb{E}_{t,\vepsilon}\left[\vomega(t)(\Omega_{\vphi'}(\vz_t;t,\vc)-\Omega_{\vphi}(\hat{\vz}_t;t,\vc))\frac{\partial \vz_t}{\partial \vtheta}\right]
\end{align}

When we know the gradient of the loss, we can find the global minima by setting the gradient to zero, assuming the loss is convex. Therefore, VSD loss can be incorporated into training as follows,
\begin{equation}
    \tilde{\mathcal{L}}_\text{VSD}(\hat{\vx}) = \mathcal{L}_\text{MSE}[\vomega(t)\Omega_{\vphi'}(\vz_t;t,\vc), \vomega(t)\Omega_{\vphi}(\hat{\vz}_t;t,\vc)]
\end{equation}

\subsection{HF-VSD via Null-Space Projection}
We can formulate the RAW-to-RGB multi-frame super-resolution task as a linear inverse problem:
\begin{equation}
    y = A x + n
\end{equation}
where $x \in \mathbb{R}^D$ is the ground-truth high-resolution RGB image, $y \in \mathbb{R}^d$ ($d \ll D$) represents the degraded RAW burst measurements, $A$ is the forward degradation operator (blur, downsampling, mosaicing), and $n$ is sensor noise.

To sample from the posterior $p(x|y)$, score-based generative models utilize Langevin dynamics, which requires the posterior score:
\begin{equation}
    \nabla_x \log p(x|y) = \underbrace{\nabla_x \log p(y|x)}_{\text{Likelihood Score}} + \underbrace{\nabla_x \log p(x)}_{\text{Prior Score}}
\end{equation}

The forward operator $A$ induces a decomposition of the image space into a range space (observable subspace) and a null space (unobservable subspace). Let $A^\dagger$ be the pseudo-inverse of $A$. We can define the orthogonal projection matrices for the range space $P_L = A^\dagger A$ and the null space $P_H = I - A^\dagger A$. 

Any image can be decomposed as $x = P_L x + P_H x$. The likelihood score $\nabla_x \log p(y|x)$ only provides gradients for the observable subspace $P_L x$. The prior score $\nabla_x \log p(x)$ provides gradients for both subspaces. 

In our framework, the prior $p(x)$ is parameterized by an sRGB-trained latent diffusion model, while $y$ resides in the linear RAW domain. Consequently, the prior's score in the observable subspace, $P_L \nabla_x \log p(x)$, is heavily biased by sRGB tone-mapping and color processing, conflicting with the true physical measurements $y$.

To prevent this domain collision, we construct a modified posterior score that strictly isolates the prior to the null space:
\begin{equation}
    \nabla_x \log \tilde{p}(x|y) = \nabla_x \log p(y|x) + P_H \nabla_x \log p(x)
\end{equation}

In the spatial frequency domain, the degradation operator $A$ behaves as a low-pass filter. Therefore, the range space $P_L$ corresponds to low spatial frequencies, and the null space $P_H$ corresponds to high spatial frequencies. 

By applying a high-pass Fourier mask $h$ to the gradients of the Variational Score Distillation (VSD) loss, our proposed HF-VSD acts as an efficient, differentiable approximation of the null-space projection $P_H$:
\begin{equation}
    \nabla_x \mathcal{L}_{HF-VSD} \approx P_H \nabla_x \log p(x)
\end{equation}
which can be calculated using \cref{eq:loss_reg}. This demonstrates that HF-VSD is not a heuristic image-space filter, but a theoretically grounded technique to enforce complementary score matching in partially observable inverse problems.

% \subsection{Bridging the Null-Space and the Fourier Domain}

% Let the local burst degradation process be modeled as a linear system $y = Ax + n$. The operator $A$ primarily models optical blurring (lens Point Spread Function) and spatial downsampling (pixel aperture integration). 

Computing the exact null-space projection $P_H = I - A^\dagger A$ is computationally intractable for high-resolution images within a latent diffusion framework due to the massive dimensionality of $A$ and the non-rigid alignment modeled by our STNs.

However, if we locally approximate $A$ as a spatially invariant blur operator combined with uniform sampling, $A$ represents a discrete convolution. Under circular boundary conditions, $A$ is a Block-Circulant with Circulant Blocks (BCCB) matrix. 

A fundamental property of BCCB matrices is that they are diagonalized by the 2D Discrete Fourier Transform (DFT) matrix, denoted as $\mathcal{F}$. Thus, we can decompose $A$ as:
\begin{equation}
    A = \mathcal{F}^{-1} \Lambda \mathcal{F}
\end{equation}
where $\Lambda$ is a diagonal matrix containing the Fourier coefficients of the degradation kernel (the Optical Transfer Function, OTF).

Because optical blur and pixel integration act as low-pass filters, the diagonal elements of $\Lambda$ approach zero at high spatial frequencies. The pseudo-inverse projection operator onto the observable range space becomes:
\begin{equation}
    P_L = A^\dagger A = \mathcal{F}^{-1} (\Lambda^\dagger \Lambda) \mathcal{F}
\end{equation}
The term $(\Lambda^\dagger \Lambda)$ is a diagonal matrix with ones at low frequencies (where the signal is observable) and zeros at high frequencies (where the OTF attenuates the signal below the noise floor). 

Consequently, the projection onto the unobservable null-space $P_H$ is:
\begin{equation}
    P_H = I - P_L = \mathcal{F}^{-1} (I - \Lambda^\dagger \Lambda) \mathcal{F}
\end{equation}
The matrix $(I - \Lambda^\dagger \Lambda)$ is exactly a high-pass frequency mask in the Fourier domain. 

Therefore, while a fixed Fourier mask $h$ does not capture the complex, spatially varying local deformations of handheld motion, it serves as the mathematically optimal and computationally tractable approximation of the true null-space projection for camera sensor degradations. By applying this mask to the score gradients, HF-VSD effectively restricts the generative prior to the unobservable high-frequency subspace without requiring the intractable computation of the true spatially varying pseudo-inverse.

\section{Model description and training}
\label{sec:training}

\subsection{Model details}
We use the pretrained Stable Diffusion 2.1 model as the generator U-net $\Omega_\theta$ and trainable regularizer $\Omega_\phi$. Selected layers of these U-nets and the encoder ${E}_\theta$ are trained using Low Rank Adapters (LoRAs). The decoder $\mathcal{D}_\theta$ was frozen and used for decoding the reconstructed latents. The frozen regularizer $\Omega_{\phi'}$ is initialized from Stable Diffusion 2.1 and kept frozen.
The high-pass filter is initialized with $\alpha=0.8$,$\beta=0.2$, and $\gamma=4$. This filter is used in place of the Gaussian filter to provide a controllable limit on the maximum allowed high frequency. 

\subsection{Frame alignment for fidelity improvement.}
A critical challenge in multi-frame RAW processing is temporal misalignment caused by hand tremor. Prior generative models operate on single frames and fail catastrophically when presented with misaligned bursts. To solve this, we embed Spatial Transformer Networks (STNs) directly within the multi-frame encoder $E_{\vtheta}$. The distortion is modeled by a homography transformation relative to a base frame (e.g., the first frame). The encoder's input layer is replaced with a convolutional layer with a channel input size of $33$ to accommodate multiple RAW frames. 

To accommodate the 3-channel prior of the pre-trained sRGB encoder while preserving the physical properties of the sensor data, we apply a deterministic linear channel extraction to the Bayer RAW inputs. Specifically, the Red and Blue channels are extracted and linearly upsampled ($2\times$) to the full spatial resolution. The two Green channels are similarly extracted, upsampled, and averaged. These are stacked to form a 33-channel input volume for the 11-frame burst. 
We explicitly note that while this constitutes a naive linear demosaicing, this 3-channel representation strictly retains the radiometric linearity, original bit-depth, and uncompressed dynamic range of the raw sensor signal. By doing so, we successfully reduce the dimensional domain gap for the generative VAE while completely bypassing the destructive, non-linear tone-mapping, gamma correction, and heavy spatial denoising applied by standard mobile ISPs. 

The trainable encoder ($E_{\vtheta}$) and diffusion networks ($\Omega_{\vtheta}, \Omega_{\vphi}$) are finetuned by introducing trainable LoRA \cite{hu2022lora} layers. The localization networks of Spatial Transformer Networks (STNs) \cite{jaderberg2015spatial} in $E_{\vtheta}$ are initialized to give an identity homography matrix at the start and are then trained from scratch. While training a custom RAW-domain VAE might seem intuitive, doing so would fundamentally alter the latent space, disconnecting it from the pre-trained Stable Diffusion prior. Because VSD relies on the pre-trained prior ($\Omega_{\phi'}$) to provide score estimates, our encoder must map the 33-channel RAW burst directly into the standard SD latent space. Therefore, we freeze the standard decoder $D_\theta$ and force the encoder $E_\theta$ to bridge the domain gap.

% Because the STNs are applied to the deep feature embeddings extracted by the initial convolutional layers rather than the raw pixel intensities directly, the network is able to align structural semantics without aggressively correlating the native sensor noise.

\subsection{Dataset Capture and Processing Details}
\label{supp:dataset}

\textbf{Hardware and Capture Logistics.} 
We constructed an aligned dataset using registered long and short-exposure multi-frame (MF) images captured on a mobile device mounted on a tripod, ensuring strictly static scenes to eliminate unwanted camera shake. Capturing data at the beginning of the ISP pipeline ensures that the RAW frames contain authentic, sensor-specific Poisson-Gaussian noise distributions before any destructive demosaicing occurs—a property that cannot be perfectly replicated using synthetic inverse-ISP pipelines. To maintain consistent luminance across the pairs, the brightness of the short-exposure and long-exposure images was equalized by adjusting the ISO during capture. The ground-truth high-resolution (HR) RGB images were obtained by fusing and demosaicing the pristine long-exposure RAW frames, thereby minimizing noise and eliminating blur.

\textbf{Continuous Motion Simulation.}
To simulate handheld motion without introducing the domain gaps typical of purely synthetic affine warps, we utilized a supplementary dataset of real-world handheld bursts \cite{khan2025mfsr}. Crucially, differing from \cite{khan2025mfsr} where homographies were sampled independently for each input frame, our empirical simulation extracts groups of continuous homographies from entire multi-frame captures. We apply these motion trajectories jointly to the static short-exposure frames. This modification preserves the temporal correlations inherent in actual human hand tremors, resulting in highly realistic synthetic motion sequences that accurately reflect mobile burst photography.

\textbf{Data Processing and Splits.}
During the data loading pipeline, RAW frames and ground-truth RGB images are extracted, and the handshake motion is applied. To process the motion, the RAW frames are first subsampled and resized to the original size $H \times W$ using bilinear interpolation, followed by the application of the extracted homography matrices. The RAW and RGB images are subsequently cropped to a patch size of $256 \times 256$. The cropped and subsampled RAW frames are then downsampled (via a Bayer-preserving Space-to-Depth operation) and resized back to $256 \times 256$ to construct the final input volume to the model. Out of the collected dataset, 2000 distinct scenes are utilized for training, 100 for validation, and 100 for testing.

\subsection{Training}
The model is trained on a NVIDIA H100 GPU with a batch size of 4 for around 4 days. We used the Adam optimizer with a learning rate of 5e-5. The LoRAs are initialized to give an identity at the beginning. The input layer to the encoder is initialized using random weights. The training procedure is outlined in \cref{alg:training}, where $Y$ represents the caption generator and other modules are consistent with the notations presented in the main paper.

The dataset was split into training, validation, and test sets with a ratio of 90:5:5, and we trained our model with $\lambda=1$, $\alpha=0.8$, $\beta=0.2$, and $\gamma=4$. We used the Adam optimizer with a learning rate of $5 \times 10^{-5}$. Training was performed for 500k iterations, and the checkpoint with the best MSE+LPIPS on the validation set was selected as the final checkpoint for evaluation of all the considered methods.

 \begin{algorithm}
    \caption{Training Procedure}
    \label{alg:training}
    \begin{algorithmic}[1]
        \Statex Load pretrained $\Omega_{\phi'}, E_\theta, D_\theta, Y$
        \Statex Copy weights $\Omega_\theta, \Omega_\phi \gets \Omega_{\phi'}$
        \Statex Initialize LoRA on $\Omega_\theta, \Omega_\phi, E_\theta$
        \Statex Initialize STNs on $E_\theta$

        \Procedure{Train}{Dataset $\mathcal{D}_\text{train}$}
        \For{$i$ from $0$ to $N$}
            \State $\vy, \vx \sim \mathcal{D}_\text{train}$
            \State $c \gets Y(\vy)$
            \State $\vz_T \gets E_\theta(\vy)$ \Comment{Encoding with alignment}
            \State $\hat{\vv}_0 \gets \Omega_\theta(\vz_T; T, \vc)$
            \State $\hat{\vz} \gets \frac{1}{\sqrt{\Bar{\alpha}_T}}(\vz_T - \sqrt{1-\Bar{\alpha}_T}\hat{\vv}_0)$
            \State $\hat{\vx} \gets D_\theta(\hat{\vz})$
            \State $\mathcal{L}_\text{data}(\hat{\vx}, \vx) \gets \mathcal{L}_\text{MSE}(\hat{\vx}, \vx) + \lambda_\text{LPIPS}\mathcal{L}_\text{LPIPS}(\hat{\vx}, \vx)$
            \Statex
            \State $\vm \gets \mathcal{F}^{-1}[\vh \cdot \mathcal{F}[\hat{\vz}]]$ \Comment{Latent space HF mask}
            \State $\vepsilon \sim \mathcal{N}(0, \mI)$
            \State $t \sim [20, 980]$
            \State $\hat{\vz}_t \gets \alpha_t \hat{\vz} + \beta_t\vepsilon$
            \State $\vv_\phi \gets \Omega_\phi(\hat{\vz}_t; t, \vc)$
            \State $\vv_{\phi'} \gets \Omega_{\phi'}(\hat{\vz}_t; t, \vc)$
            \State $\vomega \gets \frac{1}{\text{mean}(\|\hat{\vz} - \vv_{\phi}\|^2)} $
            \State $\mathcal{L}_\text{VSD-HF}(\hat{\vx}) = \mathcal{L}_\text{MSE}(\hat{\vz}_t, \hat{\vz}_t - \vm \odot \vomega(\vv_{\phi'} - \vv_\phi))$
            \Statex
            \State $\vepsilon \sim \mathcal{N}(0, \mI)$
            \State $t \sim [0, 1000]$
            \State $\hat{\vz}_t \gets \alpha_t \hat{\vz} + \beta_t\vepsilon$
            \State $\mathcal{L}_\text{diff}  \gets \mathcal{L}_\text{MSE}(\Omega_{\vphi}(\hat{\vz}_t;t, \vc),\vepsilon)$
            \Statex
            \State Update $\theta$ using $\mathcal{L}_\text{data} + \lambda\mathcal{L}_\text{VSD-HF}$
            \State Update $\phi$ using $\mathcal{L}_\text{diff}$
        \EndFor
        \EndProcedure
    \end{algorithmic}
\end{algorithm}

\subsection{Inference}
We disregard the models used for regularization during the inference. Thus, to obtain the final reconstructed image, we need only $\Omega_\theta$, $E_\theta$, and $D_\theta$ modules. The inference procedure is given in \cref{alg:inference}.

\begin{algorithm}
    \caption{Inference Procedure}
    \label{alg:inference}
    \begin{algorithmic}[1] 
        \Statex Load trained $\Omega_{\theta}, E_\theta, D_\theta, Y$
        
        \Procedure{Infer}{Dataset $\mathcal{D}_\text{test}$}
            
            \State $\vy \sim \mathcal{D}_\text{test}$
            \State $c \gets Y(\vy)$
            \State $\vz_T \gets E_\theta(\vy)$
            \State $\hat{\vv}_0 \gets \Omega_\theta(\vz_T; T, \vc)$
            \State $\hat{\vz} \gets \frac{1}{\sqrt{\Bar{\alpha}_T}}(\vz_T - \sqrt{1-\Bar{\alpha}_T}\hat{\vv}_0)$
            \State $\hat{\vx} \gets D_\theta(\hat{\vz})$
            \State \Return $\hat{\vx}$
        \EndProcedure
    \end{algorithmic}
\end{algorithm}

\section{Additional experiments}
\label{sec:exp}
\subsection{Inference-Time Scale Agnosticism:}
We discussed the primary results for the $4\times$ SR case in \cref{tab:main_comparison}. However, our model is not restricted to a fixed SR scale factor. To demonstrate this scale-agnostic capability, we illustrate qualitative results in \cref{fig:qualitative_gan}, and further details are given in the supplementary material, where our method again shows the best fidelity compared to the competing generative baselines. The illustration in \cref{fig:qualitative_gan} further supports this statement by demonstrating improved texture generation from our method compared to GANs, particularly in the higher frequencies.
\begin{figure}
    \begingroup
      \setlength{\tabcolsep}{0.1pt} % Default value: 6pt
      \renewcommand{\arraystretch}{1} % Set array row spacing to 0
      % \resizebox{\linewidth}{!}{%
      \begin{tabular}{cccc}
        % --- First Row with Vertical Caption ---
        \rotatebox{90}{\hspace*{2em}{$4\times$ SR}} &
        \begin{subfigure}{0.3\linewidth}
            \centering
            \begin{tikzpicture}
                \node[anchor=south west, inner sep=0] (image) at (0,0)
                    {\includegraphics[width=\textwidth,trim={0 0 0 0}]{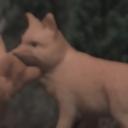}};
                \begin{scope}[x={(image.south east)}, y={(image.north west)}]
                    % \draw[red, thick, -{Latex[length=2mm]}] (0.85,0.45) -- (0.65,0.35);
                \end{scope}
            \end{tikzpicture}
        \end{subfigure} &
        \begin{subfigure}{0.3\linewidth}
            \centering
            \begin{tikzpicture}
                \node[anchor=south west, inner sep=0] (image) at (0,0)
                    {\includegraphics[width=\textwidth,trim={0 0 0 0}]{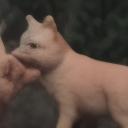}};
                \begin{scope}[x={(image.south east)}, y={(image.north west)}]
                    % \draw[red, thick, -{Latex[length=2mm]}] (0.85,0.45) -- (0.65,0.35);
                \end{scope}
            \end{tikzpicture}
        \end{subfigure} &
        \begin{subfigure}{0.3\linewidth}
            \centering
            \begin{tikzpicture}
                \node[anchor=south west, inner sep=0] (image) at (0,0)
                    {\includegraphics[width=\textwidth,trim={0 0 0 0}]{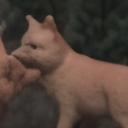}};
                \begin{scope}[x={(image.south east)}, y={(image.north west)}]
                    % \draw[green, thick, -{Latex[length=2mm]}] (0.85,0.45) -- (0.65,0.35);
                \end{scope}
            \end{tikzpicture}
        \end{subfigure} \\
        
        % Add a vertical space between the two rows
        \noalign{\vspace{-3.4pt}}
        % \hfill
        % --- Second Row with Vertical Caption ---
        \rotatebox{90}{\hspace*{3em}{$3\times$ SR}} &
        % First figure in the second row
        \begin{subfigure}{0.3\linewidth}
          \centering
            \begin{tikzpicture}
                \node[anchor=south west, inner sep=0] (image) at (0,0)
                    {\includegraphics[width=\textwidth,trim={0 0 0 0}]{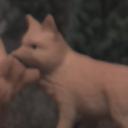}};
                \begin{scope}[x={(image.south east)}, y={(image.north west)}]
                    % \draw[red, thick, -{Latex[length=2mm]}] (0.85,0.45) -- (0.65,0.35);
                \end{scope}
            \end{tikzpicture}
          \caption{NAFNet}
        \end{subfigure} &
        % Second figure in the second row
        \begin{subfigure}{0.3\linewidth}
            \centering
            \begin{tikzpicture}
                \node[anchor=south west, inner sep=0] (image) at (0,0)
                    {\includegraphics[width=\textwidth,trim={0 0 0 0}]{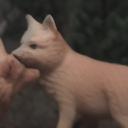}};
                \begin{scope}[x={(image.south east)}, y={(image.north west)}]
                    % \draw[red, thick, -{Latex[length=2mm]}] (0.85,0.45) -- (0.65,0.35);
                \end{scope}
            \end{tikzpicture}
          \caption{VSD}
        \end{subfigure} &
        % Third figure in the second row
        \begin{subfigure}{0.3\linewidth}
            \centering
            \begin{tikzpicture}
                \node[anchor=south west, inner sep=0] (image) at (0,0)
                    {\includegraphics[width=\textwidth,trim={0 0 0 0}]{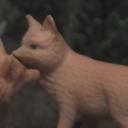}};
                \begin{scope}[x={(image.south east)}, y={(image.north west)}]
                    % \draw[green, thick, -{Latex[length=2mm]}] (0.85,0.45) -- (0.65,0.35);
                \end{scope}
            \end{tikzpicture}
          \caption{Ours}
        \end{subfigure} \\
      \end{tabular}
      % }% End of \resizebox
      \endgroup % End the group
      \caption{Qualitative comparison of our method with adversarially-trained baseline (NAFNet) and generative baseline (OSEDiff) for $4\times$ and $3\times$ super-resolution (SR). Our model demonstrates better feature reconstruction under different SR factors.}
      \label{fig:qualitative_gan}
\end{figure}

\subsection{Comparison with related works}
\begin{figure*}
\centering
\begingroup
\setlength{\tabcolsep}{1pt} % Default value: 6pt
\renewcommand{\arraystretch}{1} % Default value: 1
\resizebox{\linewidth}{!}{
\begin{tabular}{m{1.0cm} m{2.2cm} m{2.2cm} m{2.2cm} m{2.2cm} m{2.2cm} m{2.2cm} m{2.2cm} m{2.2cm} m{2.2cm} m{2.2cm}}
    & 
    \multicolumn{3}{c}{\includegraphics[width=0.35\linewidth]{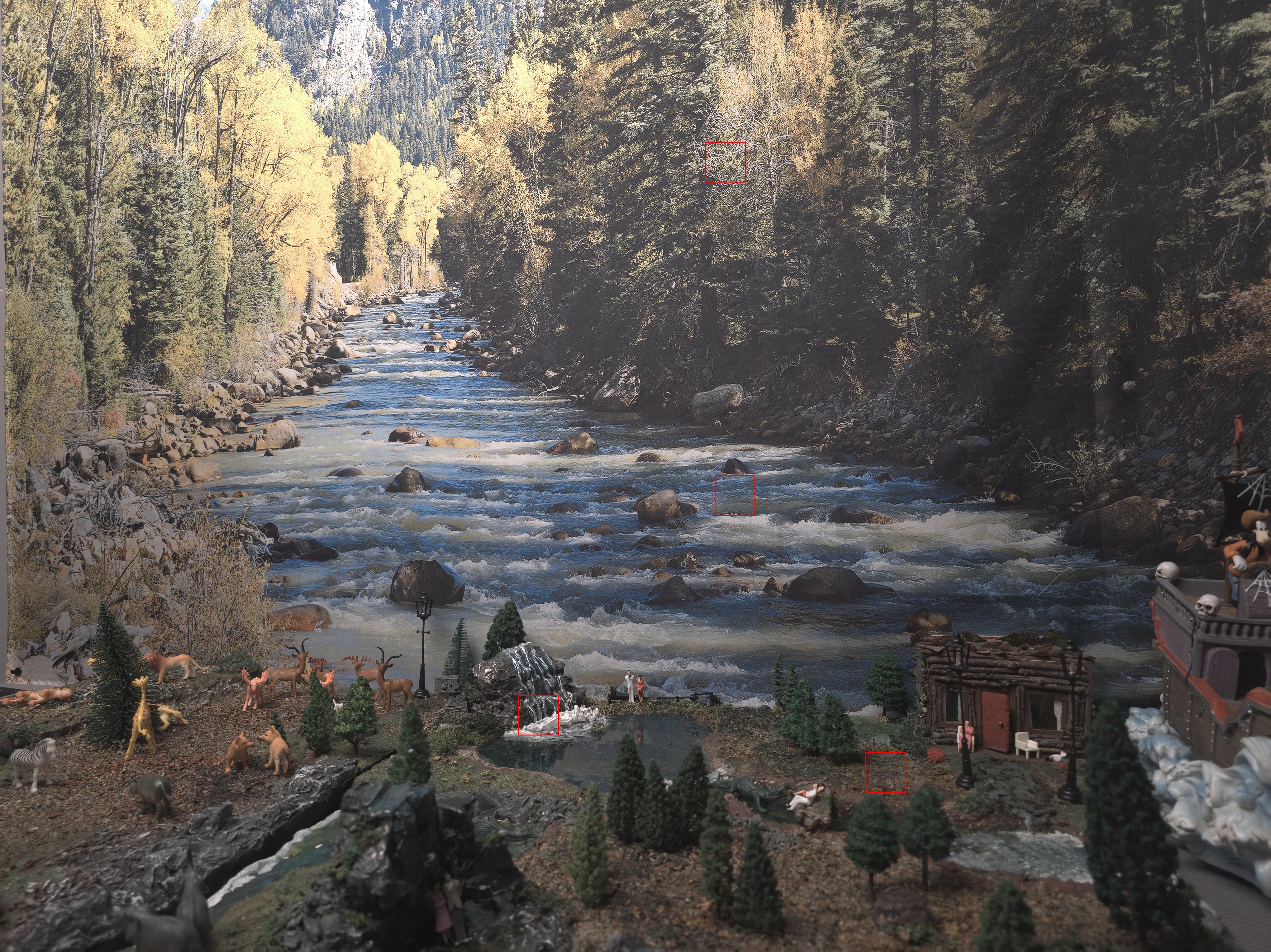}} &
    \multicolumn{3}{c}{\includegraphics[width=0.35\linewidth]{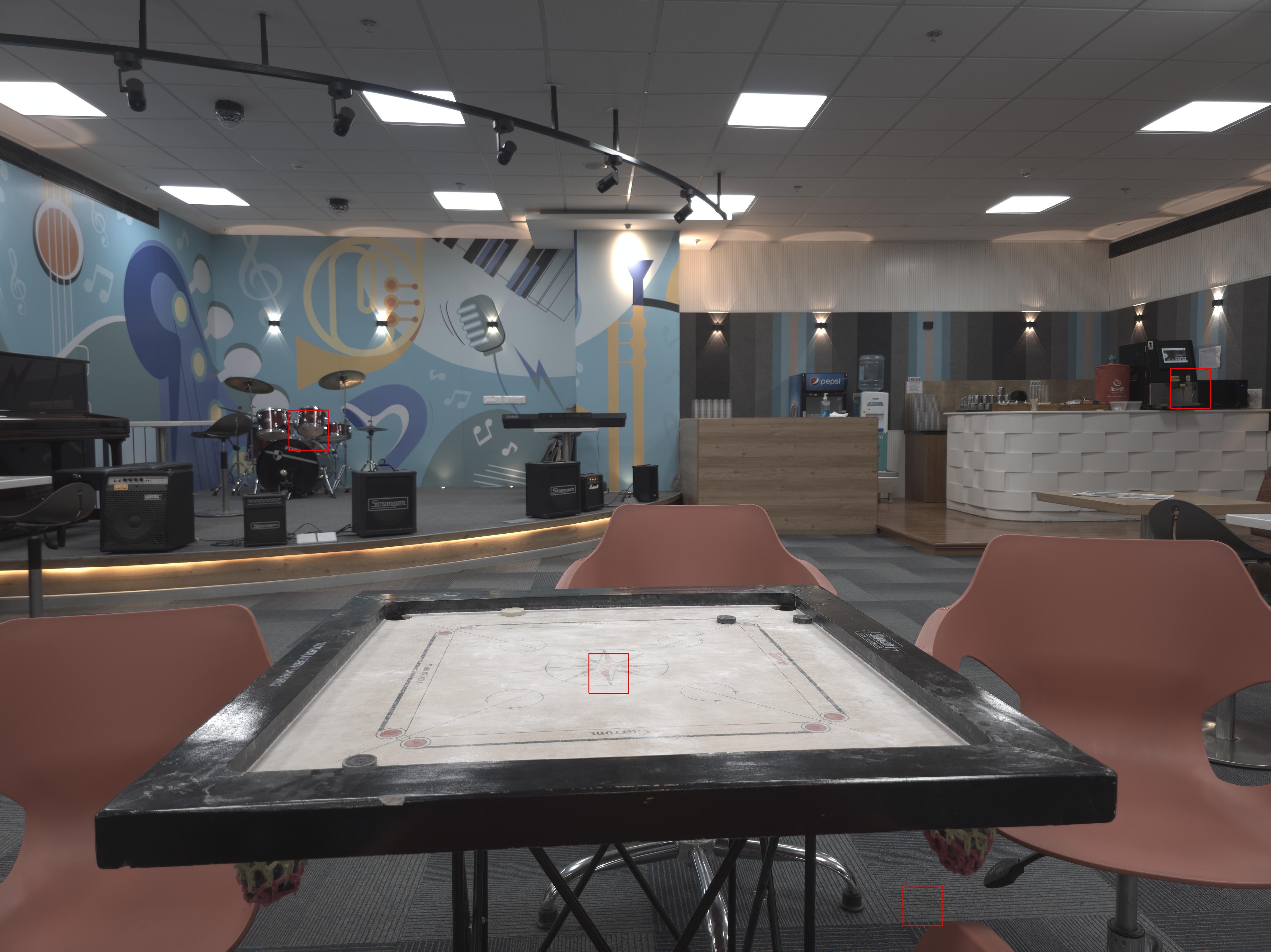}} &
    \multicolumn{3}{c}{\includegraphics[width=0.35\linewidth]{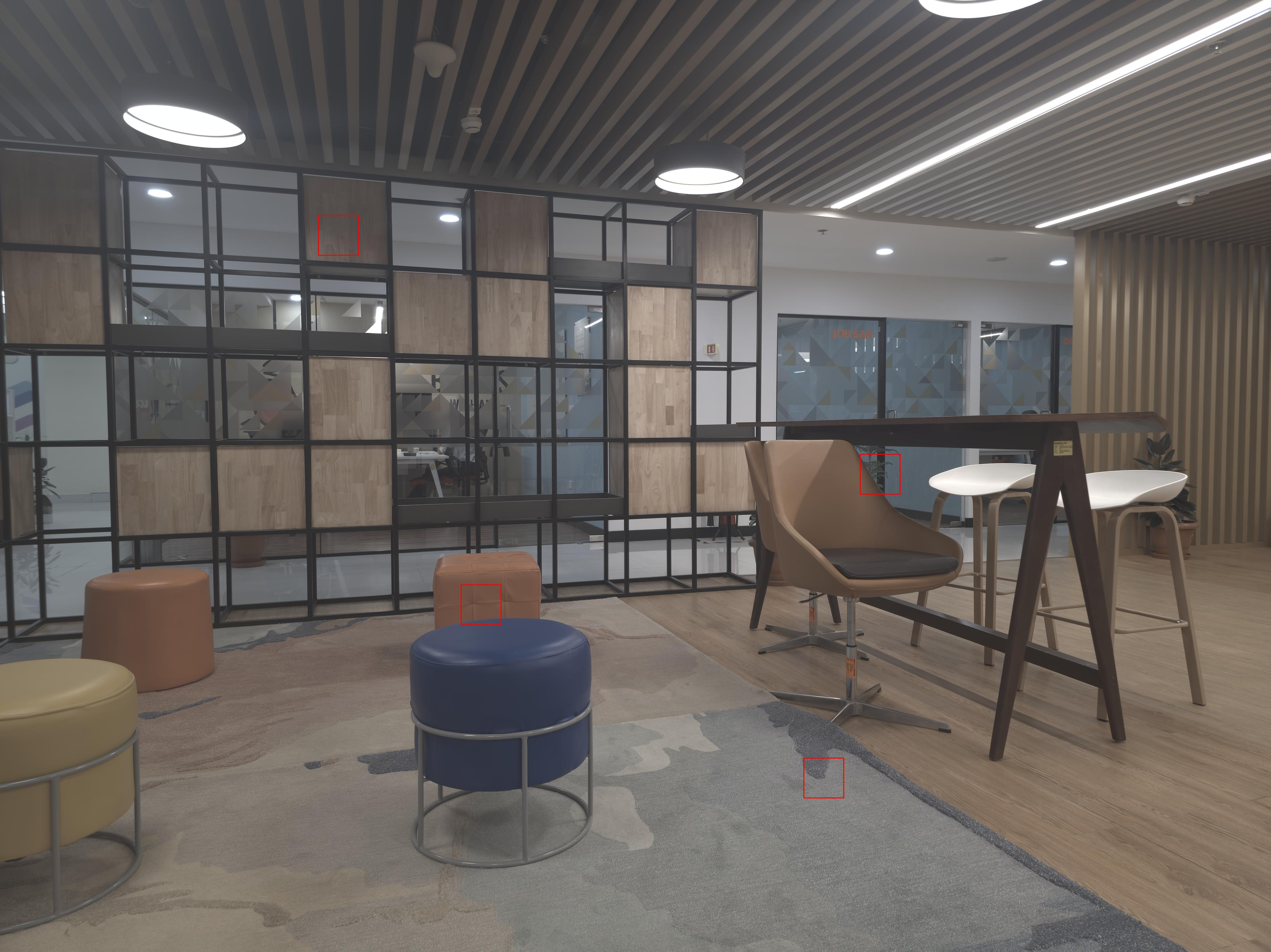}} \\

    & 
    \multicolumn{3}{c}{(a)} &
    \multicolumn{3}{c}{(b)} &
    \multicolumn{3}{c}{(c)} \\ [0.25cm]
    
    \multirow{4}{*}[0pt]{(a)}
    % &
    % \includegraphics[width=\linewidth]{figs/supp/1/0_raw_mean.jpg} & 
    % \includegraphics[width=\linewidth]{figs/supp/1/0_nafnet_0.5x.jpg} &
    % \includegraphics[width=\linewidth]{figs/supp/1/0_swinunet_0.5x.jpg} &
    % \includegraphics[width=\linewidth]{figs/supp/1/0_burstomer_0.5x.jpg} &
    % \includegraphics[width=\linewidth]{figs/supp/1/0_Restormer_0.5x.jpg} &
    % \includegraphics[width=\linewidth]{figs/supp/1/0_fbanet_0.5x.jpg} &
    % \includegraphics[width=\linewidth]{figs/supp/1/0_mprnet_0.5x.jpg} &
    % \includegraphics[width=\linewidth]{figs/supp/1/0_bipnet_0.5x.jpg} &
    % \includegraphics[width=\linewidth]{figs/supp/1/0_ours_0.5x.jpg} &
    % \includegraphics[width=\linewidth]{figs/supp/1/0_gt.jpg} \\
    
     &
    \includegraphics[width=\linewidth]{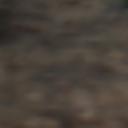} & 
    \includegraphics[width=\linewidth]{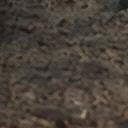} &
    \includegraphics[width=\linewidth]{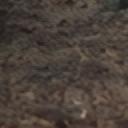} &
    \includegraphics[width=\linewidth]{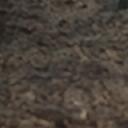} &
    \includegraphics[width=\linewidth]{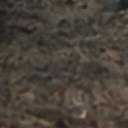} &
    \includegraphics[width=\linewidth]{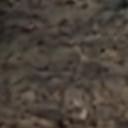} &
    \includegraphics[width=\linewidth]{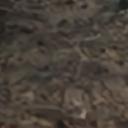} &
    \includegraphics[width=\linewidth]{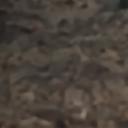} &
    \includegraphics[width=\linewidth]{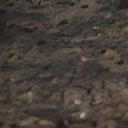} &
    \includegraphics[width=\linewidth]{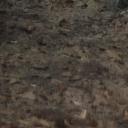} \\
    
     &
    \includegraphics[width=\linewidth]{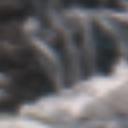} & 
    \includegraphics[width=\linewidth]{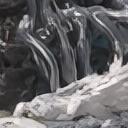} &
    \includegraphics[width=\linewidth]{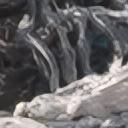} &
    \includegraphics[width=\linewidth]{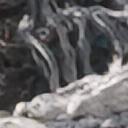} &
    \includegraphics[width=\linewidth]{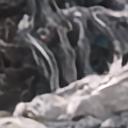} &
    \includegraphics[width=\linewidth]{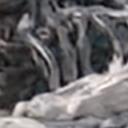} &
    \includegraphics[width=\linewidth]{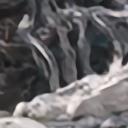} &
    \includegraphics[width=\linewidth]{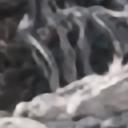} &
    \includegraphics[width=\linewidth]{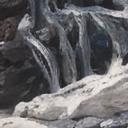} &
    \includegraphics[width=\linewidth]{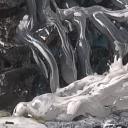} \\
    
     &
    \includegraphics[width=\linewidth]{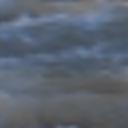} & 
    \includegraphics[width=\linewidth]{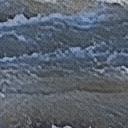} &
    \includegraphics[width=\linewidth]{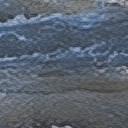} &
    \includegraphics[width=\linewidth]{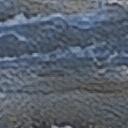} &
    \includegraphics[width=\linewidth]{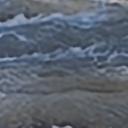} &
    \includegraphics[width=\linewidth]{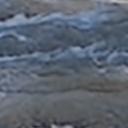} &
    \includegraphics[width=\linewidth]{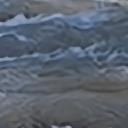} &
    \includegraphics[width=\linewidth]{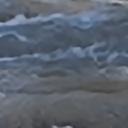} &
    \includegraphics[width=\linewidth]{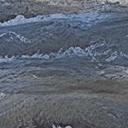} &
    \includegraphics[width=\linewidth]{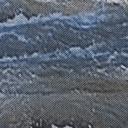} \\[1.5cm]

    \multirow{4}{*}[0pt]{(b)} 
    % &
    % \includegraphics[width=\linewidth]{figs/supp/2/0_raw_mean.jpg} & 
    % \includegraphics[width=\linewidth]{figs/supp/2/0_nafnet_0.5x.jpg} &
    % \includegraphics[width=\linewidth]{figs/supp/2/0_swinunet_0.5x.jpg} &
    % \includegraphics[width=\linewidth]{figs/supp/2/0_burstomer_0.5x.jpg} &
    % \includegraphics[width=\linewidth]{figs/supp/2/0_Restormer_0.5x.jpg} &
    % \includegraphics[width=\linewidth]{figs/supp/2/0_fbanet_0.5x.jpg} &
    % \includegraphics[width=\linewidth]{figs/supp/2/0_mprnet_0.5x.jpg} &
    % \includegraphics[width=\linewidth]{figs/supp/2/0_bipnet_0.5x.jpg} &
    % \includegraphics[width=\linewidth]{figs/supp/2/0_ours_0.5x.jpg} &
    % \includegraphics[width=\linewidth]{figs/supp/2/0_gt.jpg} \\
    
     &
    \includegraphics[width=\linewidth]{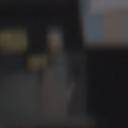} & 
    \includegraphics[width=\linewidth]{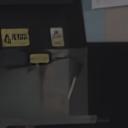} &
    \includegraphics[width=\linewidth]{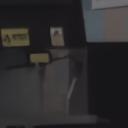} &
    \includegraphics[width=\linewidth]{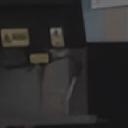} &
    \includegraphics[width=\linewidth]{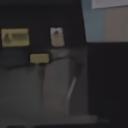} &
    \includegraphics[width=\linewidth]{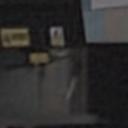} &
    \includegraphics[width=\linewidth]{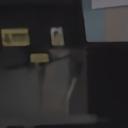} &
    \includegraphics[width=\linewidth]{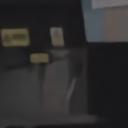} &
    \includegraphics[width=\linewidth]{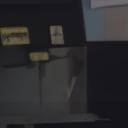} &
    \includegraphics[width=\linewidth]{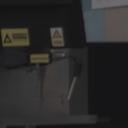} \\
    
     &
    \includegraphics[width=\linewidth]{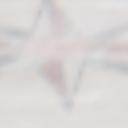} & 
    \includegraphics[width=\linewidth]{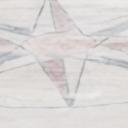} &
    \includegraphics[width=\linewidth]{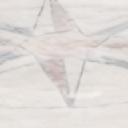} &
    \includegraphics[width=\linewidth]{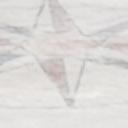} &
    \includegraphics[width=\linewidth]{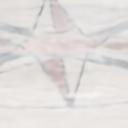} &
    \includegraphics[width=\linewidth]{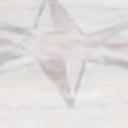} &
    \includegraphics[width=\linewidth]{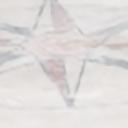} &
    \includegraphics[width=\linewidth]{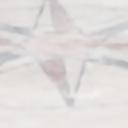} &
    \includegraphics[width=\linewidth]{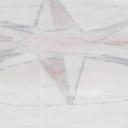} &
    \includegraphics[width=\linewidth]{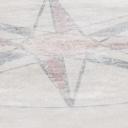} \\
    
     &
    \includegraphics[width=\linewidth]{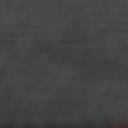} & 
    \includegraphics[width=\linewidth]{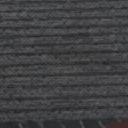} &
    \includegraphics[width=\linewidth]{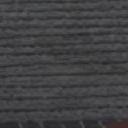} &
    \includegraphics[width=\linewidth]{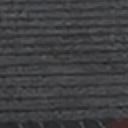} &
    \includegraphics[width=\linewidth]{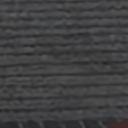} &
    \includegraphics[width=\linewidth]{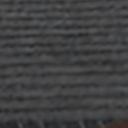} &
    \includegraphics[width=\linewidth]{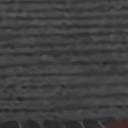} &
    \includegraphics[width=\linewidth]{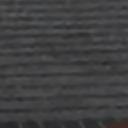} &
    \includegraphics[width=\linewidth]{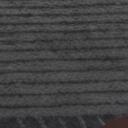} &
    \includegraphics[width=\linewidth]{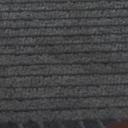} \\[1.5cm]

    \multirow{4}{*}[0pt]{(c)} &
    \includegraphics[width=\linewidth]{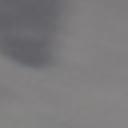} & 
    \includegraphics[width=\linewidth]{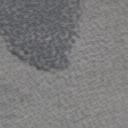} &
    \includegraphics[width=\linewidth]{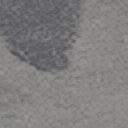} &
    \includegraphics[width=\linewidth]{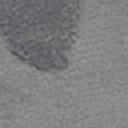} &
    \includegraphics[width=\linewidth]{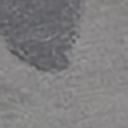} &
    \includegraphics[width=\linewidth]{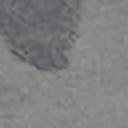} &
    \includegraphics[width=\linewidth]{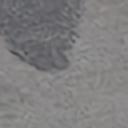} &
    \includegraphics[width=\linewidth]{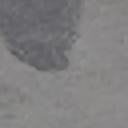} &
    \includegraphics[width=\linewidth]{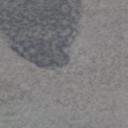} &
    \includegraphics[width=\linewidth]{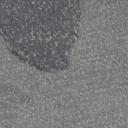} \\
    
     &
    \includegraphics[width=\linewidth]{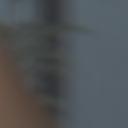} & 
    \includegraphics[width=\linewidth]{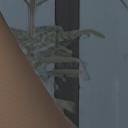} &
    \includegraphics[width=\linewidth]{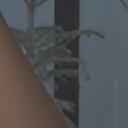} &
    \includegraphics[width=\linewidth]{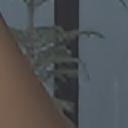} &
    \includegraphics[width=\linewidth]{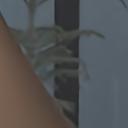} &
    \includegraphics[width=\linewidth]{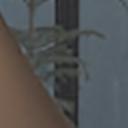} &
    \includegraphics[width=\linewidth]{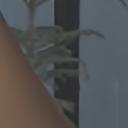} &
    \includegraphics[width=\linewidth]{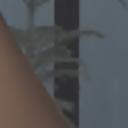} &
    \includegraphics[width=\linewidth]{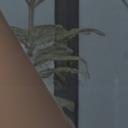} &
    \includegraphics[width=\linewidth]{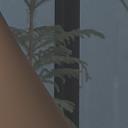} \\
    
    %  &
    % \includegraphics[width=\linewidth]{figs/supp/3/2_raw_mean.jpg} & 
    % \includegraphics[width=\linewidth]{figs/supp/3/2_nafnet_0.5x.jpg} &
    % \includegraphics[width=\linewidth]{figs/supp/3/2_swinunet_0.5x.jpg} &
    % \includegraphics[width=\linewidth]{figs/supp/3/2_burstomer_0.5x.jpg} &
    % \includegraphics[width=\linewidth]{figs/supp/3/2_Restormer_0.5x.jpg} &
    % \includegraphics[width=\linewidth]{figs/supp/3/2_fbanet_0.5x.jpg} &
    % \includegraphics[width=\linewidth]{figs/supp/3/2_mprnet_0.5x.jpg} &
    % \includegraphics[width=\linewidth]{figs/supp/3/2_bipnet_0.5x.jpg} &
    % \includegraphics[width=\linewidth]{figs/supp/3/2_ours_0.5x.jpg} &
    % \includegraphics[width=\linewidth]{figs/supp/3/2_gt.jpg} \\
    
     &
    \includegraphics[width=\linewidth]{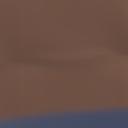} & 
    \includegraphics[width=\linewidth]{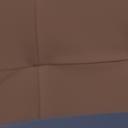} &
    \includegraphics[width=\linewidth]{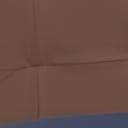} &
    \includegraphics[width=\linewidth]{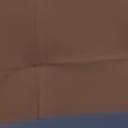} &
    \includegraphics[width=\linewidth]{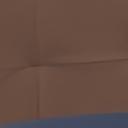} &
    \includegraphics[width=\linewidth]{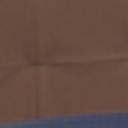} &
    \includegraphics[width=\linewidth]{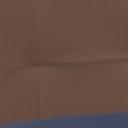} &
    \includegraphics[width=\linewidth]{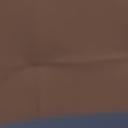} &
    \includegraphics[width=\linewidth]{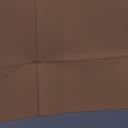} &
    \includegraphics[width=\linewidth]{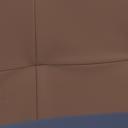} \\[1.5cm]
    
     & Input mean & NAFNet \cite{chen2022nafnet} & SwinUnet \cite{liu2021swin} & Burstormer \cite{dudhane2023burstormer} & Restormer \cite{zamir2022restormer} & FBANet \cite{wei2023fbanet} & MPRNet \cite{zamir2021mprnet} & BIPNet \cite{dudhane2022burst} & Ours  & GT\\
\end{tabular}
}
\endgroup
\caption{Qualitative comparison of different adversarial learning based postprocessing algorithms for generating RGB images using multi-frame RAW frames as input. Best viewed zoomed in.}
\label{fig:qualitative_adv_supp}
\end{figure*}

\begin{figure*}
\centering
\begingroup
\setlength{\tabcolsep}{1pt} % Default value: 6pt
\renewcommand{\arraystretch}{1} % Default value: 1
\resizebox{0.8\linewidth}{!}{
\begin{tabular}{m{1.0cm} m{2.8cm} m{2.8cm} m{2.8cm} m{2.8cm} m{2.8cm} m{2.8cm}}
    & 
    \multicolumn{3}{c}{\includegraphics[width=0.25\linewidth]{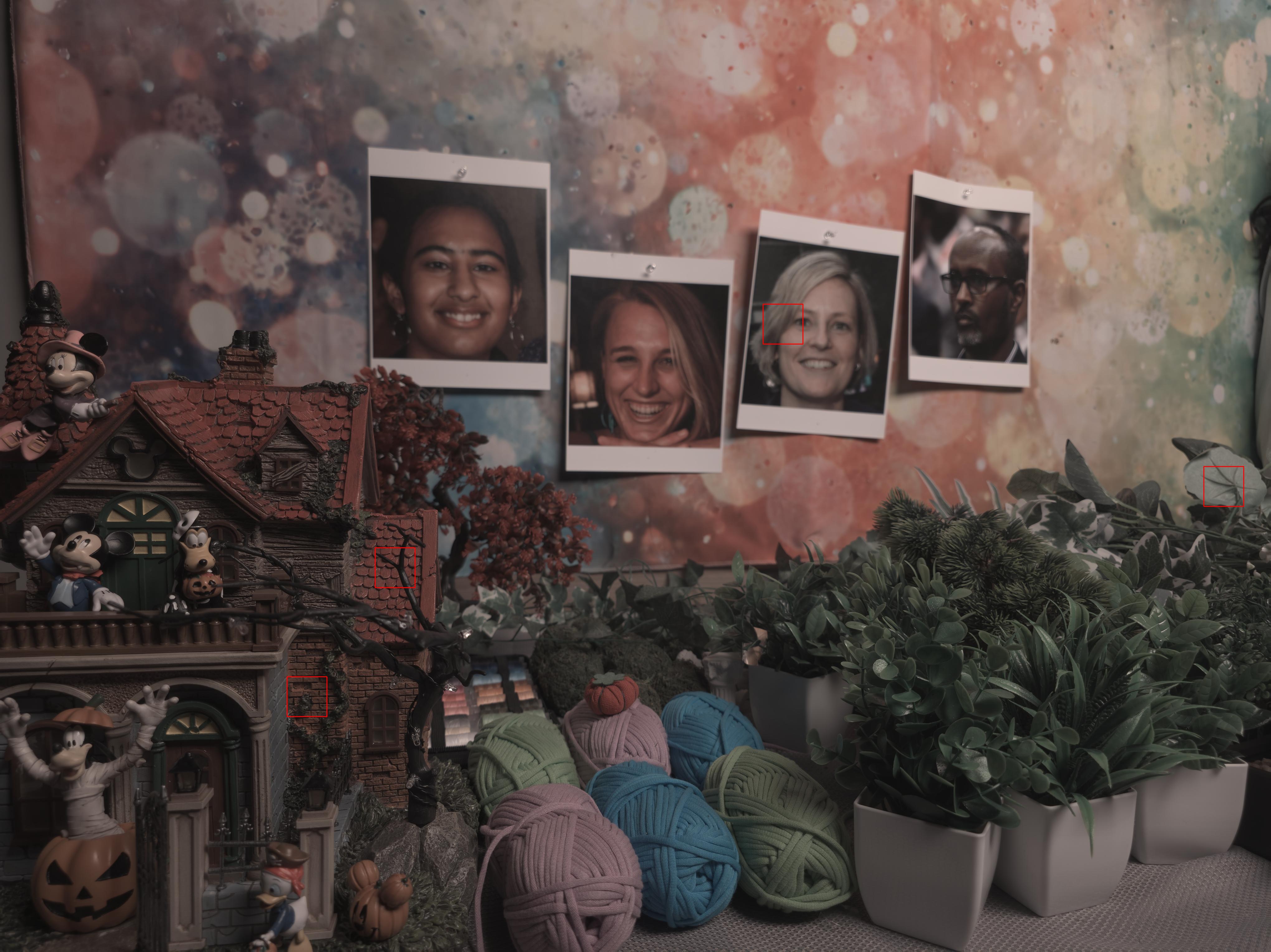}} &
    \multicolumn{3}{c}{\includegraphics[width=0.25\linewidth]{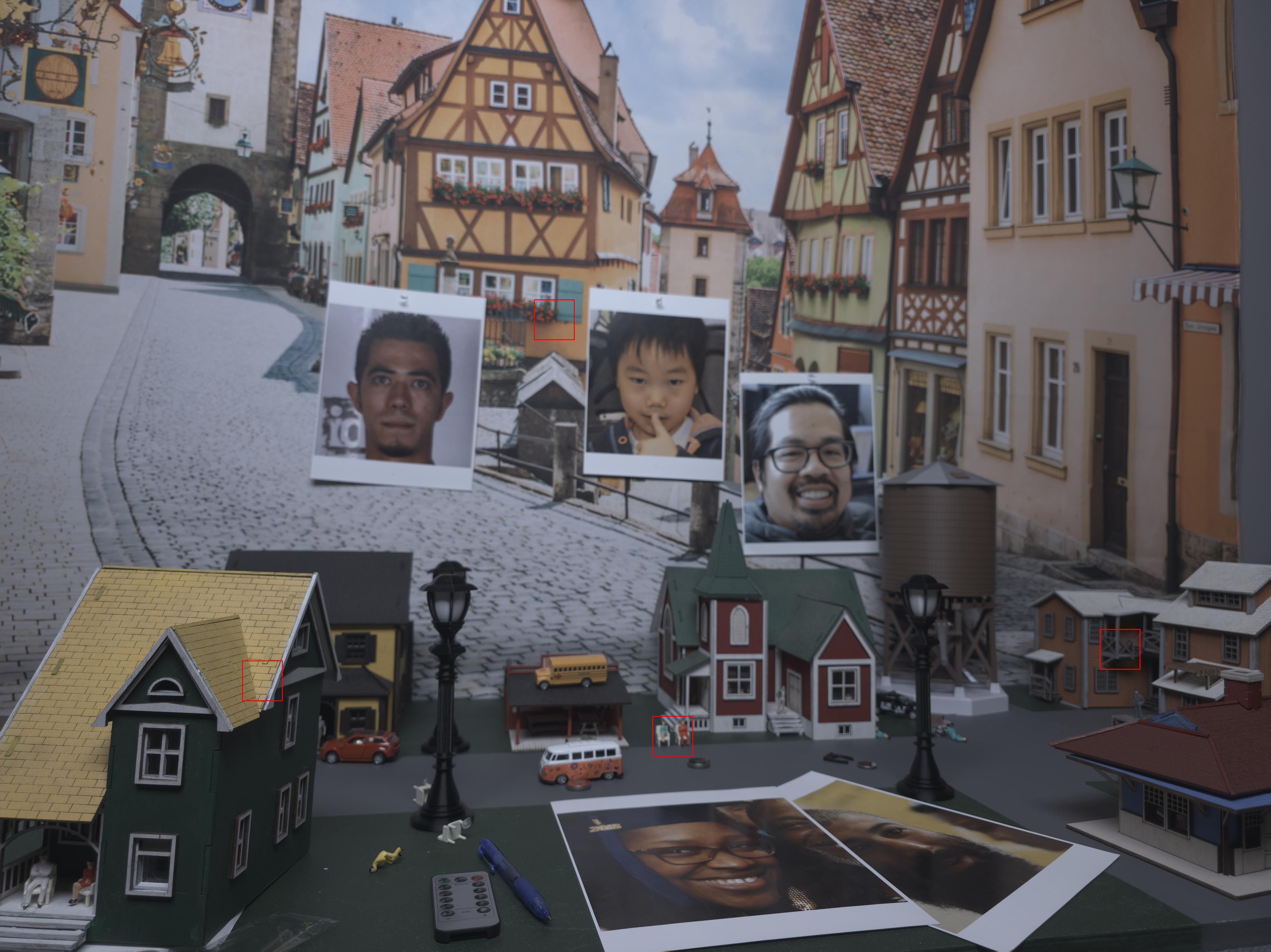}} \\

    & 
    \multicolumn{3}{c}{(a)} &
    \multicolumn{3}{c}{(b)} \\ 
    
    \multirow{4}{*}[0pt]{(a)} &
    \includegraphics[width=\linewidth]{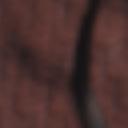} & 
    \includegraphics[width=\linewidth]{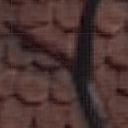} &
    \includegraphics[width=\linewidth]{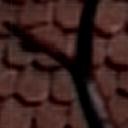} &
    \includegraphics[width=\linewidth]{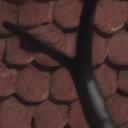} &
    \includegraphics[width=\linewidth]{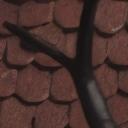} &
    \includegraphics[width=\linewidth]{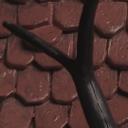} \\
    &
    \includegraphics[width=\linewidth]{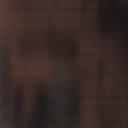} & 
    \includegraphics[width=\linewidth]{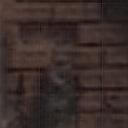} &
    \includegraphics[width=\linewidth]{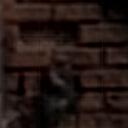} &
    \includegraphics[width=\linewidth]{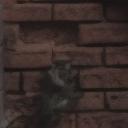} &
    \includegraphics[width=\linewidth]{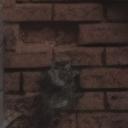} &
    \includegraphics[width=\linewidth]{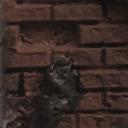} \\
    &
    \includegraphics[width=\linewidth]{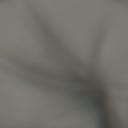} & 
    \includegraphics[width=\linewidth]{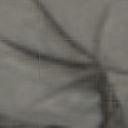} &
    \includegraphics[width=\linewidth]{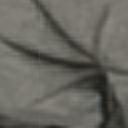} &
    \includegraphics[width=\linewidth]{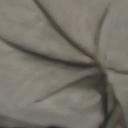} &
    \includegraphics[width=\linewidth]{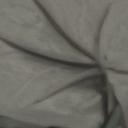} &
    \includegraphics[width=\linewidth]{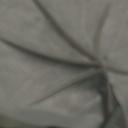} \\
    &
    \includegraphics[width=\linewidth]{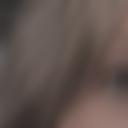} & 
    \includegraphics[width=\linewidth]{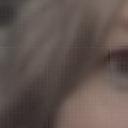} &
    \includegraphics[width=\linewidth]{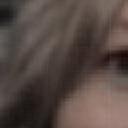} &
    \includegraphics[width=\linewidth]{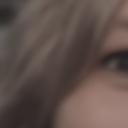} &
    \includegraphics[width=\linewidth]{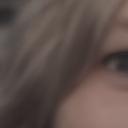} &
    \includegraphics[width=\linewidth]{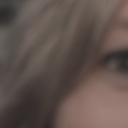} \\[1.5cm]
    
    \multirow{4}{*}[0pt]{(b)} &
    \includegraphics[width=\linewidth]{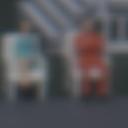} & 
    \includegraphics[width=\linewidth]{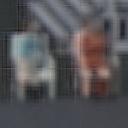} &
    \includegraphics[width=\linewidth]{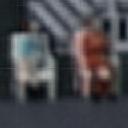} &
    \includegraphics[width=\linewidth]{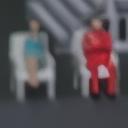} &
    \includegraphics[width=\linewidth]{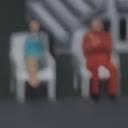} &
    \includegraphics[width=\linewidth]{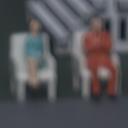} \\
    &
    \includegraphics[width=\linewidth]{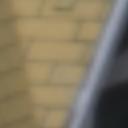} & 
    \includegraphics[width=\linewidth]{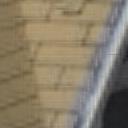} &
    \includegraphics[width=\linewidth]{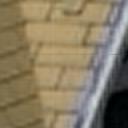} &
    \includegraphics[width=\linewidth]{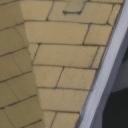} &
    \includegraphics[width=\linewidth]{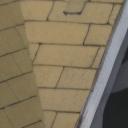} &
    \includegraphics[width=\linewidth]{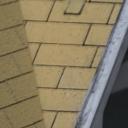} \\
    &
    \includegraphics[width=\linewidth]{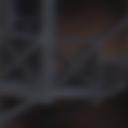} & 
    \includegraphics[width=\linewidth]{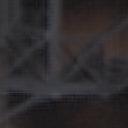} &
    \includegraphics[width=\linewidth]{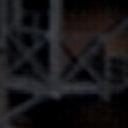} &
    \includegraphics[width=\linewidth]{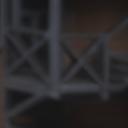} &
    \includegraphics[width=\linewidth]{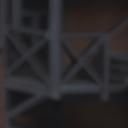} &
    \includegraphics[width=\linewidth]{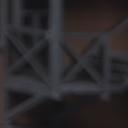} \\
    &
    \includegraphics[width=\linewidth]{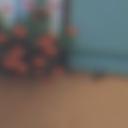} & 
    \includegraphics[width=\linewidth]{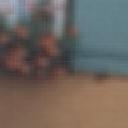} &
    \includegraphics[width=\linewidth]{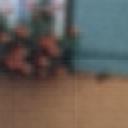} &
    \includegraphics[width=\linewidth]{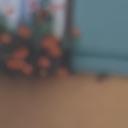} &
    \includegraphics[width=\linewidth]{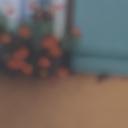} &
    \includegraphics[width=\linewidth]{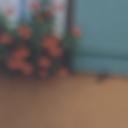} \\[1.5cm]
     & Input mean & InDI \cite{delbracio2023inversion} & ResShift \cite{yue2023resshift} & OSEDiff \cite{wu2024neurips_osediff} & Ours  & GT\\
\end{tabular}
}
\endgroup
\caption{Qualitative comparison of different diffusion-based postprocessing algorithms for generating RGB images using multi-frame RAW frames as input. Best viewed zoomed in.}
\label{fig:qualitative_diff_supp}
\end{figure*}

In addition to the qualitative comparison methods given in the main paper, we show the qualitative results from BIPNet \cite{dudhane2022burst}, FBANet \cite{wei2023fbanet}, MPRNet \cite{zamir2021mprnet}, Restormer \cite{zamir2022restormer}, Burstormer \cite{dudhane2023burstormer}, and ResShift \cite{yue2023resshift} in \cref{fig:qualitative_adv_supp} and \cref{fig:qualitative_diff_supp}.

\subsection{Super resolution agnostism}
\begin{figure}
    \centering
    \setlength{\tabcolsep}{1pt} % Default value: 6pt
    \renewcommand{\arraystretch}{1} % Default value: 1
    \resizebox{0.8\linewidth}{!}{
    \begin{tabular}{m{2.8cm} m{2.8cm} m{2.8cm}}
        \includegraphics[width=\linewidth]{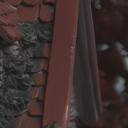} &
        \includegraphics[width=\linewidth]{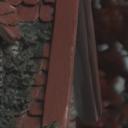} &
        \includegraphics[width=\linewidth]{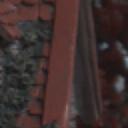} \\
        \includegraphics[width=\linewidth]{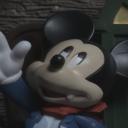} &
        \includegraphics[width=\linewidth]{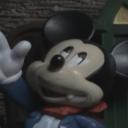} &
        \includegraphics[width=\linewidth]{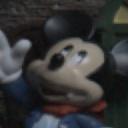} \\
        \includegraphics[width=\linewidth]{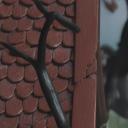} &
        \includegraphics[width=\linewidth]{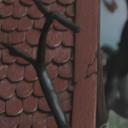} &
        \includegraphics[width=\linewidth]{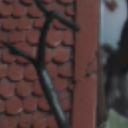} \\
        \includegraphics[width=\linewidth]{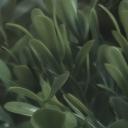} &
        \includegraphics[width=\linewidth]{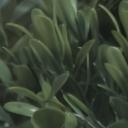} &
        \includegraphics[width=\linewidth]{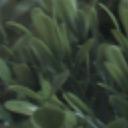} \\
        \includegraphics[width=\linewidth]{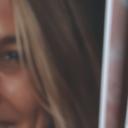} &
        \includegraphics[width=\linewidth]{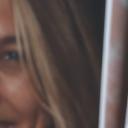} &
        \includegraphics[width=\linewidth]{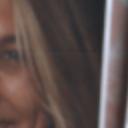} \\
        
        (a) $3\times$ SR & (b) $4\times$ SR & (c) $8\times$ SR \\
    \end{tabular}
    }
    \caption{Qualitative results of our method on $3\times$, $4\times$, and $8\times$ super resolution.}
    \label{fig:sr_agno}
\end{figure}

During the inference, we can use different upsampling scales since the model is trained to improve particular high frequencies. Thus, we show some zero-shot results from $3\times$ and $8\times$ super-resolution using the same model, which is trained to predict the $4\times$ super-resolution. The qualitative results in \cref{fig:sr_agno} demonstrate that our method achieves comparable performance across various super-resolution scales. 

% \begin{table}[h]
%     \centering
%     \caption{Performance metrics for $3\times$ super-resolution case.}
%     \begin{tabular}{lcccc}
%     \hline
%     \textbf{Method} & \textbf{LPIPS} & \textbf{FID} & \textbf{PSNR} & \textbf{SSIM} \\ \hline
%     GAN (NAFNet)        & 0.18 & 32.95 & 24.21 & 0.69 \\
%     GAN (SwinU)         & 0.18 & 41.72 & 23.95 & 0.67 \\ \hline
%     ResShift (NAFNet)   & & & & \\ 
%     InDI (NAFNet)       & & & & \\ 
%     VSD                 & 0.19 & 45.9 & 23.83 & 0.68 \\ 
%     Ours (w/o HF loss)  & 0.19 & 45.64 & 23.71 & 0.67 \\
%     Ours                & & & & \\ \hline
%     \end{tabular}
%     \label{tab:1.5x_comparison}
% \end{table}

\subsection{Other techniques for alignment}
\begin{figure}
    \centering
    \includegraphics[width=\linewidth]{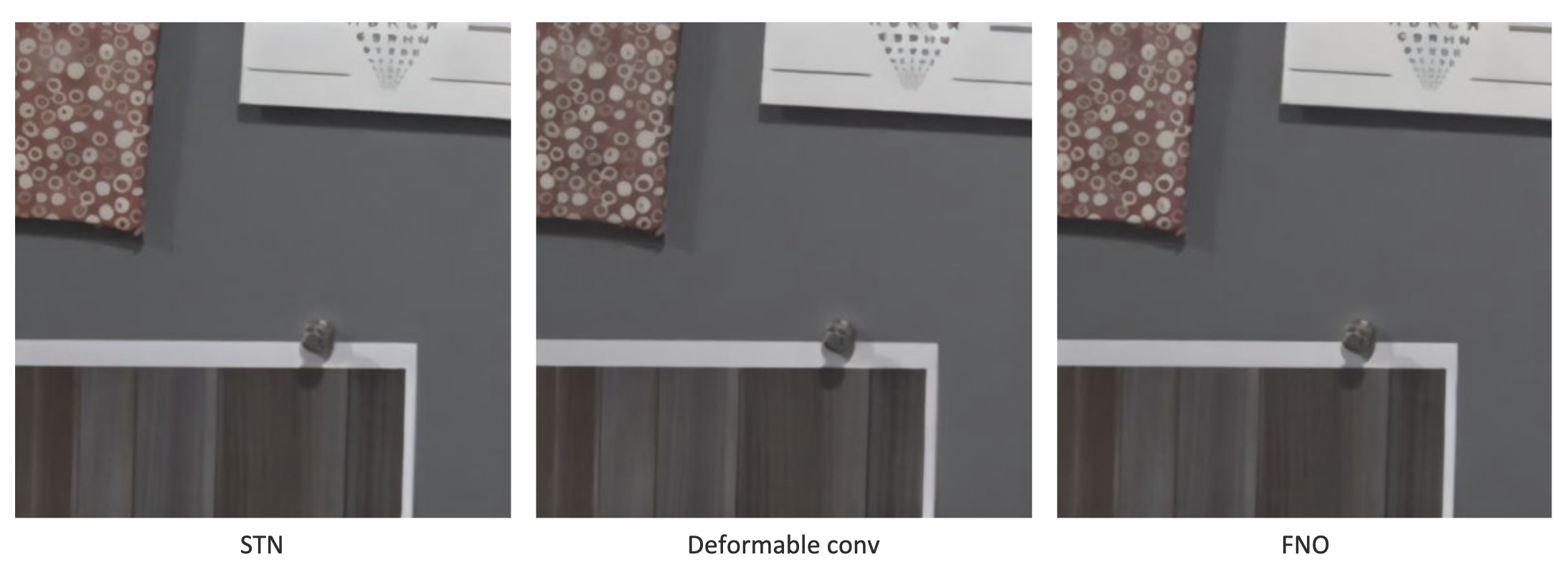}
    \caption{Performance of different alignment techniques used for RAW multi-frame alignment. STNs demonstrate slightly better visual quality when compared to other methods.}
    \label{fig:alignment}
\end{figure}

We trained the Variational Auto Encoder (VAE) separately to verify the quality of encoding using different alignment approaches. First, we used deformable convolution layers \cite{dai2017deformable} in intermediate steps of the encoder. However, this method is difficult to implement on hardware. We also tried to align RAW frames using Fourier Neural Operators (FNO) \cite{li2020fourier}, where its Fourier Layer performs a global convolution using the Fourier transform. We used multiple stacked FNO layers in the encoder for frame alignment. However, the STNs performed better and had less complexity when aligning frames, and thus, we used STNs as the alignment operator in the encoder. The VAE reconstruction performance of the STNs, Deformable convolution layers, and FNOs is illustrated in \cref{fig:alignment} where these VAEs are trained independently after initializing the weights using a pretrained Stable Diffusion 2.1 \cite{rombach2021highresolution} VAE. 

\section{User study}
\label{sec:user_study}
\begin{figure}
    \centering
    \setlength{\tabcolsep}{1pt} % Default value: 6pt
    \renewcommand{\arraystretch}{1} % Default value: 1
    \resizebox{0.7\linewidth}{!}{
    \begin{tabular}{m{2.8cm} m{2.8cm} m{2.8cm}}
        \includegraphics[width=\linewidth]{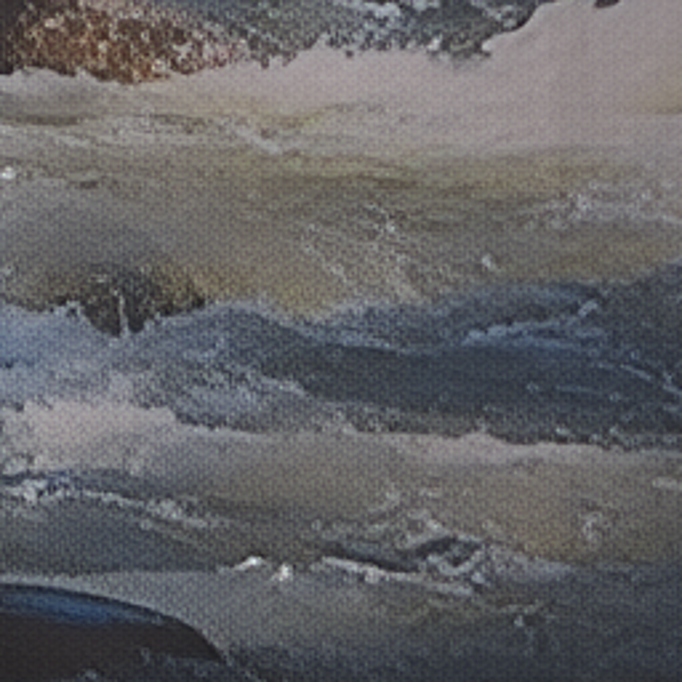} &
        \includegraphics[width=\linewidth]{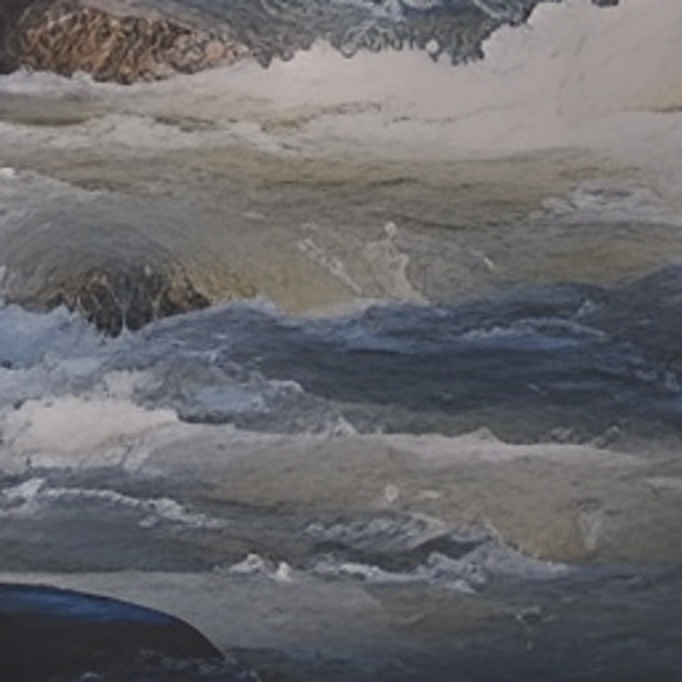} &
        \includegraphics[width=\linewidth]{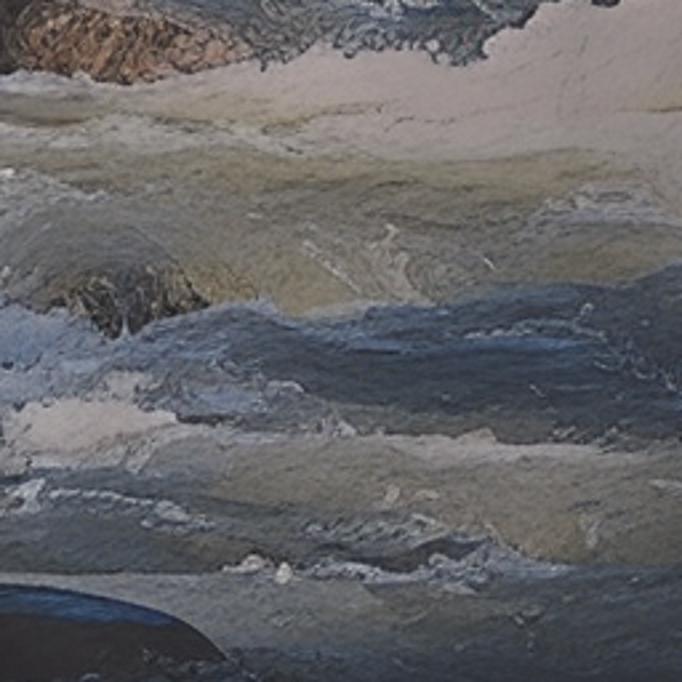} \\
        
        \includegraphics[width=\linewidth]{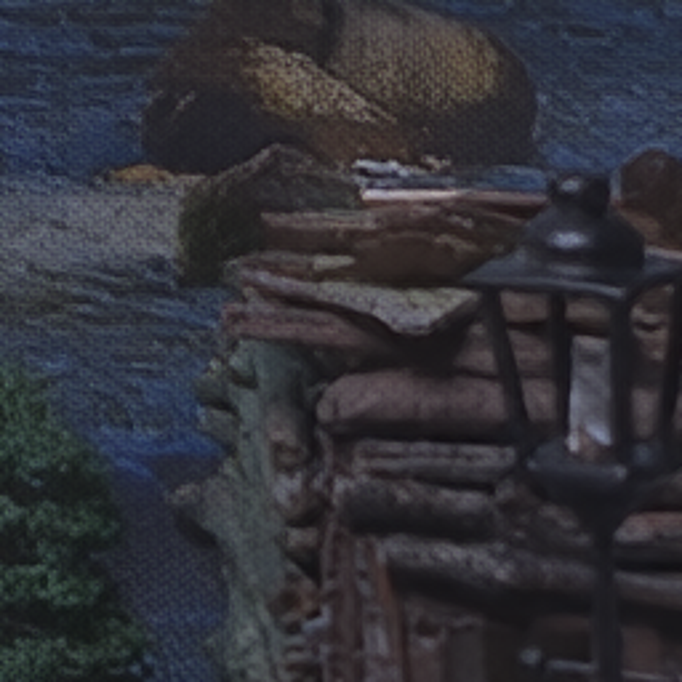} &
        \includegraphics[width=\linewidth]{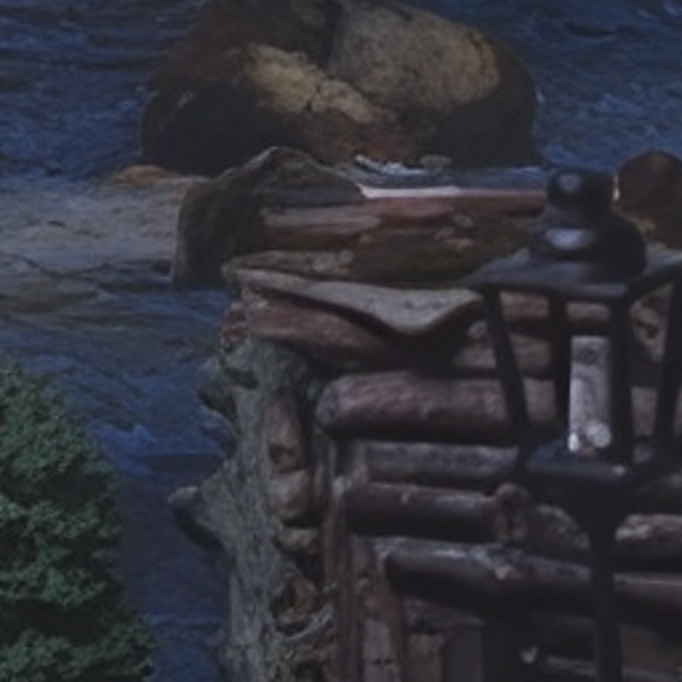} &
        \includegraphics[width=\linewidth]{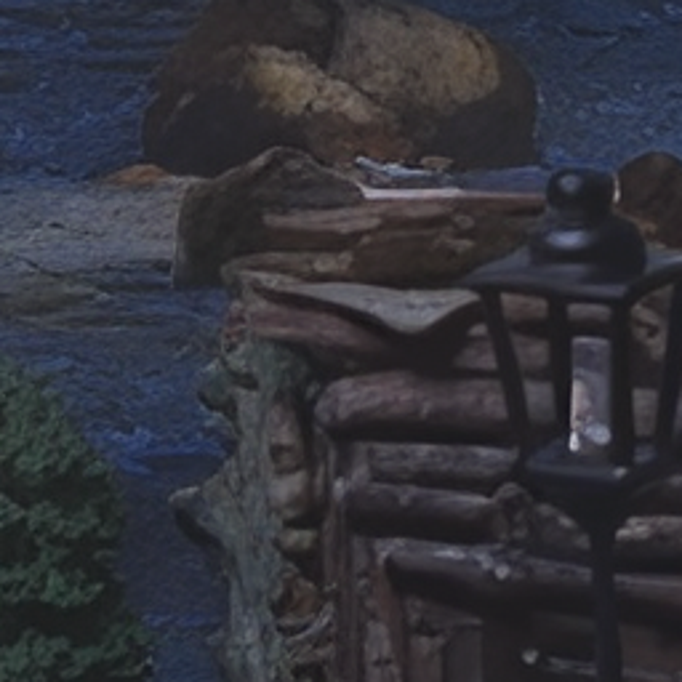} \\
        
        \includegraphics[width=\linewidth]{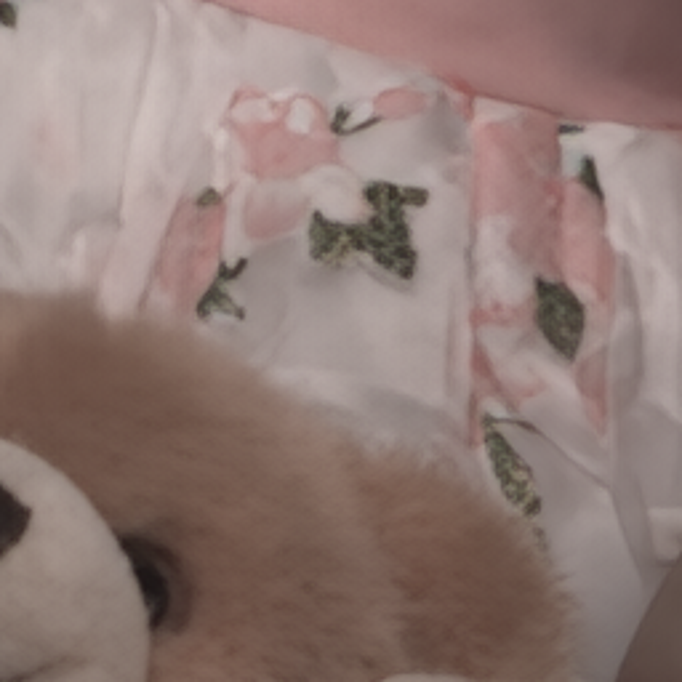} &
        \includegraphics[width=\linewidth]{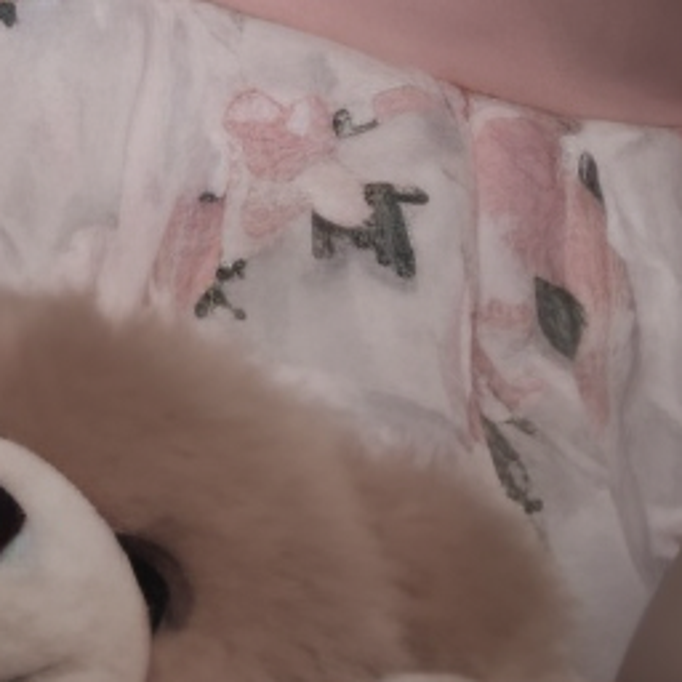} &
        \includegraphics[width=\linewidth]{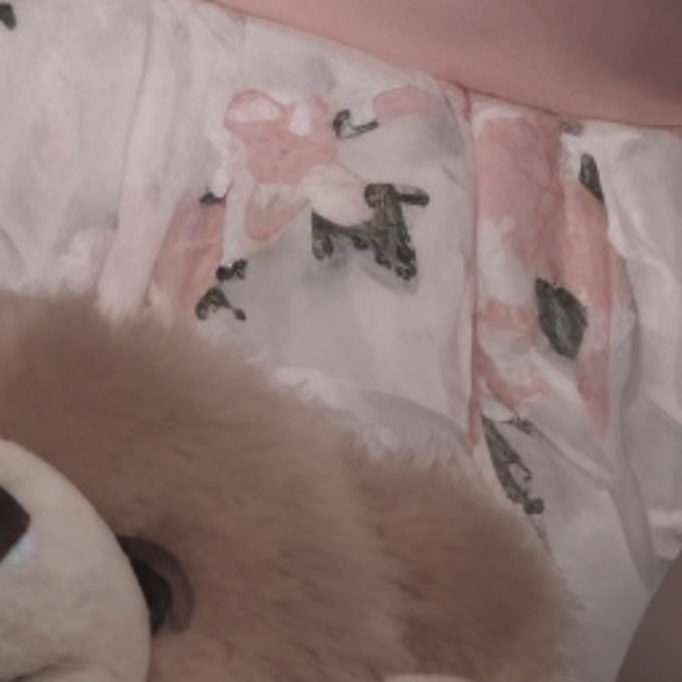} \\
        
        \includegraphics[width=\linewidth]{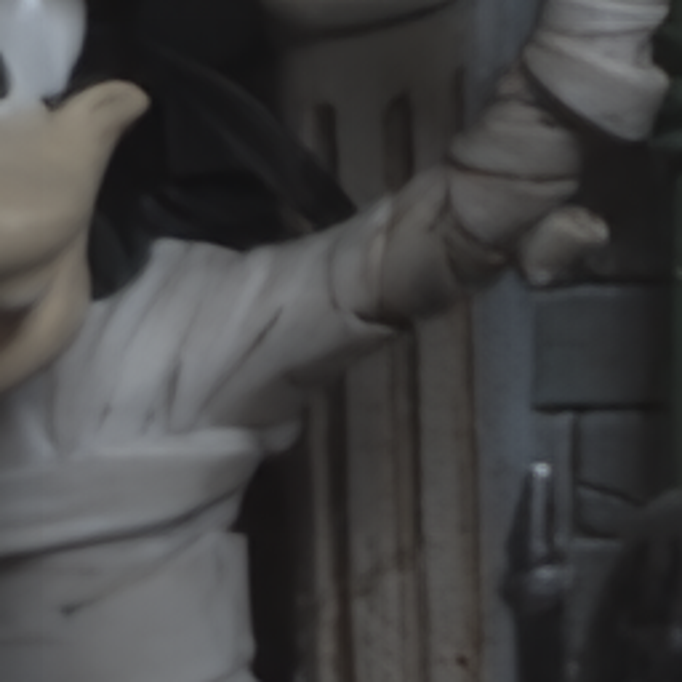} &
        \includegraphics[width=\linewidth]{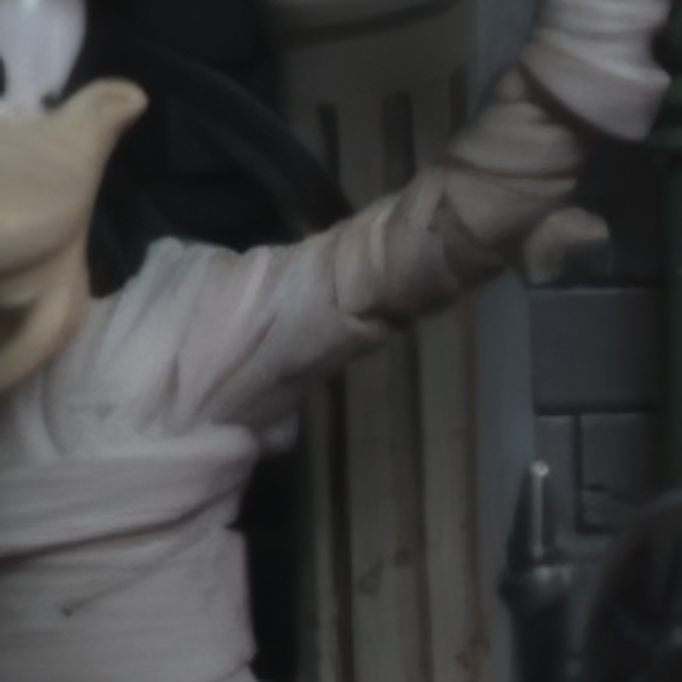} &
        \includegraphics[width=\linewidth]{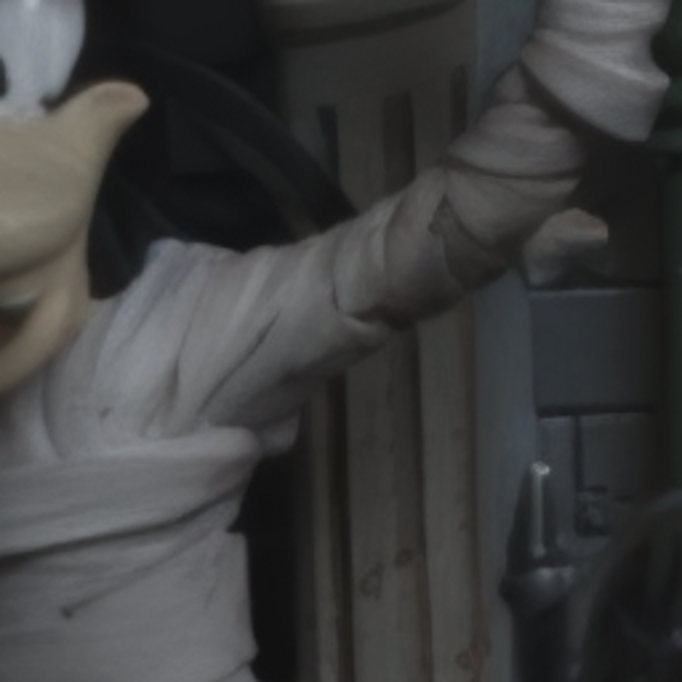} \\
        
        \includegraphics[width=\linewidth]{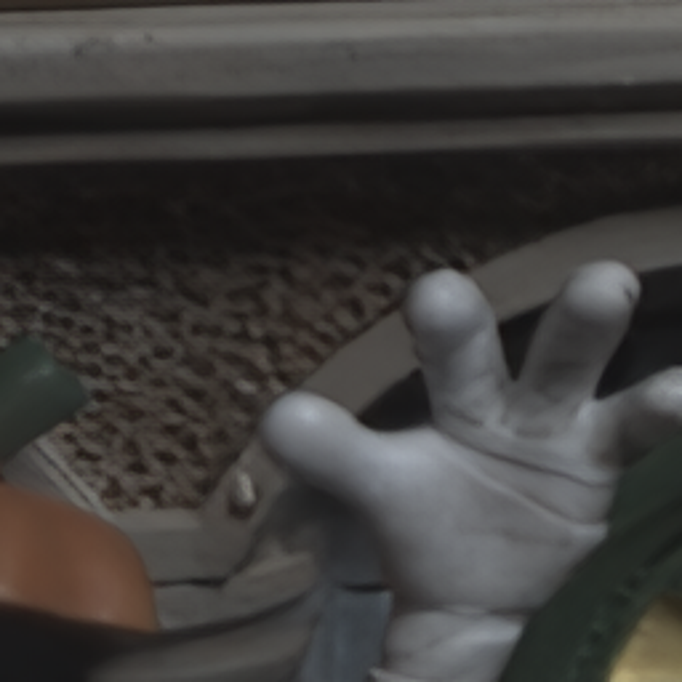} &
        \includegraphics[width=\linewidth]{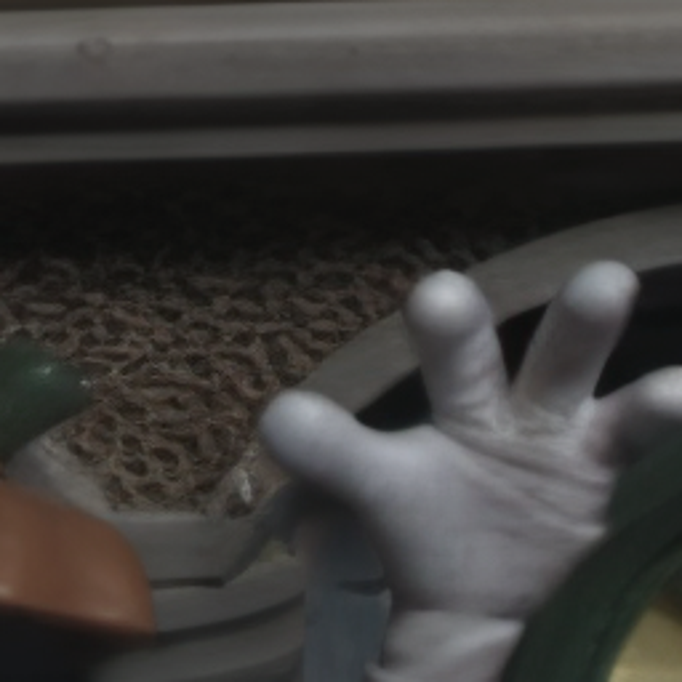} &
        \includegraphics[width=\linewidth]{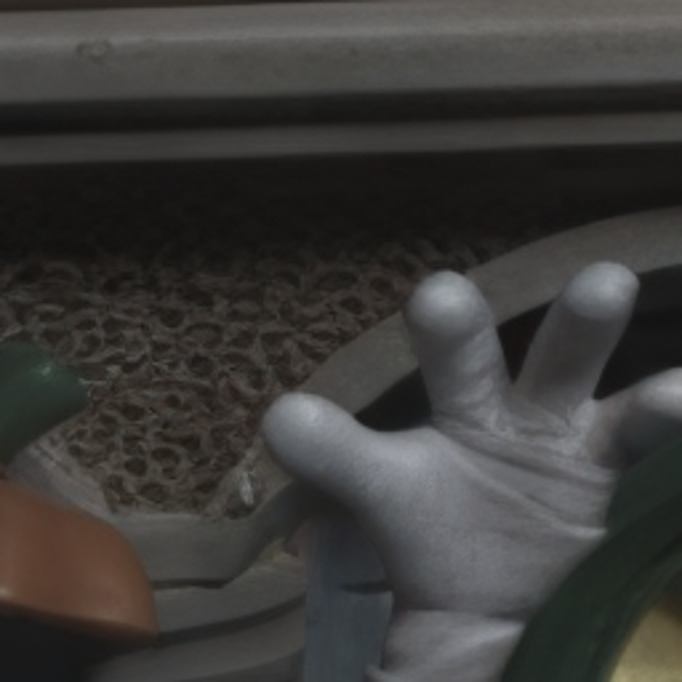} \\
        
        \includegraphics[width=\linewidth]{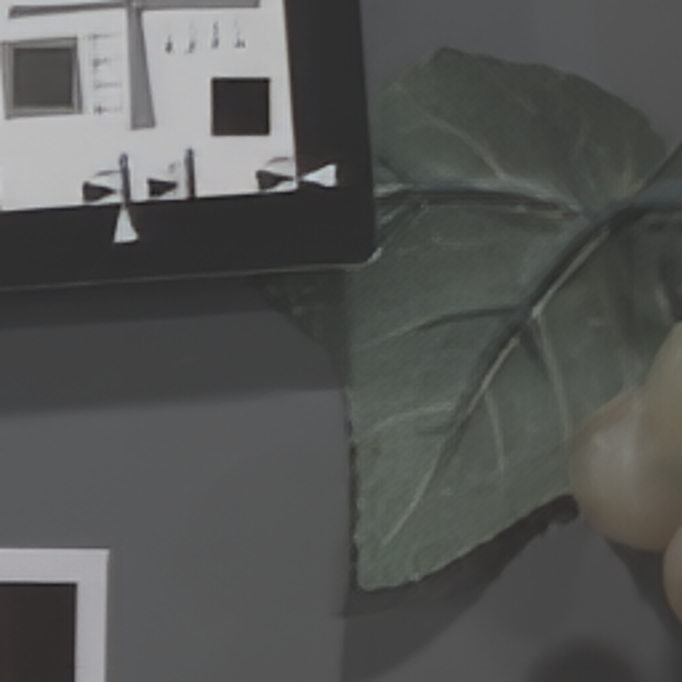} &
        \includegraphics[width=\linewidth]{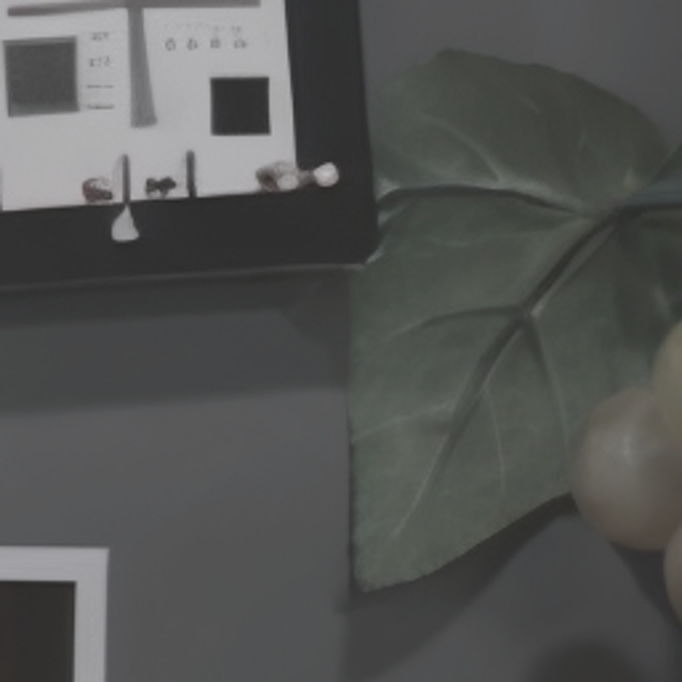} &
        \includegraphics[width=\linewidth]{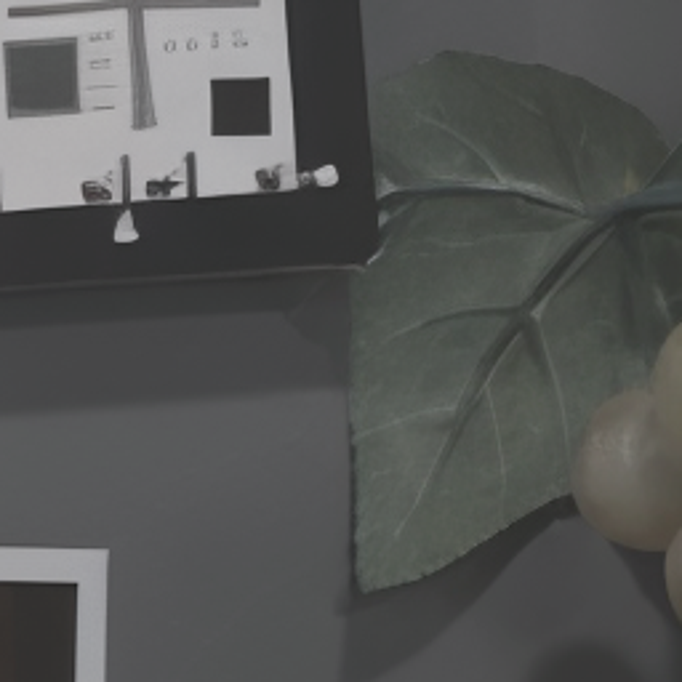} \\

        (a) NAFNet & (b) VSD & (c) Ours \\
    \end{tabular}
    }
    \caption{Samples of the figures used in the user study}
    \label{fig:user_study}
\end{figure}

For the user study, we used images from NAFNet (yielding the best quantitative results from adversarially trained methods), OSEDiff (utilizing VSD loss for image super-resolution), and our own method. 
We employed a two-level randomization design to ensure the study's validity. First, the presentation order of the 30 randomly selected crops was randomized for each subject. This standard procedure mitigates rater fatigue, a known decline in evaluator precision and consistency over time, and prevents sequential bias, which could unfairly penalize a method consistently shown at the end of the study. Second, within each trial, the spatial order of the three outputs (NAFNet, VSD, and GenMFSR) was also randomized. This prevents positional bias, where a rater might develop a preference for a specific on-screen location (e.g., left, middle, or right) or memorize a pattern.
The images used in this study are shown in \cref{fig:user_study}.  

\section{Q\&A}
\begin{enumerate}
    \item \textbf{Clarification on the contributions.}
    The core contribution of this paper is the system-level implementation of the burst RAW-to-RGB problem. We utilize STNs for alignment and propose a new loss for high-frequency detail generation. We moved the VSD loss descriptions to the supplement. to avoid distracting the reader.
    \item \textbf{Further details about baseline architecture modifications to accommodate multi-frame inputs.}
    Changes to baselines are described in the experiments section.
    \item \textbf{There is no comparison with the RGB-to-RGB performance of these methods.}
    Raw-to-RGB vs RGB-to-RGB is an irrelevant question. Once the image is converted to the RGB domain, some information is already lost at the ISP stage, defeating the whole purpose of improving RAW-to-RGB conversion.
    \item \textbf{There are not enough details about the dataset.}
    Dataset collection details are available in the methodology section.
    \item \textbf{While the paper highlights the use of “single-step diffusion” to enhance computational efficiency, it does not present concrete metrics such as inference-time latency, FLOPs, or memory usage.}
    Refer to \cref{tab:params_flops} in the main manuscript.
    % \item The experimental comparisons could be expanded to include a broader range of recent generative diffusion models.
    \item \textbf{There is a substantial gap in fidelity metrics compared to regression-based methods.} 
    The authors agree with the accusation. However, the purpose of the generative model is to generate sub-pixel information, which reduces the performance on fidelity metrics as shown in the literature. We show better qualitative results to counter this argument and also with better perception quality metrics further reinforce this claim.
    \item \textbf{The visual evidence provided in the ablation studies is unconvincing.}
    We have highlighted the visual results and quantitative results in the ablation study section.     
    % \item The architectural novelty appears incremental and relies heavily on component reuse. Rather than introducing an architectural breakthrough, the proposed framework primarily integrates existing, well-established modules, such as Spatial Transformer Networks (STNs) for alignment and Variational Score Distillation (VSD) for generation.
    % • We emphasize more on the “First Gen MFSR” claim. Our novelty is the system. Several people suggested using this as the main contribution.
    \item \textbf{The presentation quality and diagram clarity require improvement.}
    Improved aesthetic polish of the figures.
\end{enumerate}

\end{document}